\documentstyle[12pt]{article}
\makeatletter
\@addtoreset{equation}{section}
\makeatother

\newcommand{\be}{\begin{equation}}
\newcommand{\ee}{\end{equation}}
\newcommand{\bee}{\begin{eqnarray}}
\newcommand{\eee}{\end{eqnarray}}
\newcommand{\hsa}{$hu (2^{\N-1},2^{\N-1}|8)\quad$}
\newcommand{\csa}{$cu (2^{\N-1},2^{\N-1}|8)\quad$}
\newcommand{\hsan}{$hu_\ga (2^{\N-1},2^{\N-1}|8)\quad$}
\newcommand{\hsao}{$hu_0 (2^{\N-1},2^{\N-1}|8)\quad$}
\newcommand{\ga}{\alpha}
\newcommand{\gb}{\beta}
\newcommand{\gga}{\gamma}
\newcommand{\hgo}{\hat{\omega}}
\newcommand{\gd}{\delta}

\newcommand{\gep}{\epsilon}
\newcommand{\gvep}{\varepsilon}
\newcommand{\gs}{\sigma}
\newcommand{\go}{\omega}

\newcommand{\by}{{\bar{y}}}
\newcommand{\ova}{\tilde{a}}
\newcommand{\ovs}{\tilde{s}}
\newcommand{\ovt}{\tilde{t}}
\newcommand{\ovh}{\tilde{h}}
\newcommand{\ovj}{\tilde{j}}
\newcommand{\ovb}{\tilde{b}}
\newcommand{\q}{\,,\qquad}
\newcommand{\ls}{\!\!\!\!\!\!}

\newcommand{\dga}{{\dot{\alpha}}}
\newcommand{\dgb}{{\dot{\beta}}}
\newcommand{\dgga}{{\dot\gamma}}
\newcommand{\dgd}{{\dot\delta}}

\newcommand{\nn}{\nonumber}
\newcommand{\half}{\frac{1}{2}}
\newcommand{\ptl}{\partial}
\newcommand{\ep}{\mbox{$\wedge$}}

\newcommand{\un}{{\underline{n}}}
\newcommand{\unu}{{\underline{\nu}}}
\newcommand{\um}{{\underline{m}}}

\newcommand{\ui}{{\underline{i}}}
\newcommand{\uga}{{\underline{\alpha}}}
\newcommand{\udgb}{{\underline{\dgb}}}
\newcommand{\uj}{{\underline{j}}}
\newcommand{\uA}{{\underline{A}}}

\newcommand{\uhga}{{\underline{\hga}}}
\newcommand{\uhgb}{{\underline{\hgb}}}
\newcommand{\p}{\partial}

\newcommand{\N}{{\cal N}}
\newcommand{\D}{{\cal D}}

\newcommand{\f}{\frac}

\newcommand{\hd}{{\hat{d}}}
\newcommand{\hga}{{\hat{\ga}}}
\newcommand{\hgd}{{\hat{\delta}}}
\newcommand{\hgb}{{\hat{\gb}}}

\newcommand{\hgga}{{\hat{\gga}}}

\newcommand{\vac}{|0\rangle\langle 0|}
\newcommand{\bvac}{{|\bar{0}\rangle\langle \bar{0}|}}
\newcommand{\uvac}{|0_u\rangle\langle 0_u|}

\begin{document}

\begin{flushright}
\vspace{1mm}
%FIAN/TD/????16--00\\

{June 2001}\\
\end{flushright}

\vspace{1cm}

\begin{center}
{\large\bf      Conformal Higher Spin Symmetries
of $4d$  Massless Supermultiplets and $osp(L,2M)$ Invariant
Equations in Generalized (Super)Space}
\\
\vglue 1  true cm

\vspace{1cm}
{\bf  M.A.~Vasiliev }\footnote{e-mail: vasiliev@lpi.
ru}  \\
\vspace{1cm}

I.E.Tamm Department of Theoretical Physics, Lebedev Physical
Institute,\\
Leninsky prospect 53, 119991, Moscow, Russia
\vspace{1.5cm}
\end{center}

\begin{abstract}
Realization of the conformal higher spin symmetry on the $4d$ massless field
supermultiplets is given. The self-conjugated supermultiplets, including the
linearized ${\cal N}=4$ SYM theory, are considered in some detail. Duality
between non-unitary field-theoretical representations
and the unitary doubleton--type representations of the $4d$ conformal algebra
$su(2,2)$ is formulated in terms of a Bogolyubov transform.  The set of $4d$
massless fields of all spins is shown to form a representation of $sp(8)$.

The obtained results are extended to the generalized superspace invariant under
$osp(L, 2M)$ supersymmetries. World line particle interpretation of the free
higher spin theories in the $osp(2\N , 2M)$ invariant (super)space is given.
Compatible with unitarity free equations of motion in the $osp(L,2M)$ invariant
(super)space are formulated. A conjecture on the chain of $AdS_{d+1}/CFT_d
\rightarrow AdS_{d}/CFT_{d-1} \rightarrow  \ldots$ dualities in the higher spin
gauge theories is proposed.

\end{abstract}

\section{Introduction}

AdS/CFT correspondence \cite{AdS/CFT,FFA,GKP,AdS/CFTW,AdS/CFTrev}
relates theories of gravity in the $d+1$-dimensional anti-de Sitter
space  $AdS_{d+1}$ to
conformal theories in the $d$-dimensional
(conformal) boundary
space. Elementary fields
in the bulk are related to the currents in the boundary theory
associated with nonlinear colorless combinations of the elementary
boundary fields.

{}From the $d=4$ example it is known \cite{FV1,more} that
gauge theories of massless fields of all spins $0\leq s \leq \infty$
admit a consistent formulation in $AdS_4$  (see \cite{Gol,rev}
for more details and references on the higher spin gauge theories).
The cosmological constant $\Lambda = -\lambda^2$ should necessarily
be nonzero in the interacting higher spin gauge theories because it
appears in negative powers in the interaction terms that contain higher
derivatives of the higher spin gauge fields. This property is in
agreement with the fact that higher spin gauge
fields do not admit consistent
interactions with gravity in the flat background \cite{dif}.

Since the nonlinear higher spin gauge theory contains gravity and
is formulated in the AdS space-time, an interesting question
is what is its AdS/CFT dual. It was recently conjectured
\cite{W,Sun} that the boundary theories dual to the
$AdS_{d+1}$ higher spin gauge theories are free conformal theories.
These theories exhibit infinite-dimensional symmetries which
are expected to be isomorphic to the $AdS_{d+1}$ higher spin gauge
symmetries. This conjecture is in agreement with the results of
\cite{KVZ} where the conserved higher spin currents in
$d$-dimensional free scalar field theory were shown to be in the
one-to-one correspondence with the set of the 1-forms
associated with the totally symmetric higher spin gauge fields.
The AdS/CFT regime associated with the higher spin gauge theories
was conjectured \cite{W,Sun} to correspond to the limit $g^2 n \to 0$.
It is therefore opposite to the regime $g^2 n\to \infty $ underlying
the standard AdS/CFT correspondence \cite{AdS/CFT}, which relates
strongly coupled boundary theory to the classical regime of the bulk
theory.

To test the AdS/CFT correspondence for the higher spin gauge theories
it is instructive to realize the higher spin symmetries
of the bulk higher spin gauge theories in $AdS_{d+1}$
 as higher spin conformal symmetries of the free conformal fields
in $d$ dimensions. In the recent paper \cite{SV} this problem was solved
for the case of $AdS_4 /CFT_3$. In particular, it was shown in \cite{SV}
that $3d$ conformal matter fields are naturally described in terms of a
certain Fock module $F$ over the star product algebra identified \cite{Fort2}
with the $AdS_4$ higher spin algebra
\cite{FV,FrVa}. The results of \cite{SV} confirmed the conjecture
of Fradkin and Linetsky \cite{FL3} that $3d$ conformal higher spin
algebras are isomorphic to the $AdS_4$ higher spin algebras.
The non-unitary Fock module $F$ was interpreted in \cite{SV} as the
field-theoretical dual of the unitary singleton module over
$sp(4|{\bf R})$.

One of the aims of this paper is to extend the results of \cite{SV} to
 $AdS_5 /CFT_4$ higher spin correspondence which case is of
most interest from the string theory perspective. We present
a realization of the $4d$ conformal higher spin supermultiplets in
terms of the field-theoretical Fock modules (fiber bundles) dual to
the unitary doubleton \cite{dou} representations of $su(2,2)$.
The conformal equations of motion for a $4d$ massless supermultiplet
are formulated in the ``unfolded" form of the covariant constancy
conditions that makes the infinite-dimensional $4d$ conformal
higher spin symmetries manifest. We compare the obtained results with
the conjecture on the structure of the $4d$ conformal higher spin
symmetries made by Fradkin and Linetsky \cite{FLA,FL} in their analysis
 of $4d$ non-unitary higher spin conformal theories that
generalize the $C^2$ gravity, arriving at somewhat different
conclusions. Also, the obtained results are compared with
the conjecture
of the recent paper \cite{SSd} and the results of the forthcoming
papers \cite{5d,AV} on the (unitary) interacting higher spin theories
in $AdS_5$ (i.e., those referred to in the $AdS_5/CFT_4$
higher spin correspondence).

We show that the fundamental $4d$ conformal higher spin
algebras are the infinite-dimensional
algebras called $hu(m,n|8)$ in \cite{KV1}.
Here $n$ and $m$ refer to the spin 1 Yang-Mills symmetries
$u(m)\oplus u(n)$ while the label $8$ refers to the eight spinor generating
elements of the higher spin star product algebra. Let us recall the
definition of $hu(m,n|8)$. Consider the algebra of
$(m+n)\times (m+n)$ matrices
$$ \left( \begin{array}{cc}
                   A (a,b)   &   B(a,b)   \\
                   C (a,b)   &   D(a,b)
            \end{array}
    \right)
$$
\be
{}
\label{1}
\ee
with the even functions (polynomials) of the
auxiliary spinor variables $a_\hga$ and $b^\hga$
($\hga,\hgb =1\div 4$) in the
diagonal $m\times m$ block $A (a,b)$
and the $n\times n$ block $D (a,b)$,
\be
\label{p1}
A(-a, -b) = A(a,b)\,,\qquad
D(-a, -b) = D(a,b)\,,
\ee
and  odd functions in
the off-diagonal $m\times n$ block $B (a,b)$
and  $n\times m$ block $C (a,b)$,
\be
\label{p2}
B(-a, -b) = -B(a,b)\,,\qquad C(-a, -b) = - C(a,b)\,.
\ee
Consider the associative algebra of matrices of the form
(\ref{1}) with the associative
star product law for the functions of
the spinor variables $a_\hga$ and $b^\hgb$ defined as
\bee
\label{prod}
\!\! (f*g)(a,b)\!\!\!&=&\!\!\!\!\frac{1}{(\pi)^{8}}
\!\int\!\! d^{4}u d^{4} v d^{4}s d^{4} t
f(a\!+\!u ,b\!+\!t)g(a\!+\!s ,b\!+\!v\! )
\exp2(\!s_\hga t^\hga\!-\!u_\hga v^\hga)\nn\\
&=& e^{\half \left ( \f{\p^2}{\p s_\hga \p t^\hga} -
\f{\p^2}{\p u^\hga \p v_\hga} \right )} f (a+s , b+u )
g (a+v , b+t ) \Big |_{s=t=u=v=0}\,.
\eee
It is well-known that this star product
gives rise to the commutation relations
\be
\label{osc} [a_\hga ,b^\hgb]_*=\delta_\hga{}^\hgb \,,
\qquad [a_\hga ,a_\hgb ]_*=0\,,
\qquad [b^\hga ,b^\hgb]_*=0\,
\ee
with $[f,g]_* = f*g -g*f$.
The associative star product algebra with eight generating elements
$a_{\hat{\ga}}$
and $ b^{\hat{\gb}}$ is called Weyl algebra $A_4$ (i.e., $A_l$ for
$l$ pairs of oscillators.) The      particular star
product realization of the algebra
of oscillators we use describes the totally symmetric
(i.e., Weyl) ordering.  The matrices (\ref{1}) result from the
truncation of $A_4 \otimes Mat_{m+n}$ by the parity
conditions (\ref{p1}) and (\ref{p2}). Let us now treat this algebra
as ${\bf Z_2}$-graded algebra with even elements in the blocks $A$ and
$D$ and odd in $B$ and $C$, i.e.
\be
\label{pig}
\pi (A) = \pi (D) =0 \,,\qquad
\pi (B) = \pi (C) =1 \,.
\ee
The Lie superalgebra $hgl(m,n|8;{\bf C})$ is the
algebra of the same matrices with the product law defined via
the graded commutator
\be
[f,g]_\pm = f*g -(-1)^{\pi (f)\pi (g)} g * f \,.
\label{compm}
\ee
Note that the ${\bf Z}_2$ grading (\ref{pig}) in $hgl(m,n|8;{\bf C})$
is in accordance with the standard relationship between spin and
statistics once $a_\hga$ and $b^\hgb$ are interpreted as spinors.

The algebra $hu(m,n|8)$ is a particular real form of
$hgl(m,n|8;{\bf C})$ defined so that the
finite-dimensional subalgebra of $hu(m,n|8)$ identified
as the spin 1 Yang-Mills algebra, which is
spanned by the elements $A$ and $D$ independent of
the spinor elements $a_\hga$ and $b^\hgb$, is the compact
algebra $u(m)\oplus u(n)$.
The explicit form of the reality conditions imposed
to extract $hu(m,n|8)$ \cite{KV1} is given in section
\ref{Reality Conditions and Reduction} of this paper.

This construction is a
straightforward extension of the $3d$ conformal  $\sim\, AdS_4$
higher spin algebras $hu(m,n|4)$ via doubling of the spinor
generating elements. It is in accordance with the
conjecture of \cite{VF} that higher spin algebras in any
dimension are built in terms of the star product algebras
with spinor generating elements. The definition of $hu(m,n|2p)$
is analogous.

The Lie algebra $gl_4 $ is spanned by the bilinears
\be
\label{gl4}
T_\hga{}^\hgb = a_\hga b^\hgb \equiv \half
(a_\hga * b^\hgb  +b^\hgb * a_\hga ) I\,,
\ee
where $I$ is the unit element of the matrix part of $hu(m,n|8)$.
The central element is
\be
\label{Nb}
N_0 = a_\hga b^\hga \equiv \half ( a_\hga * b^\hga + b^\hga  * a_\hga ) I\,.
\ee
The traceless part
\be
\label{sl4}
t_\hga{}^\hgb = ( a_\hga b^\hgb -\frac{1}{4}\delta_\hga{}^\hgb N_0 )I
\ee
spans $sl_4$. The
$su(2,2)$ real form of $sl_4 ({\bf C})$ results from
 the reality conditions
\be
\label{re}
\bar{a}_\hga  =  b^\hgb C_{\hgb \hga}\,,\qquad
\bar{b}^\hga =  C^{\hga\hgb} a_\hgb \,,
\ee
where the overbar denotes complex conjugation while
$C_{\hga\hgb}=-C_{\hgb\hga}$ and $C^{\hga\hgb}=-C^{\hgb\hga}$
are some real antisymmetric matrices satisfying
\be
C_{\hga\hat{\gamma}} C^{\hgb\hat{\gamma}} = \delta _\hga^\hgb\,.
\ee

In order to incorporate supersymmetry one introduces the Clifford elements
$\phi_i$ and their complex conjugates
$\bar{\phi}^j$ ($i,j = 1\div {\cal N}$)
satisfying the commutation relations
\be
\label{fosc}
\{\phi_i , \phi_j \}_* =0\,,\qquad
\{\bar{\phi}^i , \bar{\phi}^j \}_* =0\,,\qquad
\{\phi_i , \bar{\phi}^j \}_* =\delta_i{}^j
\ee
with respect to the Clifford star product
\be
\label{cprod}
\!\!\!\!\!\! (f*g)(\phi,\bar{\phi})\!=\!\!2^\N
\!\!\int\!\! d^{\N}\psi d^{\N}\!\bar{\psi}  d^{\N}\!\chi d^{\N}\bar{\chi}
f(\phi \!+\psi ,\bar{\phi}\!+\!\bar{\chi})
g(\phi\!+\!\chi ,\bar{\phi}\!+\!\bar{\psi}\! )
\exp2(\!\psi_i \bar{\psi}^i \!-\!\chi_i \bar{\chi}^i)
\ee
with anticommuting $\phi_i$, $\bar{\phi}^i$, $\psi_i$, $\bar{\psi}^i$,
$\chi_i$ and $\bar{\chi}^i$.

The superalgebra $u(2,2|{\cal N})$ is spanned by the
$u(2,2)$ generators (\ref{gl4}) along with the supergenerators
\be
\label{Q}
Q_\hga^i = a_\hga \bar{\phi}^i \qquad \bar{Q}^\hgb_i = b^\hgb \phi_i \,
\ee
and $u({\cal N})$ generators
\be
\label{T}
T_i{}^j = \phi_i \bar{\phi}^j\,.
\ee
The central element $N_{\N} $ of $u(2,2|{\cal N})$ is
\be
\label{N}
N_{\N} = a_\hga b^\hga -\phi_i \bar{\phi}^i\,.
\ee
For $\N \neq 4$, $su(2,2|{\cal N})$=$u(2,2|{\cal N})/N_{\N}$.
The case of $\N =4$ is special because $N_{\N}$, which acts as the
unit operator on the oscillators, has trivial supertrace thus
generating an additional ideal in $su(2,2|{\cal N})$.
The corresponding simple quotient algebra is called $psu(2,2|4)$.

A natural higher spin extension of $su(2,2|{\cal N})$ is associated with
the star product algebra of even functions of superoscillators
\be
f(-a,-b;-\phi,-\bar{\phi})=f(a,b;\phi,\bar{\phi})\,.
\ee
Since the Clifford algebra with $2\N$ generating
elements is isomorphic to $Mat_{2^\N}$, one finds that the
appropriate real form of the
infinite dimensional Lie superalgebra defined this way is
isomorphic to $hu(2^{\N -1},2^{\N -1}|8)$.
Note that for $\N=4$ this gives rise to $hu(8,8|8)$.
For $\N=0$ the Clifford algebra is one-dimensional
and, therefore, $hu(2^{\N -1},2^{\N -1}|2p)$ at $\N=0$
identifies with $hu(1,0|2p)$.
The restriction of $hu(2^{\N -1},2^{\N -1}|8)$
to a particular
supermultiplet gives rise to a smaller higher spin algebra
we shall call $hu_\ga (2^{\N -1},2^{\N -1}|8)$.
$\ga$ is a number characterizing a supermultiplet.
The case of $\ga =0$ will be shown to correspond to the self-conjugated
supermultiplets. (Note that the algebra
$hu_0 (2^{\N -1},2^{\N -1}|8 )$
was called $shsc(4|\N )$  in \cite{FLA}. )
An exciting
possibility discussed at the end of this paper is that,
once there exists a phase with the whole symmetry
$hu(2^{\N -1},2^{\N -1}|8)$ unbroken, it may imply an
infinite chain of the generalized $AdS/CFT$ correspondences
\be
\label{chain}
\cdots AdS^{p+1}/CFT^p \rightarrow AdS^p /CFT^{p-1}
\rightarrow AdS^{p-1}/CFT^{p-2} \cdots\,,
\ee
resulting in a surprising generalized space-time dimension
democracy in the higher spin theories.
(Abusing notation, we use the
abbreviation $AdS^p$ for the generalized $\half p(p+1)-$dimensional
space-time defined in section \ref{$AdS / CFT$ Correspondence}).
The algebras \hfill
$hu_0 (2^{\N -1},2^{\N -1}|8 )$ associated with usual lower spin
supermultiplets and $AdS/CFT$ dualities
are argued to result from some
kind of spontaneous breakdown of the symmetries
$hu(2^{\N -1},2^{\N -1}|8)$.

 The key idea of our approach is
that the dynamics of the $4d$ higher spin massless multiplets admits
a formulation in terms of certain Fock modules over $hu(m,n|8)$
analogously to what was shown previously for
$d=2$ in \cite{2d} and for $d=3$ in \cite{SV}.
Such a formulation makes the higher spin symmetries
of the conformal systems manifest.
The field theory formalism we
work with operates with modules dual to the doubleton modules used
for the description of the unitary representations associated with the
one-particle states of the same system \cite{dou}.
(Note that these Fock modules are somewhat reminiscent of the modules
introduced for the description of  non-commutative solitons
in  string theory \cite{S}.)

In addition to the $su(2,2|\N)$ generators, the algebra
$hu(2^{\N -1},2^{\N -1}|8)$ contains the bilinear generators
\be
U_{\hga\hgb} = a_\hga a_\hgb \,,\qquad
V^{\hga\hgb} = b^\hga b^\hgb \,,
\ee
\be
\label{UVij}
U_{ij} = \phi_i \phi_j\,,\qquad
\bar{V}^{ij} = \bar{\phi}^i \bar{\phi}^j\,
\ee
and supergenerators
\be
R_{\hga i} = a_\hga \phi_i \,,\qquad \bar{R}^{\hgb i} = b^\hgb \bar{\phi^i}\,,
\ee
which extend $u(2,2;\N)$ to $osp(2\N, 8 )$.
(Recall that one can define $osp(p,2q)$ as the superalgebra
spanned by various bilinears built from $p$ fermionic oscillators
and $q$ pairs of bosonic oscillators; see, e.g., \cite{BG} for
more details on the oscillator realizations of simple superalgebras.)
$u(2,2;\N)$ is spanned by
the bilinears in oscillators that commute to the operator $N_\N$, i.e.
$u(2,2;\N)$ is the centralizer of $N_\N$ in $osp(2\N,8)$\footnote{I
am grateful to M.G\"unaydin for drawing my attention to this fact.}.
An important consequence of this simple fact
is that
\be
su(2,2;\N) \subset osp(2\N ;8)
 \subset
hu(2^{\N -1},2^{\N -1}|8)\,.
\ee
As a result, once the higher spin algebra
$hu(2^{\N -1},2^{\N -1}|8)$ is shown to admit a realization
on the conformal supermultiplets of massless fields, it follows
that the same
is true for its finite-dimensional subalgebra $osp(2\N ;8)$. Indeed,
we shall show explicitly how the $osp(2\N ;8)$ transformations link
together different massless (super)fields,
requiring infinite sets of massless supermultiplets to be involved. This
result is  the field-theoretical counterpart of the fact that
the singleton representation of $osp(2\N ;8)$ decomposes into
all doubleton representations of $su(2,2;\N)$. Note that
the field-theoretical realization of $osp(2\N ;8)$
will be shown to be local.

This result confirms the conjecture of \cite{Fp,VF} that the
algebras $osp(L , 2^p)$ may play a distinguished role in
the higher spin gauge theories in higher dimensions.
More generally, it was first suggested in \cite{HP}
that algebras of this class result from the
supersymmetrization of conformal and $AdS$ space-time
symmetry algebras. In  \cite{AF} the contraction of
$osp(1,32)$ was applied for the description of the eleven-dimensional
superalgebra. Somewhat later it was found out that
the algebras $osp(L,2M)$ (in most cases with $M=2^q$)
and their contractions appear naturally in
the context of M-theory dualities and brane charges
\cite{Tow,Ba,GM,G,FP}. One of the messages of this paper is that
these symmetries can be unbroken in the phase in which all higher spin
fields are massless. An immediate speculation is that not only do massive
higher spin modes in fundamental strings result from some spontaneous
breaking of the higher spin symmetries, but also branes are built from
the higher spin gauge fields.

This raises
the important question of what are the higher-dimensional geometry and
dynamics that supports  $osp(L , 2^p)$  symmetries.
Generally, there is no genuine reason to believe that a higher dimensional
geometry should necessarily be Riemannian and, in particular, that
the bosonic coordinates are necessarily Lorentz vectors.
We shall call this presently dominating belief ``Minkowski track".
An alternative option, which looks more natural from various points of
view, is that higher-dimensional bosonic and fermionic dimensions
beyond $d=4$ may be associated with certain coset superspaces
built from $osp(L , 2M)$. We call this alternative  ``symplectic
track". An important advantage of this alternative
is due to supersymmetry. Indeed the main reason why supersymmetry
singles out some particular dimensions in the Minkowski track is
the mismatch between the numbers of bosonic and fermionic
coordinates in higher dimensions as a result of the fact that the
dimension of the spinor representations of the Lorentz algebra
increases with the space-time dimension
as $2^{[\f{d}{2}]}$ while the dimensions of its tensor representations
increase polynomially. Only for some lower dimensions $d\leq11$
where the number of spinor coordinates is not too high due to
some Majorana and/or Weyl conditions the matching can be restored.

Some ideas on a possible structure of alternative to Minkowski space-times have
appeared in both the field-theoretical
\cite{Wein,Fp,MC,GJ,Az,cas,GGHT,dWN,ZhLi,Pir} and world particle dynamics
\cite{ES,SR,BL,BLS} contexts. In particular, important algebraic and geometric
insights most relevant to the subject of this paper were elaborated by Fronsdal
in the pioneering work \cite{Fp}.  Further extensions with higher rank tensor
coordinates were discussed in \cite{Sez,Dev}. The nontrivial issue, however, is
that it is not {\it a priori} clear whether a particular $osp(L ,2M)$ invariant
symplectic track equation allows for quantization compatible with unitarity for
$M>2$. This point is tricky. On the one hand, a Lorentz invariant interval
built
from the ``central charge coordinates" associated with $sp(2^p )$ has  many
time-like directions which, naively, would imply ghosts. On the other hand, it
is well-known \cite{BG} that $osp(L , 2M)$ admits unitary lowest weight
representations (by lowest weight we mean that it is a quotient of a Verma
module) thus indicating that some its quantum-mechanically consistent
field-theoretical realizations have to exist.

Here is where the power of the ``unfolded formulation" dynamics
\cite{plfda,Ann,unf} plays a crucial role. Because this
approach suggests a natural Bogolyubov transform duality between
the field-theoretical unfolded equations and lowest weight
unitary modules \cite{SV}, which, in fact, implies
quantization, it allows us to solve the
problem by identifying  the differential equations that give rise
to the field-theoretical module dual to an appropriate unitary
module. This is achieved by solving a certain cohomology problem.
One of the central results of this paper consists of the explicit
formulation of the $osp(L , 2M/R)$ invariant equations of motion
in the symplectic track space associated with the
massless unitary lowest weight modules of $osp(L , 2M/R)$ via a
Bogolyubov duality transform. Let us note that for the particular
case of $sp(8)$
two simple equations in the symplectic track space for scalar and
svector (i.e., a vector of the symplectic algebra interpreted as a spinor
in the Minkowski track) fields encode all massless equations
in the usual $4d$ Minkowski space.  This
opens an exciting new avenue to higher dimensional physics
in the framework of the symplectic track.
To put it short, the right geometry is going in all cases
to be associated with  symplectic twistors
while  for some lower dimensions we happened to live in it
turns out to be equivalent to the usual  Minkowski geometry.

The rest of the paper is organized as follows. In section
\ref{Generalities} we summarize the general approach to unfolded
dynamics with the emphasize on the cohomological interpretation
of the dynamical fields and equations of motion. In section
\ref{4d Vacuum} we identify the vacuum gravitational field and
discuss the global higher spin symmetries. 4d free equations for
massless fields of all spins in the unfolded form are studied in
 section \ref{$4d$ Conformal Field Equations}. In  subsection
\ref{Fock Space Realization} we reformulate the free massless
equations of motion for $4d$ massless fields of all spins in
terms of flat sections of an appropriate  Fock fiber bundle and
identify various types of the $4d$ higher spin conformal algebras.
Generic solution of these equations in the flat space-time is
presented in  subsection \ref{Generic Solution}. The reality
conditions are defined in subsection
\ref{Reality Conditions and Reduction}. The reduction to
self-conjugated supermultiplets based on a certain antiautomorphism
and the corresponding reduced higher spin algebras are
discussed in subsection
\ref{Self-conjugated  Supermultiplet Reduction}.
In  section \ref{4d Conformal Higher Spin Symmetries}
we explain how the  formulas for any global
conformal higher spin symmetry transformation of the massless
fields can be derived and present explicit formulas for the
global $osp(2\N , 8)$
transformations.  The duality between the field-theoretical
Fock module and unitary ($sp(8)$ - singleton) module
is discussed in section \ref{Field Theory - Doubleton Duality}.
The dynamics of the $4d$ conformal massless fields is reformulated
in the $osp(2\N , 8)$ invariant (super)spaces in section \ref{CD}.
We start in subsection (\ref{Superspace}) with the example of
usual superspace. The compatible with unitarity  unfolded equations in the
$sp(2M)$ invariant space-time  are derived
in subsection \ref{$sp(2M)$ Covariant Equations}.
The unfolded dynamics in the $osp(L ,2M)$ invariant superspaces
is formulated in subsection \ref{???}.
Further extension of the equations to the infinite-dimensional
higher spin superspace is given in subsection
\ref{Higher Spin (Super)Space}.
The worldline particle interpretation
of the obtained massless equations of motion is discussed in section
\ref{World Line Particle Interpretation} where some new twistor-like
particle models are presented. The AdS/CFT correspondence in the
framework of the higher spin gauge theories is the subject of
section \ref{$AdS / CFT$ Correspondence} where, in particular,
a possibility of the infinite chain of $AdS/CFT$ dualities
in the higher spin gauge theories is discussed.
Finally,  section \ref{Conclusions and Perspectives} contains a
summary of the main results of the paper and discussion
of some perspectives.

\section{Unfolded Dynamics}
\label{Generalities}
As usual in the higher spin theory framework,
we shall use the ``unfolded formulation'' approach
\cite{plfda,Ann,unf} which allows one to
reformulate any dynamical equations in the form
\be
\label{dFw}
dw^A = F^A (w)
\ee
($d=dx^\un \frac{\partial}{\partial x^\un}$; underlined indices
$\um$, $\un =0\div d-1$ are used for
the components of differential forms)
with some set of differential forms $w$ and a function
$F^A (w)$ built from $w$ with the help of the exterior product and
satisfying the compatibility condition
\be
\label{dF}
F^B (w) \f{\delta F^A (w)}{\delta w^B}         =0.
\ee
In the linearized approximation, i.e. expanding near
some particular solution $w_0$ of (\ref{dFw}), one finds
that nontrivial dynamical equations are associated with
null-vectors of the linearized part $F_1$ of $F$.

For example, consider the system of equations
\be
\label{dc}
\p_\un C_{a_1 \ldots a_n}(x) + h_\un{}^b C_{b a_1 \ldots a_n}(x) =0\,,
\ee
\be
\label{dh}
dh^a =0
\ee
with the set of 0-forms $C_{a_1 \ldots a_n}$ with all
$n = 0,1,2,\ldots \infty$ and the 1-form $h^a =dx^\un h_\un{}^a$
($a,b,\ldots = 0\div d-1$ are fiber vector indices).
This system is obviously consistent in the sense of (\ref{dF}).
Assuming that $h_{\un}{}^a$ is a nondegenerate matrix
(in fact, flat space-time frame), say, choosing
$h_{\un}{}^a = \delta_{\un}{}^a$ as a particular solution of
(\ref{dh}), one finds that the system is dynamically empty,
just expressing the highest components $C_{a_1 \ldots a_n}$
via highest derivatives of $C$
\be
\label{dC}
C_{a_1 \ldots a_n}(x) = (-1)^n \p_{a_1}\ldots \p_{a_n} C(x)\,.
\ee
However, once some of the components of $C_{a_1 \ldots a_n}$
are missed in a way consistent with the compatibility condition
(\ref{dF}), this will impose the differential restrictions on
the ``dynamical field'' $C(x)$. In particular, this happens if the
tensors are required to be traceless
\be
\label{tr}
C^b{}_{ba_3 \ldots a_n}=0\,.
\ee
In accordance with (\ref{dC}) this implies the Klein-Gordon
equation
\be
\label{box}
\Box C (x) =0
\ee
and, in fact,  no other independent conditions.

An important point is that
any system of differential equations can be reformulated
in the form (\ref{dFw}) by
virtue of introducing enough (usually, infinitely many)
auxiliary fields. We call such a reformulation ``unfolding''.
In many important cases the  linearized equations have the form
\be
\label{geneq}
(\D +\sigma_- + \sigma_+ )C = 0 \,,
\ee
where $C$ denotes some (usually infinite) set of fields (i.e.,
a section of some linear fiber bundle over the space-time
with a fiber space $V$) and
the operators $\D$ and $\sigma_\pm$ have the properties
\be
\label{ds}
(\sigma_\pm )^2 =0\,,\qquad \D^2 + \{\sigma_- , \sigma_+ \} =0\,,
\qquad \{ \D , \sigma_\pm \} =0\,.
\ee
It is  assumed that only the operator $\D$ acts nontrivially
(differentiates) on the space-time coordinates while $\sigma_\pm$
act in the fiber $V$. It is also assumed that there exists
a gradation operator $G$ such that
\be
[G , \D ] = 0\,,\qquad [G ,\sigma_\pm ] = \pm \sigma_\pm\,,
\ee
$G$ can be diagonalized in the fiber space $V$
and the spectrum of $G$ in $V$ is bounded from below.
In the example above $\D = d$, $\sigma_+ = 0$,
$\sigma_- (C)_{a_1 \ldots a_n} = h^b C_{b a_1 \ldots a_n}$.
The gradation operator $G$ counts a number of indices
$G (C)_{a_1 \ldots a_n}  = n C_{a_1 \ldots a_n}$.

The important observation is (see, e.g., \cite{SVsc}) that
the nontrivial dynamical equations hidden in (\ref{geneq})
are in  one-to-one correspondence with the nontrivial
cohomology classes of $\sigma_-$. For the case under
consideration with $C$ being a
0-form, the relevant cohomology group is $H^1 (\sigma_- )$.
For the more general situation with $C$ being a $p$-form,
the relevant cohomology group is $H^{p+1} (\sigma_- )$
(in a somewhat implicit form this analysis for
the case of 1-forms was applied in \cite{LV,VF}).

Indeed, consider the decomposition of the space of fields $C$
into the direct sum of eigenspaces of $G$. Let a
field having definite eigenvalue $k$ of $G$ be
denoted $C|_k$, $k= 0,1,2 \ldots$. Suppose that the dynamical
content of the equations (\ref{geneq}) with the eigenvalues
$k \leq k_q $ is found. Applying the operator $\D +\sigma_+$
to the left hand side
of  the equations (\ref{geneq}) at  $k \leq k_q $ we obtain
taking into account (\ref{ds})
that
\be
\label{sdc}
\sigma_- \Big ( (\D +\sigma_- +\sigma_+ ) (C) \Big |_{k_q+1} \Big ) =0\,.
\ee
Therefore $(\D +\sigma_- +\sigma_+ )( C)\Big|_{k_q +1} $ is
$\sigma_-$ closed. If the group $H^1 (\sigma_- )$ is trivial
in the grade $k_q+1$
sector,  any solution of (\ref{sdc}) can be written in
the form
$(\D +\sigma_- +\sigma_+ )( C)\Big |_{k_q+1}
= \sigma_- (\tilde{C}|_{k_q +2}$)
for some  field $\tilde{C}|_{k_q +2}$. This,
in turn, is equivalent to the statement that one can adjust
$C|_{k_q +2}$ in such a way
that $\tilde{C}|_{k_q +2} =0$ or, equivalently, that the part of the
equation (\ref{geneq}) of the grade $k_q+1$ is some constraint
that expresses $C|_{k_q +2}$ in terms of the derivatives of $C|_{k_q +1}$
(to say that this is a constraint we have used the assumption that
the operator $\sigma_-$ is algebraic
in the space-time sense, i.e. it does not contain space-time derivatives.)
If $H^1 (\sigma_- )$ is nontrivial, this means that the equation
(\ref{geneq}) sends the corresponding cohomology class to zero and,
therefore, not only
expresses the field $C|_{k_q+2}$ in terms of  derivatives of
$C|_{k_q+1}$ but also imposes some additional
differential conditions on $C|_{k_q+1}$.  Thus, the nontrivial
space-time differential equations described by (\ref{geneq})
are classified by the cohomology group $H^1 (\sigma_- )$.

The nontrivial dynamical fields are associated with
$H^0 (\sigma_-)$ which is always non-zero because it at least
contains a nontrivial subspace of $V$ of minimal
grade. As follows from the $H^1 (\sigma_- )$
analysis of the dynamical equations, all
fields in $V/H^0 (\sigma_-)$ are auxiliary, i.e. express via the
space-time derivatives of the dynamical fields by virtue
of the equations (\ref{geneq}).

For the scalar field example  one finds \cite{SVsc} that
$H^0 (\sigma_- )$ is spanned by the linear space of the rank-zero
tensors associated with the scalar field.
For the case with the fiber $V$ realized by all symmetric tensors
$H^1 (\sigma_- )=0$ and, therefore, the corresponding system is
dynamically empty. For the case of $V$ spanned by traceless symmetric
tensors $H^1 (\sigma_- )$ turns out to be one-dimensional with the
1-form representative
\be
\label{coho}
\kappa h_\un{}^a
\ee
taking values in the subspace of rank 1 tensors
(i.e., vectors). Indeed, it is obvious that
any element of the form (\ref{coho}) is $\sigma_-$ closed. It is not
 $\sigma_-$ exact because $h_\un{}_b \neq h_\un{}^a C_{ab}$ with
some symmetric traceless $C_{ab}$. As a result, the only nontrivial
equation contained in (\ref{dc}) is its trace part at $n=1$, which is
just the Klein-Gordon equation (\ref{box}).

Let us note that the ``unfolded equation" approach is to some
extent analogous to the coordinate free formulation of gravity
by Penrose \cite{Pen} and the concept of exact sets of fields
(see \cite{PR} and references therein)
in which the dynamical equations are
required to express all space-time derivatives of the
fields in terms of the  fields themselves.
The important difference between these two approaches is that
``unfolded dynamics" operates in terms of
differential forms thus leaving  room
for gauge potentials and gauge symmetries that in
most cases are crucial for the interaction problem.
In some sense, the exact sets of fields  formalism
corresponds to the particular case of the unfolded
dynamics in which all fields are described as 0-forms.

\section{Vacuum and Global Symmetries}
\label{4d Vacuum}

Let us now consider the four-dimensional case
introducing $4d$ index notation. We will use two pairs of
two-component spinors $a_\ga$, $b^\ga$, $\ova{}_\dga$ and $\ovb{}^\dgb$.
The basis commutation relations become
\be
\label{br}
[ a_\ga ,b^\gb ]_* = \delta_\ga^\gb\,,\qquad
[\ova{}_\dgga , \ovb{}^\dgb ]_* = \delta_\dgga^\dgb\,.
\ee
The $4d$ identification of the elements of $su(2,2)$ is as follows.
\be
\label{L}
L_\ga{}^\gb = a_\ga b^\gb -\half\delta_\ga{}^\gb  a_\gga b^\gga \,, \qquad
\bar{L}_\dga{}^\dgb = \ova{}_\dga \ovb{}^\dgb
-\half\delta_\dga{}^\dgb \ova_{{\dgga}} \ovb^{\dgga}
\ee
are Lorentz generators.
\be
\label{D}
D = \half( a_\ga b^\ga -\ova{}_\dga \ovb{}^\dga )
\ee
is the dilatation generator.
\be
\label{P}
P_\ga{}^\dgb = a_\ga  \ovb{}^\dgb
\ee
and
\be
\label{K}
K_\dga{}^\gb = \ova{}_\dga  b^\gb
\ee
are the generators of
$4d$ translations and special conformal transformations, respectively.
The complex conjugation rules
\bee
\bar{a}_\ga = \ovb{}_\dga\,,\qquad \bar{b}^\ga = \ova{}^\dga\,,\qquad
\bar{\ova}{}_\dga = b{}_\ga\,,\qquad \bar{\ovb}{}^\dga = a{}^\ga\,
\eee
are in accordance with (\ref{re}) with the
antisymmetric matrix $C^{\hga\hgb}$ having nonzero components
\be
C^{\ga\dgb} = \gvep^{\ga\dgb}\,,\qquad C^{\dgga\gb} =\gvep^{\dgga\gb}\,,
\ee
where $\gvep^{\ga\gb}$ is the $2\times 2$ antisymmetric matrix
normalized to $\gvep^{12} =1$.

Let $\go(a,b;\phi,\bar{\phi}|x)$ be a 1-form  taking values
in the higher spin algebra\hfil\\
\hsa, i.e. $\go$ is the generating function
of the conformal higher spin gauge fields
\bee
&{}&\go(a,b;\phi,\bar{\phi}|x)=\nn\\
&{}&\ls\ls\ls\sum_{m,n=0}^\infty \sum_{k,l=0}^{\cal N}
\frac{1}{m!n!k!l!}\go_{\hga_1\ldots \hga_m,}\!{}^{\hgb_1 \ldots \hgb_n}
{}_{i_1 \ldots i_k}{}^{j_1 \ldots j_l}(x)
b^{\hga_1}\!\!\ldots
b^{\hga_m} a_{\hgb_1}\!\!\ldots a_{\hgb_n}\phi^{i_1}\!\! \ldots\phi^{i_k}
\bar{\phi}_{j_1}\! \ldots \bar{\phi}_{j_l} .
\eee

In the cases of interest the general equation (\ref{dFw})
admits a solution with all fields equal to zero except for some
1-forms $\go_0$ taking values in an appropriate Lie (super)algebra
$h$ (in the case under consideration $h=$\hsa$\!\!\!)$.
The equation (\ref{dFw}) then reduces to the zero-curvature equation
on $\go_0$. To describe nontrivial
space-time geometry one has to require $h$ to contain an appropriate
space-time symmetry algebra whose gauge fields identify with the
background gravitational
fields.  In particular, the components of $\go_0$ in the sector of
translations are identified with the gravitational frame field
which is supposed to be non-degenerate.  Let $\go_0$ be such a
solution of the zero-curvature equation
\be
d\go_0=\go_0\ep *\go_0\,.
\label{ezc}
\ee
The equation (\ref{ezc}) is invariant under the gauge transformations \be
\gd\go_0=d\gep-{[}\go_0,\gep{]}_*\,, \label{gtw}
\ee where $\gep (a,b;\phi,\bar{\phi}|x)$ is an infinitesimal symmetry parameter
being a 0-form.  Any vacuum solution $\go_0$ of the equation (\ref{ezc}) breaks
the local higher spin symmetry to its stability subalgebra with the
infinitesimal parameters $\gep_0 (a,b;\phi,\bar{\phi}|x)$ satisfying the
equation
\be
d\gep_0-{[}\go_0 ,\gep_0{]}_*=0\,.  \label{ipgs}
\ee
Consistency of this equation is guaranteed by
the zero-curvature equation (\ref{ezc}).

Locally, the equation (\ref{ezc}) admits a pure gauge solution
\be
\go_0 =-g^{-1}* dg\,.
\label{wvg}
\ee
Here $g (a,b;\phi,\bar{\phi}|x)$ is some invertible element of the
associative algebra, i.e.
$g^{-1}*g=g*g^{-1}=1$. For $\go_0$  (\ref{wvg}), one finds that the
generic solution of (\ref{ipgs}) is
\be
\gep_0 (a,b;\phi,\bar{\phi}|x)=
g^{-1}(a,b;\phi,\bar{\phi}|x) *\xi (a,b;\phi,\bar{\phi}) *
g(a,b;\phi,\bar{\phi}|x)\,,
\label{e0}
\ee
where $\xi (a,b;\phi,\bar{\phi})$
is an arbitrary $x$--independent element that plays a role of
the ``initial data" for the equation (\ref{ipgs}).
\be
\gep_0 (a,b;\phi,\bar{\phi}|x)|_{x=x_0}=\xi (a,b;\phi,\bar{\phi})
\ee
for such a point $x_0$ that $g(x_0)=1$. Since
$[\gep_0^1 , \gep_0^2 ]_*$ has the same form with
$\xi^{12} = [\xi^1 , \xi^2 ]_* $, it is clear that the global symmetry
algebra is \hsa.

As usual, the gravitational fields (i.e., frame and Lorentz connection)
are associated with the generators of translations and Lorentz rotations
in the Poincare or AdS subalgebras of the conformal algebra.
For $AdS_4$ one sets
\be
\label{goads}
\go_0 = \go_0{}^\ga{}_\gb (x) L_\ga{}^\gb +
\bar{\go}_0{}^\dga{}_\dgb (x)\bar{L}_\dga{}^\dgb
+h_0{}^\ga{}_\dgb (x) ( P_\ga{}^\dgb + \lambda^2 K^\dgb{}_\ga )\,,
\ee
where $-\lambda^2$ is the cosmological constant. The indices of
$K_\dgb{}_\ga$ have been raised and lowered with the aid of the
Lorentz invariant antisymmetric
forms $\gvep^{\ga\gb}$ and $\gvep^{\dga\dgb}$
according to the rules
\be
A^\ga = \gvep^{\ga\gb} A_\gb \,,\qquad  A_\gb = \gvep_{\ga\gb} A^\ga \,,
\qquad
A^\dga = \gvep^{\dga\dgb} A_\dgb \,,\qquad  A_\dgb = \gvep_{\dga\dgb} A^\dga \,
\ee
which, as expected for the $AdS_4$ space having a dimensionful radius,
breaks down the scaling symmetry of the ansatz (\ref{goads}).
The condition that the ansatz (\ref{goads}) solves the zero-curvature
equation (\ref{ezc}) along with the condition that $h_0{}^\ga{}_\dgb (x)$
is nondegenerate implies that $\go_0{}^\ga{}_\gb (x)$,
$\bar{\go}_0{}^\dga{}_\dgb (x)$ and $h_0{}^\ga{}_\dgb (x)$ describe
$AdS_4$ Lorentz connection and the frame field, respectively.
(Note that the generator $P_\ga{}^\dgb + \lambda^2 K^\dgb{}_\ga $
describes the embedding of the $AdS_4$ translations
 into the conformal algebra $su(2,2)$.)

For the $4d$ flat Minkowski space one can choose
\be
\go_0 =d x^\un \gs_\un{}^{\ga}{}_{\dgb}a_\ga \ovb{}^\dgb\,,
\label{fc}
\ee
thus setting all fields  equal to zero
           except for the flat space vierbein
associated with the translation generator.
Here $\gs_\un{}^{\ga\dgb}$ is the set of $2\times 2$ Hermitian
matrices normalized to
\be
\gs_\un{}^{\ga\dgb} \gs_\um{}_{\ga\dgb}=\eta_{\un\um}\,,\qquad
\gs_\un{}^{\ga\dgb} \gs_\um{}_{\gga\dot{\delta}}\eta^{\un\um}
= \delta^\ga_\gga\delta^\dgb_{\dot{\delta}}\,,
\ee
where $\eta_{\un\um}$ is the flat Minkowski metric tensor.
The function $g$ that gives rise to the flat gravitational
field (\ref{fc}) is
\be
g=\exp(-x^{\ga}{}_{\dgb}a_\ga \ovb{}^\dgb)\,,
\label{gfc}
\ee
where
\be
x^{\ga\dgb}=x^\un\gs_\un{}^{\ga\dgb}\,,
\qquad x^\un=\gs^\un{}_{\ga\dgb}x^{\ga\dgb}\,.
\ee

\section{$4d$ Conformal Field Equations}
\label{$4d$ Conformal Field Equations}

As shown in  \cite{plfda,Ann}, the equations of motion for
massless fields in $AdS_4$ admit a
formulation in terms of the generating function
\be
\label{fe}
C(y,\bar{y}|x)=\sum_{m,n=0}^\infty\frac{1}{m!n!}
c_{\ga_1\ldots\ga_m\,,\dgb_1 \ldots \dgb_n} (x)
y^{\ga_1}\ldots y^{\ga_m} \bar{y}^{\dgb_1}\ldots \bar{y}^{\dgb_n}
\ee
with the auxiliary spinor variables $y^\ga$ and $\bar{y}^\dgb$.
$C(y,\bar{y}|x)$  is the generating
function for all on-mass-shell nontrivial spin $s\geq 1$ gauge invariant
curvatures and matter fields of spin 0 and 1/2. Every spin $s$
massless field appears in two copies because the generating
function $C(y,\bar{y}|x)$ is complex. It forms the twisted adjoint
representation of the algebra $hu(1,1|4)$. The associated covariant
derivative reads
\be
\label{dtwist}
DC=dC -\go * C +C * \tilde{\go }\,,
\ee
where $\go(y,\bar{y}|x)$ is the generating function for higher spin
gauge fields taking values in $hu(1,1|4)$, $*$ denotes Moyal star
product induced by the Weyl (i.e., totally symmetric) ordering of the
oscillators $y^\ga$ and $\bar{y}^\dga$ with the basis commutation relations
\be
[y^\ga , y^\gb ]_* = 2i\varepsilon^{\ga\gb}\,\qquad
[\bar{y}^\dga , \bar{y}^\dgb ]_* = 2i\varepsilon^{\dga\dgb}\,,
\ee
 and  tilde denotes the involutive automorphism of the algebra
 $\tilde{\go}(y,\bar{y}|x)=\go(-y,\bar{y}|x)$\footnote{The
covariant derivative of the complex conjugated field $\bar{C}$
is analogous with the roles of dotted and undotted indices
interchanged. Note that the twisted adjoint representation is most
conveniently described with the help of Klein operators \cite{Ann}
(see also \cite{Gol}).}.
Fluctuautions of the higher spin gauge fields are linked to the invariant
field strengths by virtue of their own  field equations
\cite{plfda,Ann}. The sector of higher spin gauge fields plays
important role in the analysis of higher spin interactions and
Lagrangian higher spin dynamics, very much as the Lagrangian
form of Maxwell theory is formulated in terms of potentials rather
than field strengths. In this paper we however  confine ourselves to
consideration of free field equations formulated in terms of
field strengths with $\go = \go_0$ being a fixed vacuum gravitational
field
taking values in the gravitational $sp(4|{\bf R})$ subalgebra of $hu(1,1|4)$
and  satisfying zero curvature equation for the higher spin algebra.
Note that, as explained in Introduction,
 $sp(4|{\bf R})$ belongs to the finite-dimensional
subalgebra $ osp(2,4)$ of $hu(1,1|4)$, i.e. the system
under consideration exhibits $\N=2$ supersymmetry.
(Also note that $ osp(2,4)\oplus u(1)$ with $u(1)$ factor
associated with the unit element of the star product algebra
is the maximal finite dimensional subalgebra of $hu(1,1|4)$.)

As shown in \cite{plfda,Ann}
the free equations of motion for massless fields of all spins
have the form
\be
\label{AEQ}
D_0 (C)=0
\ee
where $D_0$ is the  covariant
derivative (\ref{dtwist}) with respect to
the vacuum field $\go_0$. Since the equations (\ref{ezc}) and
(\ref{AEQ}) are invariant under the gauge transformations (\ref{gtw})
and
\be
\label{delC}
\delta C = \epsilon *C - C*\tilde{\epsilon}
\ee
from the general argument of section \ref{4d Vacuum}
it follows that, for some fixed vacuum field $\go_0$ satisfying
(\ref{ezc}),
the equation (\ref{AEQ}) is
 invariant under  global symmetry transformations with the
parameters (\ref{e0}),
that form the $AdS_4$ higher spin algebra $hu(1,1|4)$. This
realization of the higher spin field equations
therefore makes manifest
the $AdS_4$ symmetry $sp(4|{\bf R})\subset hu(1,1|4)$, while
the conformal symmetry $su(2,2)$ of free massless equations
remains hidden.

For the reader's
convenience let us analyse the content of the equations (\ref{AEQ}) in
somewhat more details.
Upon some rescaling of fields
the free massless equations of motion for all spins in $AdS_4$
of \cite{plfda,Ann} acquire the form
\be
\label{deqa}
D_0^LC(y,\by |x)=-h_0{}^{\ga\dgb}\left (
\frac{1}{\ptl y^\ga\ptl \bar{y}^\dgb} +\lambda^2 y_\ga\bar{y}_\dgb \right )
C(y,\by |x)\,,
\ee
where $D_0^L$ is the background Lorentz covariant derivative
\be
D_0^L = d- \left (\go_0{}^\ga{}_\gb (x) y^\gb \f{\p}{\p y^\ga} +
\bar{\go}_0{}^\dga{}_\dgb (x)\bar{y}^\dgb \f{\p}{\p \bar{y}^\dga}
\right )\,.
\ee
It gives a particular realization of (\ref{geneq}) with
\be
\D = D_0^L \q \sigma_- =  h {}^{\ga}{}^{\dgb}
\frac{1}{\ptl y^\ga\ptl \bar{y}^\dgb}\q \sigma_+ =\lambda^2
h {}^{\ga}{}^{\dgb} y_\ga\bar{y}_\dgb \,.
\ee
The gradation operator is
\be
\label{GG}
G =
\half\left ( y^\ga \f{\p}{\p y^\ga } +
\bar{y}^\dga \f{\p}{\p \bar{y}^\dga }\right )\,.
\ee
The equation (\ref{deqa}) decomposes into the infinite set of
subsystems associated with the eigenvalues
of the operator
\be
\sigma =\half \left (y^\ga \f{\p}{\p y^\ga}
-\by^\dga \f{\p }{\p \by^\dga}\right )\,
\ee
identified with spin
\be
\label{sig}
\sigma C(y ,\by |x) = \pm s C(y ,\by |x)
\ee
(the fields associated with the eigenvalues that differ by sign
are conjugated).

The flat limit of the
free equations of motion for the integer and half-integer
spin massless fields of \cite{plfda,Ann} has the form
\be
\label{deq}
dC(y,\by |x)+dx^\un
\gs_\un{}^{\ga\dgb}\frac{1}{\ptl y^\ga\ptl \bar{y}^\dgb}C(y,\by |x)=0
\,,
\ee
which provides a particular realization of
(\ref{geneq}) with
\be
\D = d\q \sigma_- = dx^\un \sigma_\un{}^\ga{}^\dgb
\frac{1}{\ptl y^\ga\ptl \bar{y}^\dgb}\q \sigma^+ =0\,.
\ee
Let us note that the fact that the
free equations of motion of $4d$ massless fields
in the flat space admit
reformulation in the form (\ref{deq}) was also observed in \cite{Eis}.

The dynamical fields associated with $H^0 (\sigma_- )$ identify
with the lowest degree eigenspaces of $G$ for various
eigenvalues of $\sigma$. These are
analytic fields $C(y,0 |x)$  and their conjugates
$C(0,\by |x)$. Some standard examples
are provided with spin 0
\be
C(0,0|x )= c (x) \,,
\ee
spin 1/2
\be
C(y,0|x )= y^\ga c_\ga (x)\,,\qquad
C(0,\by |x)= \by^\dga \bar{c}_\dga (x)\,,
\ee
spin 1
\be
 C(y,0|x )= y^\ga y^\gb c_{\ga\gb} (x)\,,\qquad
C(0,\by |x) = \by^\dga \by^\dgb \bar{c}_{\dga \dgb } (x)\,,
\ee
spin 3/2
\be
C(y,0|x )= y^{\ga_1} y^{\ga_2}y^{\ga_3}
c_{\ga_1 \ga_2 \ga_3} (x)\,,\qquad
C(0,\by |x) =\by^{\dga_1}\by^{\dga_2} \by^{\dga_3}
\bar{c}_{\dga_1 \dga_2 \dga_3  } (x)\,,
\ee
and spin 2
\be
C(y,0|x )= y^{\ga_1} \ldots y^{\ga_4}
c_{\ga_1 \ldots \ga_4} (x)\,,\qquad
C(0,\by |x) =\by^{\dga_1}\ldots \by^{\dga_4}
\bar{c}_{\dga_1\ldots \dga_4 } (x)\,.
\ee
All fields $C(y,\by|x) $ starting with spin 1 are associated
with the appropriate field strengths, namely, with Maxwell field strength,
gravitino curvature and Weyl tensor for spins 1, 3/2 and 2, respectively.

The analytic fields $C(y,0 |x)$  and their conjugates
$C(0,\by |x)$ are subject to the dynamical
spin-$s$ massless equations \cite{Ann}
associated with $H^1 (\sigma_- )$.
Using the properties of two-component spinors it
is  elementary to prove that
the representatives of $H^1 (\sigma_- )$ are
\be
y_\ga h^{\ga \dgb} E_{\dgb} (y )\q
\bar{y}_\dgb h^{\ga \dgb} \bar{E}_{\ga} (\bar{y})
\q y_\ga \bar{y}_\dgb h^{\ga\dgb} \kappa\,,
\ee
where the 0-forms $E_{\dgb}(y)$ and  $\bar{E}_\ga (\bar{y})$
are, respectively, analytic and antianalytic while
 $\kappa$ is a constant. The cohomology class
parametrized by $\kappa$ corresponds to the $s=0$ massless
equation, while  the cohomology classes parametrized by
$E_{\dgb}(y)$ and $\bar{E}_\ga (\bar{y})$ are responsible for the
 field equations for spin $s>0$ massless fields. Note that
the cohomology group $H^1 (\sigma_- )$ is the same  for
the flat and  $AdS_4$ cases.
The explicit form of the flat space
dynamical massless equations resulting from (\ref{deq}) is

\bee
\ls\ls\f{\p}{\p y^\ga} \f{\p}{\p x_\ga{}^\dgb} C(y,0 |x)
 &=&0\,, \qquad
\f{\p}{\p \bar{y}^\dgb} \f{\p}{\p x^\ga{}_\dgb }
C(0 ,\bar{y} |x ) =0\,, \qquad (s\neq 0 )\,;\nn\\
0= \f{\p^2}{\p y^\ga \p \bar{y}_\dgb} \f{\p}{\p
x_\ga{}^\dgb} C(y,\bar{y} |x )|_{y=\bar{y} =0}&\longrightarrow &
\partial_\un \partial^\un C(0,0|x )=0\,\qquad (s=0)\,.
\label{4deq}
\eee
All other equations in (\ref{deq})
express the nonanalytic components of the fields $C(y,\bar{y}|x)$
via higher space-time derivatives of the dynamical massless fields
$C(0,\bar{y} |x)$  and $C(y, 0 |x )$ or reduce to identities expressing
some compatibility conditions.
Therefore,  the nonanalytic  components in  $C(y,\bar{y}|x)$
are auxiliary fields (in both the flat and  $AdS_4$ cases).

\subsection{Fock Space Realization}
\label{Fock Space Realization}

The formulation of \cite{plfda,Ann} with the
0-form $C(y,\by |x)$ taking values in the twisted adjoint
representation of the ${\rm AdS}_4$ higher spin algebra
made the symmetry $hu(1,1|4)$ manifest.  Let us now show that
the same equation (\ref{deq})  admits a
realization in the Fock space that makes the
higher spin conformal symmetries of the system manifest.

Let us introduce the Fock vacuum $\vac$ defined by the relations
\be
\label{xx}
a_\ga * \vac =0\,, \qquad \ovb{}^\dgb *\vac =0\,,
\qquad \phi_i * \vac =0\,.
\ee
It can be realized as the element of the star product
algebra
\be
\label{vac}
\vac =2^{4-\N} \exp{2\Big (\ova_\dga \ovb^\dga
-a_\ga b^\ga      +\phi_i \bar{\phi}^i }\Big)\,,
\ee
which also satisfies
\be
\vac * \ova^\dga =0 \,,\qquad \vac * {b}_\ga =0\,
,\qquad \vac * \bar{\phi}^i =0\,.
\ee

As a result, the vacuum  is bi-Lorentz invariant
\be
\label{linv}
L_\ga{}^\gb *\vac =0\,,\qquad \bar{L}_\dga{}^\dgb *\vac =0\,,
\ee
\be
\vac *L_\ga{}^\gb  =0\,,\qquad \vac* \bar{L}_\dga{}^\dgb  =0\,,
\ee
bi-$su(\N )$ invariant
\be
T_i{}^j *\vac = \vac *T_i{}^j
= \half \delta_i^j \vac
\ee
and has conformal weight 1
\be
\label{conv}
D * \vac  = \vac *D = \vac \,.
\ee
Also, it is left Poincare invariant
\be
P_\ga{}^\dgb *\vac =0
\ee
and supersymmetric
\be
Q_\ga^i *\vac =0 \,,\qquad \bar{Q}^{\hat{\beta}}_i *\vac =0\,.
\ee

Note that $\vac$ is a projector
\be
\vac*\vac =\vac \,
\ee
and space-time constant
\be
d \vac =0\,.
\ee

Let us now consider the left module over the algebra \hsa spanned
by the states
\be
\label{PC}
|\Phi(\ova , b,\bar{\phi} |x) \rangle =
C(\ova , b,\bar{\phi} |x )* \vac \,,
\ee
where
\be
C(\ova , b ,\bar{\phi}|x)
= \sum_{m,n,k=0}^\infty \frac{1}{m!n!k!}
c_{\gb_1 \ldots \gb_m }{}^{\dga_1 \ldots \dga_n }{}_{j_1 \ldots j_k} (x)
\ova{}_{\dga_1} \ldots \ova{}_{\dga_n} b^{\gb_1} \ldots b^{\gb_m}
\bar{\phi}^{j_1} \ldots \bar{\phi}^{j_k}\,.
\label{cdef}
\ee
Note that
\be
\label{uf}
C(\ova , b,\bar{\phi} |x )* \vac =
C(2\ova ,2 b, 2\bar{\phi} |x )\,
2^{4-\N} \exp{2\Big (\ova_\dga \ovb^\dga
-a_\ga b^\ga      +\phi_i \bar{\phi}^i }\Big)\,.
\ee

The system of equations
\be
d|\Phi \rangle -\go_0  * |\Phi \rangle =0
\label{ffe}
\ee
concisely encodes all $4d$ massless field equations
provided that
the equation (\ref{ezc}), which guarantees the formal consistency
of (\ref{ffe}), is true. Indeed, the
choice of $\go_0$ in the form (\ref{fc}) makes the equation
(\ref{ffe}) equivalent to (\ref{deq}) upon the identification
of $b^\ga$ with $y^\ga$ and $\ova_\dgb$ with $\bar{y}_\dgb$
(for every $su(\N )$ tensor structure).
Analogously, choosing $\go_0$ in the form (\ref{goads}),
one finds that the equation (\ref{ffe}) describes massless
fields in $AdS_4$.
Let us note that the equations on the component fields
 are Lorentz and scale invariant due to Lorentz invariance (\ref{linv})
and definite scaling (\ref{conv}) of the vacuum $\vac$.
The dynamical components identify with the holomorphic and
antiholomorphic parts
\be
\label{cc}
c_{\gb_1 \ldots \gb_n}{}_{i_1 \ldots i_k} (x)=
\f{\p^n}{\p b^{\gb_1}\ldots \p b^{\gb_n}}
\f{\p^k}{\p \bar{\phi}^{i_1}\ldots \p \bar{\phi}^{i_k}}
C(0 , b,\bar{\phi} |x )\Big |_{ b^\ga = \bar{\phi}^j =0}\,,
\ee
\be
\label{ccc}
c^{\dga_1 \ldots \dga_m}{}_{i_1 \ldots i_k} (x) =
\f{\p^m}{\p \ova{}_{\dga_1}\ldots \p \ova{}_{\dga_m}}
\f{\p^k}{\p \bar{\phi}^{i_1}\ldots \p \bar{\phi}^{i_k}}
C(\ova,0,\bar{\phi} |x )\Big
|_{ \ova_\dga = \bar{\phi}^j =0}\,.
\ee

Recall that the equation (\ref{ffe}) imposes
the dynamical massless equations of motion on the components
(\ref{cc}) and (\ref{ccc}) and expresses all other components in
$C(\bar{a}_\dga , b^\gb ,\bar{\phi}|x)$ via their derivatives
according to (\ref{deq}) rewritten in the form
\be
\label{deqr}
\frac{\p^2}{\ptl b^\ga\ptl \ova_\dgb}
C(\ova ,b,\bar{\phi} |x) =
\gs^\un{}_{\ga}{}^{\dgb}\partial_\un C(\ova ,b,\bar{\phi} |x) \,,
\ee
or, equivalently,
\be
\label{xeq}
\frac{\p^2}{\ptl b^\ga\ptl \ova_\dgb}
C(\ova ,b,\bar{\phi} |x) =-\f{\p}{\p x^\ga{}_\dgb}
C(\ova ,b,\bar{\phi} |x) \,.
\ee

As discussed in more detail in section
\ref{4d Conformal Higher Spin Symmetries}
the system of massless equations in the form (\ref{ffe})
is manifestly invariant under the higher spin global symmetry
$hu(1,1|8)$. Note that
the formulation we use is in a certain sense dual to
the usual construction of induced representations \cite{mack}.
The difference is that the module we use is realized in the
auxiliary Fock space, while the space-time dependence is reconstructed
by virtue of the dynamical equation (\ref{ffe}) that links
the dependence on the space-time coordinates to the dependence
on the auxiliary coordinates. This module  is
induced from the vacuum annihilated by the translation generator
$P_\ga{}^\dgb$ that acts on the auxiliary
spinor coordinates, while in the construction of
\cite{mack} the vacuum state is assumed to be annihilated
by the generators $K^\un$ of the special conformal transformations
acting directly on the dynamical relativistic
fields.  (Let us stress that this is not just a matter of
notation since   $P_\un$ is eventually identified with the $\p_\un$
by virtue of (\ref{ffe}).)

Because $N_\N$ commutes to the generators of $su(2,2|\N )$,
the Fock module $F$ of $su(2,2|\N)$ decomposes into submodules $F_\ga$
of $su(2,2|\N)$ classified by eigenvalues of $N_\N$, i.e. spanned
by the vectors satisfying
\be
\label{eig}
N_\N* |\Phi \rangle  = \ga |\Phi \rangle\,.
\ee
According to the definition (\ref{N}),
the vacuum has definite eigenvalue of $N_\N$
\be
N_\N *\vac =-\f{\N}{2}\,.
\ee
Because
\be
\label{Nf1}
[N_\N ,f ]_* = (N_b  + N_{\bar{\phi}}-N_a - N_{\phi} )f\,,
\ee
where
\be
\label{Nab}
N_a = a_\hga \f{\p}{\p a_\hga }\,,\qquad N_b = b^\hga \f{\p}{\p b^\hga }\,,
\ee
\be
\label{Nphibphi}
N_{\phi} = \phi_i \f{\p}{\p \phi_i }\,,\qquad N_{\bar{\phi}} =
\bar{\phi}^j \f{\p}{\p \bar{\phi}^j }\,,
\ee
the eigenvalue  in (\ref{eig}) takes values
\be
\label{n}
\ga=m- \f{\N}{2} \,,\qquad  m\in {\bf Z}\,,
\ee
i.e., $\ga$ is an arbitrary half-integer for odd $\N$
and an arbitrary integer for even $\N$.

{}From (\ref{Nf1}) it follows that the fields contained in $F_\ga$
are $C(\ova_\dga , b^\gb ,\bar{\phi}|x)$ with
\be
(N_a - N_b - N_{\bar{\phi}} +m )
C(\ova_\dga , b^\gb ,\bar{\phi}|x) =0\,.\,
\ee
{}From (\ref{sig}) it follows that the relationship between
the number of inner indices and spin $s$ of a field in the
supermultiplet is
\be
s = \half  |  N_{\bar{\phi}} - m  | \,.
\ee
Let, for definiteness, $m$ be some non-negative integer.
Then the following dynamical massless fields appear
in the multiplet
\bee
c_{\ga_1 \ldots \ga_m (x)}\,,\quad&{}&
c_{\ga_1 \ldots \ga_{m-1}},{}_i (x)\,,\quad \ldots \quad
c_{\ga_1 \ldots \ga_{m-k}},{}_{i_1 \ldots i_k} (x)\,,\ldots \nn\\
c_{i_1 \ldots i_m} (x)\,,\quad     &{}&\ldots\quad
c^{\dgb_1\ldots \dgb_k},{}_{i_1 \ldots i_{m+k }} (x)\,,\ldots
c^{\dgb_1\ldots \dgb_{\N -m}},{}_{i_1 \ldots i_{\N}} (x)\,.
\eee

The modules $F_\ga$ describe various
supermultiplets of $su(2,2|\N )$ with the type of a
conformal supermultiplet characterized by  $\ga$.
The most interesting case is
$\ga=0$. According to (\ref{n}) $\ga =0$ requires $\N$ to be even.
Let us show that  the $\ga=0$ supermultiplets are
self-conjugated conformal supermultiplets. These include
$\N =2$ hypermultiplet and $\N=4$ Yang-Mills supermultiplet.

{}From (\ref{n}) it follows that
$\ga =0$ implies $m= \f{\N}{2}$
and, therefore, the set of dynamical massless
fields in the supermultiplet contains
\be
c_{\ga_1 \ldots \ga_{\f{\N}{2}}} (x)\,,\quad
c_{\ga_1 \ldots \ga_{\f{\N}{2}-1}},{}_i (x)\,,\ldots
c_{\ga_1 \ldots \ga_{\f{\N}{2}-k}},{}_{i_1 \ldots i_k} (x)\,,\ldots
c_{i_1 \ldots i_{\f{\N}{2}}} (x)\,,
\ee
along with
\be
c^{\dgb_1},{}_{i_1 \ldots i_{\f{\N}{2}+1 }} (x)\,,\ldots
c^{\dgb_1\ldots \dgb_k},{}_{i_1 \ldots i_{\f{\N}{2}+k }} (x)\,,\ldots
c^{\dgb_1\ldots \dgb_{\f{\N}{2}}},{}_{i_1 \ldots i_{\N}} (x)\,.
\ee

In particular, for the case $\N =0$ we obtain a single scalar
field. For $\N=2$ the hypermultiplet appears
\be
c_\ga (x)\,,\qquad c_i (x)\,,\qquad c^\dgb,{}_{ij} \,.
\ee
For $\N=4$ we find the $\N=4$ Yang-Mills multiplet
\be
\label{N4mul}
c_{\ga\gb}(x) \,,\qquad c_{\ga},{}_i (x)\,,\qquad c_{ij} (x)\,,\qquad
c^{\dga},{}_{ijk} (x)\,,\qquad
c^{\dga\dgb},{}_{ijkl} (x)\,.
\ee

The algebra
\hsa contains the infinite-dimensional subalgebra\hfil\\
\csa being the
centralizer of $N_\N$ in \hsa, i.e.\hfil\\
\csa is spanned by the elements
$f\in$ \hsa that commute to $N_\N$
\be
\label{Nf}
[N_\N , f ]_* =0\,.
\ee
This is equivalent to
\be
\label{en}
(N_a + N_{\phi} )   f = (N_b +N_{\bar{\phi}}) f\,.
\ee
Because of (\ref{Nf}), the algebra \csa is not simple,
containing ideals $I_\ga$ spanned by the
elements of the form $h=(N_\N -\ga)*f$, $[f , N_\N]_* =0$.
Now we observe that
the operator $N_\N -\ga$ trivializes on the module $F_\ga$. Therefore,
$F_\ga$ forms a module over the quotient
algebra \hsan = \csa$\!/I_\ga$.
Thus, different $\ga$ correspond to different subsectors
(quotients) of \csa$\ls$ associated with different supermultiplets.

Let us note that in \cite{FLA}
the algebra \csa was called
$shsc^\infty (4|\N)$, while the algebra \hsan
was called $shsc_\ga^0 (4|\N )$. It was argued in \cite{FL} that
it is the algebra \csa that plays a role of the $4d$
higher spin conformal algebra, while the algebra
$shsc_\ga^0 (4|\N )$
is unlikely to allow consistent conformal higher spin interactions.
The conclusions of the present paper are somewhat opposite.
We will argue that consistent conformal theories
exhibiting the higher spin conformal symmetries may correspond
to the simple (modulo
the trivial center associated with the unit element)
algebras \hsa or \hsan and their further simple
reductions of orthogonal or symplectic type (see subsection
\ref{Self-conjugated  Supermultiplet Reduction}).
Note that in \cite{5d} it is shown that
the $\N=0$ algebra $hu_0 (1,0|8)$  admits consistent cubic higher spin
interactions in $AdS_5$. An interesting problem for the future is to
extend the proposed conformal form of massless field equations to
the case with dynamical conformal higher spin gauge fields
(1-forms) included. Taking into account that the conformal higher spin
gauge theory framework allows for off-mass-shell formulation
of higher spin constraints for higher spin gauge fields \cite{FL,FMV}
such an  extension is expected to
be of crucial importance for the construction of nonlinear
off-mass-shell higher spin dynamics.

Finally, let us note that it is straightforward to introduce color
indices by allowing the Fock vacuum to be a column
$$
|\Phi \rangle = \left( \begin{array}{cc}
                   E^p (\ova ,b|x) *\vac     \\
                   O^r (\ova ,b|x) *\vac
            \end{array}
    \right)\,,
$$
\be
\label{FC}
{}
\ee
where $E^p (\ova , b)$ and $O^r (\ova , b)$  are, respectively,
even and odd functions of the spinor variables $\ova_\dga$
and $b^\ga$
\be
E^p (-\ova ,- b|x) = E^p (\ova , b|x)\,,\qquad
O^r (-\ova ,- b|x) =- O^r (\ova , b|x)
\ee
and $p=1\div m$, $r= 1\div n$. The algebra $hu(m,n|8)$
realized by the matrices
(\ref{1}) acts naturally on such a column. It is clear
that the fermionic Fock states due to the Clifford
variables $\phi_i $ and $\bar{\phi}^i$ give rise to a particular
realization of this construction. Most of the content of this
paper applies equally well to the both constructions.
We will mainly use the Clifford realization because, although
being less general, it has larger supersymmetries explicit.
Note that the algebras $hu(m,n|8)$ are not supersymmetric for generic
$m$ and $n$ (i.e. they do not contain the usual supersymmetry algebras
as finite-dimensional subalgebras). They are $\N =1$ conformal
supersymmetric however for the case $m=n$ and acquire more
supersymmetries when $m=n$ are multiples of $2^q$.
The superalgebras $hu(n2^{\N-1},n2^{\N-1}|8)$ and their orthogonal
and symplectic reductions $ho(n2^{\N-1},n2^{\N-1}|8)$
and $husp(n2^{\N-1},n2^{\N-1}|8)$ act on the set of $n$ copies
of $\N-$extended conformal supersymmetry multiplets.
In this notation it is the $n\to\infty$ limit that plays
a crucial role in the string theory AdS/CFT correspondence
\cite{AdS/CFT,GKP,AdS/CFTW,AdS/CFTrev}.
(For more detail on the properties of $hu(m,n|8)$ we refer
the reader to \cite{KV1}. See also section
\ref{Self-conjugated  Supermultiplet Reduction}.)

\subsection{Generic Solution}
\label{Generic Solution}

Once the massless equations are reformulated in the form
(\ref{ffe})  and the vacuum background field $\go_0$ is
represented in the pure gauge form (\ref{wvg}),
generic solution of the massless equations acquires the
form
\be
\label{sol}
|\Phi (\ova, b,\bar{\phi} |x )\rangle = g^{-1}(\ova, b;\phi, \bar{\phi} |x ) *
|\Phi_0 (\ova, b,\bar{\phi})\rangle \,,
\ee
where $|\Phi_0 (\ova, b,\bar{\phi})\rangle =
|\Phi_0 (\ova, b,\bar{\phi}|x_0)\rangle$ at such a point $x_0$
that $g(x_0 ) = 1$.
For the gauge function $g$ (\ref{gfc}) one obtains
with the help of (\ref{uf}) the general solution in the form
\bee
\label{solu}
C(\ova ,b,\bar{\phi} |x) &=& \f{1}{(2\pi)^2 } \int d^2 \ovs d^2 \ovt
C_0 (\ova+\ovs  ,b^\ga +x^\ga{}_\dgb \ovt^\dgb,\bar{\phi} )
\exp \ovs_\dga \ovt^\dga
\nn\\
&=&\exp \Big (-x^\ga{}_\dgb \f{\p^2}{\p b^\ga \p \ova_\dgb}\Big )
C_0 (\ova  ,b,\bar{\phi} )\,.
\eee
Here $C_0 (\ova  ,b,\bar{\phi} )$ is an arbitrary function
of the  variables $\ova_\dga$, $b^\ga$, and $\bar{\phi}^i$.
It provides ``initial data" for the problem. Choosing
$C_0 (\ova  ,b,\bar{\phi} )$ in the form
\be
C_0 (\ova  ,b,\bar{\phi} ) = c_0 (\bar{\phi})
\exp (b^\ga\eta_\ga + \bar{\eta}^\dgb \ova_\dgb )\,,
\ee
where $\eta_\ga$ and $\bar{\eta}^\dgb$ are (commuting) spinor parameters,
one obtains the plane wave solution
\be
C(\ova ,b,\bar{\phi} |x) = c_0 (\bar{\phi})
\exp (b^\ga\eta_\ga + \bar{\eta}^\dgb \ova_\dgb -\bar{\eta}^\dgb
x^\ga{}_\dgb \eta_\ga )\,
\ee
with the light-like wave vector
\be
k_{\ga\dgb} = \eta_\ga \bar{\eta}_\dgb\,.
\ee

Let us note that our approach exhibits deep similarity with the
twistor theory \cite{P1,P2,PR}. The conformal spinors
$a_\hga$ and $b^\hgb$, which play a key role in the construction
as the generating elements of the star product algebra,
are analogous to the quantum twistors of \cite{P2}.
An important difference however is that we do not assume
that $x_\hga{}^\hgb$ maps one pair of twistors to another. In our
construction $x$-space is treated as the base manifold while the
spinor variable generate the Fock-space fiber.
At the first stage the field variables (sections of the
vector fiber bundle)
are arbitrary functions of the variables $x_\hga{}^\hgb$,
$a_\hga$, and $b^\hgb$ so that there is no direct
relationship between the two sectors. They are
linked to each other by the equations of motion (\ref{ffe})
which imply that solutions of the massless equations are flat sections
of the Fock fiber bundle over space-time. This allows one to solve
the field equations using the star product techniques as explained
in this section, thus providing a counterpart of the twistor
contour integral formulas.
 Typical twistor combinations of the coordinates
and spinors (such as, e.g. the combination $x^\ga{}_\dgb \tilde{t}^\dgb$
in (\ref{solu})) then appear as a result of insertion of the
gauge function $g$ (\ref{gfc}) that reproduces
Cartesian coordinates in the flat space.  Another difference mentioned
at the end of the section \ref{Generalities} is due to systematic use of
the language of $x-$ space differential forms in our approach.
In fact, this allows us to
handle higher spin gauge symmetries in a systematic way
that is of key importance for the analysis of interactions.

Note that our approach can be used in any other
coordinate system by choosing other forms of $g$.
Provided that the higher spin symmetry algebra contains
conformal subalgebra (as is the case in this paper),
analogously to the twistor theory,
it works for any conformally flat
geometry because conformally flat gravitational
fields satisfy the zero curvature equations of the
conformal algebra. For example, it can be
applied to the $AdS_4$ space. The generic solution
of the massless field equations
in  $AdS_4$ was found by a similar method in \cite{BV,Gol}.

\subsection{Reality Conditions}
\label{Reality Conditions and Reduction}
So far we considered complex fields.
The conjugated multiplet is described by the right module
formed
by the states
\be
\label{con1}
\langle{\Psi}(\ova , {b}^\gb , \phi_ j|x)|  =
\bvac * G( b,  \ova ,{\phi}_j |x )\,,
\ee
where the vacuum
$\bvac$ is defined by the conditions\footnote{Let us note that
the vacua $\vac$ and $\bvac$
belong to algebraically distinct sectors of the star product
algebra: the computation of $\vac * \bvac $ leads to a
divergency.}
\be
\label{ccr}
\bvac * a_\ga =0 \,,\qquad \bvac * \ovb^\dgb =0\,,\qquad
\bvac * \bar{\phi}^j =0\,,
\ee
i.e.,
\be
\label{bvac}
\bvac
=2^{4-\N} \exp{2\Big (
a_\ga b^\ga - \ova_\dga \ovb^\dga    +\phi_i \bar{\phi}^i }\Big)\,.
\ee
In components,
\be
G({b} , \ova ,{\phi}|x)
= \sum_{m,n=0}^\infty \sum_{k=0}^\N\frac{1}{m!n!k!}
g_{\ga_1 \ldots \ga_n }{}^{\dgb_1 \ldots \dgb_m }
{}^{j_1 \ldots j_k} (x)
b^{\ga_1} \ldots b^{\ga_n} \ova{}_{\dgb_1} \ldots \ova{}_{\dgb_m}
{\phi}_{j_1} \ldots {\phi}_{j_k}\,.
\label{cdefc}
\ee
Analogously, one can consider the row representation of $hu(m,n|8)$.

The dynamical equation for $\langle {\Psi} |$ is
\be
d \langle {\Psi} | +\langle {\Psi}| *\go_0 =0 \,.
\label{cffe}
\ee

To impose the reality conditions let us define the
involution $\dagger$ by the relations
\be
\label{inv}
(a_\ga )^\dagger = i \ovb_\dga\,,\qquad
(b^\ga )^\dagger = i \ova^\dga\,,\qquad
(\ova_\dga )^\dagger = i b_\ga\,,\qquad
(\ovb_\dga )^\dagger = i a_\ga\,,
\ee
\be
 (\phi_i )^\dagger = \bar{\phi}^i\,,\qquad
 (\bar{\phi}^i )^\dagger = {\phi}_i\,.
\ee
Since an involution is defined to reverse an order of product
factors
\be
\label{ord}
(f*g)^\dagger = g^\dagger *f^\dagger\,
\ee
and conjugate complex numbers
\be
\label{alin}
(\mu f)^\dagger = \bar{\mu} f^\dagger\,,\qquad \mu \in {\bf C}\,,
\ee
one can see that $\dagger$ leaves invariant the defining
relations (\ref{osc}) and (\ref{fosc}) of the star product
algebra and has the involutive property $(\dagger )^2 = Id$.
By (\ref{ord}) the action of $\dagger$ extends to
an arbitrary element $f$ of the star product algebra. Since
the star product we use corresponds to the totally (anti)symmetric
(i.e. Weyl) ordering of the product factors, the result is simply
\be
\label{dag}
(f(a,\ova, b, \ovb ;\phi, \bar{\phi}))^\dagger =
\bar{f}^r (i\ovb ,ib ,i\ova ,ia ; \bar{\phi}, \phi )\,,
\ee
where $f^r $ implies reversal of the order of the Grassmann
factors $\phi$ and $\bar{\phi}$, i.e.
$f^r = (-1)^{\half n(n-1)} f $ if $f$ is an
order-$n$ polynomial in $\phi$ and $\bar{\phi}$.
One can check directly with the formulas (\ref{prod})
and (\ref{cprod}) that (\ref{dag}) defines an involution of
the star product algebra.

Let us note that in the general case of $hu(m,n |8)$ the
involution $\dagger$ is defined  by (\ref{inv})
along with the usual hermitian conjugation in the matrix sector.
The column (\ref{FC}) is mapped to the
appropriate conjugated row vector
\be
\label{Psi}
\langle\Psi | = \left( \bvac *
                   \bar{E}_p (\ova ,b|x) \,,\quad
                  \bvac * \bar{O}_r (\ova ,b|x)
    \right)\,.
\ee

The reality conditions on the elements of the higher spin algebra
have to be imposed in a way consistent with the form of the
zero curvature equations (\ref{ezc}). This is equivalent to singling
out a real form of the higher spin Lie superalgebra.
With the help of any involution $\dagger$ this is achieved
by imposing the reality conditions
\be
\label{reco}
f^\dagger = - i^{\pi (f)} f
\ee
($\pi (f) =0$ or 1). This condition defines the real higher spin
algebra $hu(m,n|2M)$ for $M$ pairs of oscillators. For the
Clifford realization of the matrix part one arrives at the
real algebra \hsa.

Let us stress that the condition (\ref{reco})
extracts a real form of the Lie superalgebra built from the
star product algebra but not of the associative star product
algebra itself. The situation is very much the same as for
the Lie algebra $u(n)$ singled out from the complex Lie algebra
of $n\times n$ matrices by the condition (\ref{reco}) ($\pi =0$
for the purely bosonic case) with $\dagger$ identified with the
hermitian conjugation. Antihermitian matrices form the Lie algebra
but not an associative algebra. In fact, the relevance of the
reality conditions of the form (\ref{reco}) is closely related to
this matrix example because it guarantees that the spin 1
(i.e., purely Yang-Mills) part of the higher spin algebras is
compact. More generally, these reality conditions guarantee
that the higher spin symmetry admits appropriate unitary highest weight
representations (see section \ref{Field Theory - Doubleton Duality}).
Note that in the sector of
the conformal algebra $su(2,2)$ the reality condition (\ref{reco})
is equivalent to (\ref{re}).

Now one observes that
\be
( \vac )^\dagger = \bvac \,.
\ee
Imposing the reality condition analogous to (\ref{reco}) on the
conformal matter modules
\be
(|\Phi \rangle )^\dagger = -i^{\pi(\Phi )}\langle \Psi |
\ee
equivalent to
\be
C^\dagger = - i^{\pi (C)} G\,,
\ee
one finds by (\ref{alin}) that the matter fields
$g_{\ga_1 \ldots \ga_n }{}^{\dgb_1 \ldots \dgb_m }
{}^{j_1 \ldots j_k} (x)$ are complex conjugated to
$c^{ \gb_1 \ldots \gb_m}{}_{\dga_1 \ldots \dga_n}{}_{j_1 \ldots j_k}(x)$
up to some sign factors originating from the
factors of $i$ and the reversal of the order of Grassmann factors in the
definition of $\dagger$ (\ref{dag}). For example, for the scalars we have $g(x)
= - \bar{c} (x)$, for the spin 1 field strengths $\overline{ (g_{\ga\gb} )} =
c_{\dga  \dgb} $, etc.

Let us note that the operator $N_\N$ is self-conjugate
\be
\label{Nd}
N_\N^\dagger = N_\N\,.
\ee
As a result, if $|\Phi\rangle $
satisfies (\ref{eig}) the conjugated module satisfies
\be
\langle \Psi |*  (N_\N - \ga ) =0
\ee
with the same real $\ga$.

\subsection{Antiautomorphism Reduction and
Self-Conjugated  Supermultiplets}
\label{Self-conjugated  Supermultiplet Reduction}

The algebras $hu(m,n |2p)$ were shown \cite{KV1} to admit
truncations of the orthogonal and symplectic types, $ho(m,n|2p)$
and $husp (m,n|2p)$, singled out by the
appropriate antiautomorphisms of the underlying
associative algebra. Let us recall some definitions.

Let $B$ be some algebra with the (not
necessarily associative) product law $\diamond $. A
linear invertible map $\tau $ of $B$ onto
itself is called automorphism if $\tau (a\diamond b)=\tau (a)\diamond \tau
(b)$ (i.e., $\tau $ is an isomorphism of the algebra to itself.)
A useful fact is that the subset of elements $a\in B$ satisfying
\begin{equation}
\tau (a)=a
\label{taa}
\end{equation}
spans a subalgebra $B_{\tau }\subset B$. It is customary in physical
applications to use this property to obtain reductions. In particular,
applying the boson-fermion automorphism which changes the sign of the fermion
fields, one obtains reduction to the bosonic sector. Another example is
provided by the operation $\tau (a) = - a^t$ of the Lie algebra $gl(n)$ ($t$
implies transposition). The condition (\ref{taa}) then singles out the
orthogonal subalgebra $o(n)\subset gl(n)$.

A linear invertible  map $\rho $ of an algebra onto itself is called
antiautomorphism if it reverses the order of product factors
\begin{equation}
\rho (a\diamond b)=\rho (b)\diamond \rho (a)\,.
\label{ant}
\end{equation}
One example is provided by the transposition of matrices.
More generally, let $A=Mat_{M}({\bf C})$ be the algebra of $M\times M$
matrices over the field of complex numbers,
with elements $a^{i}{}_{j}$ ($i,j=1\div M$) and the product law
\begin{equation}
(a\circ b)^{i}{}_{j}=a^{i}{}_{k}b^{k}{}_{j}\,.
\end{equation}
Let $\eta ^{ij}$ be a nondegenerate bilinear form with the inverse $\eta
_{ij}$, i.e.,
\begin{equation}
\eta ^{ik}\eta _{kj}=\delta _{j}^{i}\,.
\end{equation}
It is elementary to see that the mapping
\begin{equation}
\label{reta}
\rho _{\eta }(a)^{i}{}_{j}=\eta ^{ik}a^{l}{}_{k}\eta _{lj}  \label{antmat}
\end{equation}
is an antiautomorphism of $Mat_{M}({\bf C})$. If the bilinear form $\eta
^{ij}$ is either symmetric
\begin{equation}
\eta _{S}^{ij}=\eta _{S}^{ji}  \label{sim}
\end{equation}
or antisymmetric
\begin{equation}
\eta _{A}^{ij}=-\eta _{A}^{ji}  \label{asym}\,,
\end{equation}
the antiautomorphism $\rho _{\eta }$ is involutive, i.e. $\rho _{\eta
}^{2}=Id$. One can extend the action of $\rho$ to rows and columns
in the standard way by raising and lowering indices with the aid of
the bilinear form $\eta^{ij}$ and its inverse.

The star product algebra admits the antiautomorphism defined by the
relations
\be
\label{ant1}
\rho (a_\hga ) = ia_\hga\,,\qquad
\rho (b^\hga ) = i b^\hga\,,
\ee
\be
\label{antp}
\rho (\phi_i ) = \phi_i \,, \qquad  \rho (\bar{\phi}{}^j )
=\bar{\phi}{}^j\,.
\ee
This definition is consistent with
the property (\ref{ant}) and the basis commutation relations
(\ref{fosc}) and (\ref{osc}). For the generic element of the
star product algebra we have
\be
\label{auf}
\rho (f(a,\ova, b, \ovb ;\phi, \bar{\phi})) =
f^r (ia ,i\ova ,i b ,i\ovb ; {\phi}, \bar{\phi} )\,.
\ee

Because the product law in a
Lie superalgebra has definite symmetry properties,
any antiautomorphism $\rho$ of an
associative algebra $A$ that respects
the ${\bf Z}_2$ grading used to define the Lie superalgebra $l_A$ by
 (\ref{compm}), induces an automorphism of
$\tau_\rho $ of $l_A$ according to
\be
\tau_\rho (f) = -(i)^{\pi (f)} \rho (f)\,.
\ee
As a result, any antiautomorphism $\rho$ of the associative
algebra $A$ allows one to single out a subalgebra of $l_A$ by
imposing the condition (\ref{taa})
\be
\label{arec}
f = -(i)^{\pi (f)} \rho (f)\,.
\ee

For example,
for $A=Mat_{M}({\bf C})$, $l_{A}=gl_{M}({\bf C})$. The
subalgebras of $gl_{M}$ singled out by the condition (\ref{arec}) with $\tau
_{S}=-\rho _{S}$ and $\tau _{A}=-\rho _{A}$ are $o(M|{\bf C})$ and
$sp(M|{\bf C})$, respectively, because the condition (\ref{arec})
just implies that the form $\eta ^{ij}$ is invariant.
Note that analogously, one can define
involutions via nondegenerate hermitian forms. If $\dagger $ is such an
involution of $Mat_{M}({\bf C})$ defined via a positive-definite
Hermitian form, the resulting Lie algebra is $u(M)$.

The algebras $ho(m,n|2p)$ and $husp(m,n|2p)$ \cite{KV1}
are real Lie superalgebras
satisfying the reality conditions (\ref{reco}) and the
reduction condition (\ref{arec}) with the antiautomorphism
$\rho$ defined by the relations (\ref{ant1})
along with the definition (\ref{reta}) for the action on the
matrix indices with some
$(m+n) \times (m+n) $ bilinear form $\eta^{ij}$ that is
block-diagonal in the basis (\ref{1}) and is either symmetric,
$\eta^{ij} = \eta_S^{ij}$, or antisymmetric,
$\eta^{ij} = \eta_A^{ij}$. For $\eta_S^{ij}$ and $\eta_A^{ij}$
we arrive, respectively,  at the
algebras $ho(m,n|2p)$ and $husp(m,n|2p)$ with the spin 1
Yang-Mills subalgebras $o(m)\oplus o(n)$ and
$usp(m)\oplus usp(n)$ in the sector of elements independent of the
spinor oscillators.

For the particular case of the algebra \hsa with the Clifford
star product realization of the matrix part, the antiautomorphism
$\rho$ is defined in (\ref{antp}). As argued in \cite{KV1} this
antiautomorphism is diagonal in the basis (\ref{1}) for even $\N$
and off-diagonal for odd $\N$. To see this one can check that
the element
$$
K= \left( \begin{array}{cc}
                   I    &   0   \\
                   0    &  -I
            \end{array}
    \right)
$$
\be
\ee
identifies in terms of the Clifford algebra with the element $\Gamma$
being the product of all Clifford generating elements (in the basis
with the diagonal symmetric form in the defining Clifford relations)
so that $\Gamma^2 = I$, $\{\Gamma , \phi_i \} =0$,
 $\{\Gamma , \bar{\phi}^i \} =0$. Then one observes that
\be
\label{Ga}
\rho (\Gamma) = (-1)^{\N}\Gamma\,.
\ee
Therefore we confine ourselves to the case of even $\N$.
In fact, this case is most interesting because it admits the self-conjugated
supermultiplets.\footnote{Note that to make $\rho$ diagonal
for the case of odd $\N$ one can modify its definition
in a way that breaks
the $su(\N)$ algebra to at least $su(\N-1)$. To this end it is enough to
modify (\ref{antp}) to $\rho ({\phi_1} )= \bar{\phi^1}$,
$\rho ({\bar{\phi}^1} )= \phi_1$ leaving the definition of $\rho$ for
$\phi_j$ and $\bar{\phi}^j$ with $j>1$ intact. This will bring an
additional sign factor into (\ref{Ga}).}

Following
the analysis of \cite{KV1} one can check that the algebras
extracted by the condition (\ref{arec}) for $\N = 4p$ and $\N = 4p+2$
are isomorphic to
\be
\label{sub1}
ho(2^{4p-1},2^{4p-1}|8) \qquad\mbox{for $\N = 4p$}
\ee
and
\be
\label{sub2}
husp(2^{4p+1},2^{4p+1}|8) \qquad\mbox{for $\N = 4p+2$}\,.
\ee
In particular, for $\N = 2$ and $\N =4$ we get
$husp(2,2|8)$ and $ho(8,8|8)$, respectively.
Let us stress that the elements of the $su(2,2)$ algebra
(\ref{sl4}), (\ref{Q}), (\ref{T}) all satisfy (\ref{arec})
and, thus belong to the truncated superalgebras
$ho(2^{4p-1},2^{4p-1}|8)$ and $husp(2^{4p+1},2^{4p+1}|8)$.
The same is true for the algebra $osp(2\N , 8)$ spanned by various
bilinears of the superoscillators.

One observes that
\be
\rho( N_\N ) = - N_\N \,.
\ee
This means that the reduction (\ref{arec}) is possible for the
algebras $hu_\ga (2^{\N-1},2^{\N-1}|8)$ iff $\ga =0$.
We call the algebras resulting from the reduction of
$hu_0 (2^{\N-1},2^{\N-1}|8)$ by the antiautomorphism $\rho$
as $ho_0 (2^{4p-1},2^{4p-1}|8)$  for $\N = 4p$ and
$husp_0 (2^{4p+1},2^{4p+1}|8)$ for $\N = 4p+2$.
The algebra $ho_0 (8,8|8)$ is the minimal higher spin
conformal symmetry algebra associated with the linearized
$\N=4$ Yang-Mills supermultiplet, while the algebra
$hu(2,2|8)$ is the minimal higher spin conformal algebra
associated with the $4d$ $\N=2$ massless hypermultiplet. The minimal
purely bosonic $4d$ conformal higher spin
algebra associated with the spin-0 $4d$ massless scalar field
is $ho_0(1,0|8)$. This algebra  was recently discussed
by Sezgin and Sundell  \cite{SSd} in the context of the $AdS_5$
higher spin gauge theory (these authors denoted this
algebra $hs(2,2)$). Note that the  higher spin
gauge algebra of $AdS_5$ higher spin gauge theory
dual to the $\N=4$ $SYM$ theory is $ho_0 (8,8|8)$.

In the matter sector we define
\be
\label{fac1}
\rho (|\Phi\rangle ) =
\rho (C * {\vac}) = \f{1}{\N !}
\gvep^{j_1\ldots j_\N} \bvac * \phi_{j_1} *\ldots \phi_{j_\N} * \rho ( C)\,,
\ee
\be
\label{fac2}
\rho (\langle \Psi |) =
\rho ( {\bvac} * G) =
\f{1}{\N !}
\gvep_{i_1\ldots i_\N}\rho (G)* \bar{\phi}^{i_1} *\ldots
 *\bar{\phi}^{i_\N} * \vac\,
\ee
to make (\ref{antp}) consistent with (\ref{xx}) and (\ref{ccr}).
Now we can impose the reduction condition on the matter fields
\be
\label{rphi}
\rho (|\Phi \rangle ) = -i^{\pi (\Phi )} \langle {\Psi }|\,,
\ee
which is consistent with  (\ref{arec}).
Along with the fact that $\langle {\Psi} |$ describes the
conjugated fields subject to the hermiticity condition (\ref{reco})
this imposes the reality conditions on  the
left module $|\Phi \rangle$
\be
\rho (|\Phi \rangle ) = (|\Phi \rangle )^\dagger \,.
\ee

For the self-conjugated supermultiplets with $\ga =0$
this imposes the reality conditions on the fields of
the same multiplet. In terms of components this implies  that
\be
\bar{c}_{\gb_1 \ldots \gb_m \dga_1 \ldots \dga_n }{}^{j_1 \ldots j_k} (x)
=\f{1}{(\N -k )!} \gvep^{j_1 \ldots j_k i_{\N - k} \ldots i_1}
c_{\ga_1 \ldots \ga_n \dgb_1 \ldots \dgb_m}
{}_{i_{\N - k} \ldots i_1} (x)
\,.
\ee
In particular, for the $\N = 4$ multiplet we have
\be
\bar{c}_{\ga\gb} = \f{1}{4!} \gvep^{ijkl}
c_{\dga\dgb}{}_{ijkl}\,,
\ee
\be
\bar{c}_{\ga}{}^i = \f{1}{6} \gvep^{ijkl}
c_{\dga}{}_{jkl}\,,
\ee
\be
\bar{c}^{ij} = \f{1}{2} \gvep^{ijkl}
c_{kl}\,.
\ee
The resulting set indeed corresponds to the real
$4d$ $\N = 4$ SYM
supermultiplet with six real scalars, four Majorana
spinors and one spin 1 field strength.

The special property of the self-conjugated supermultiplets
therefore is
that the antiautomorphism $\rho$ transforms them to themselves.
In other words, they are self-conjugated with
respect to the combined action of the conjugation $\dagger$ and
the antiautomorphism $\rho$.
The infinite-dimensional superalgebras
$ho_0 (2^{4p-1},2^{4p-1}|8)$  for $\N = 4p$ and
$husp_0 (2^{4p+1},2^{4p+1}|8)$ for $\N = 4p+2$ are
therefore shown to be the algebras
of conformal higher spin symmetries acting on the
self-conjugated supermultiplets.
Finally, let us note that the whole construction extends
trivially to the case with $n$ supermultiplets described by the
algebras $hu(n2^{\N-1},n2^{\N-1}|8)$ and their further reductions
$ho(n2^{\N-1},n2^{\N-1}|8)$, $husp(n2^{\N-1},n2^{\N-1}|8)$ and
$hu_0 (n 2^{\N-1},n 2^{\N-1}|8)$,\hfil\\
$ho_0 (n 2^{4p-1},n 2^{4p-1}|8)$,  $husp_0 (n 2^{4p+1},n 2^{4p+1}|8)$
(the latter algebras are assumed to be defined as before as
the  quotients of the centralizer of $N_\N$).

\section{4d Conformal Higher Spin Symmetries}
\label{4d Conformal Higher Spin Symmetries}

The system of equations (\ref{ezc}), (\ref{ffe}) is
invariant under the infinite-dimensional local
conformal higher spin symmetries
(\ref{gtw}) and
\be
\gd |\Phi\rangle =\gep* |\Phi\rangle\,.
\label{gtc}
\ee
The reduction condition (\ref{rphi}) reduces the higher spin algebra
to the subalgebra (\ref{sub1}) or (\ref{sub2}) with the symmetry
parameters $\gep (a,b;\phi,\bar{\phi} |x )$ satisfying the condition
(\ref{arec}).

Once a  particular vacuum solution
$\go_0$ is fixed,
the local higher spin symmetry (\ref{gtc}) breaks down to the global
higher spin symmetry (\ref{e0}).
Therefore the system (\ref{ffe}) is invariant under the
infinite-dimensional algebra \hsa
of the global $4d$ conformal higher spin symmetries
\be
\label{gs}
\gd |\Phi\rangle=\gep_0*|\Phi\rangle\,,
\ee
where $\gep_0$ satisfies the equation (\ref{ipgs}) with the flat
connection (\ref{fc}).
After the higher components in $C(\ova , b,\bar{\phi}|x)$ are expressed
via the higher space-time derivatives of the dynamical massless
fields according to (\ref{deqr}) this implies invariance of the $4d$
massless equations for all
spins (\ref{4deq}) under the global conformal higher spin
symmetries.  Thus, the fact that massless equations are
reformulated in the form of the flatness conditions  (\ref{ffe})
supplemented
with the zero-curvature equation (\ref{ezc}) makes higher spin conformal
symmetries of these equations manifest.  Note that because of (\ref{deqr})
and of the quantum-mechanical nonlocality of the star product
(\ref{prod}), the higher degree of $\gep_0 (a ,b|x)$ as a polynomial of
$a$ and $b$ is, the higher space-time derivatives appear in the
transformation law. This is a particular manifestation of the well
known fact that the higher spin symmetries mix higher derivatives of the
dynamical fields.

The explicit form of the transformations can be obtained by
the substitution of (\ref{deqr}) into (\ref{gs}). In practice, it is
most convenient to evaluate the higher spin conformal transformations
for the generating parameter
\be
\label{xi}
\xi (a,\ova ,b,\ovb,\phi ,\bar{\phi} ;h,\ovh , j ,\ovj,\eta, \bar{\eta} )
=\xi
\exp(h^\ga a_{\ga}+\ovh^\dga \ova_\dga +j_\gb b^{\gb}+\ovj_\dga \ovb^\dga
+\phi_i \bar{\eta}^i +\eta_i \bar{\phi}^i )\,,
\ee
where $\xi$ is an infinitesimal parameter.
The polynomial  symmetry
parameters can be obtained via differentiation of
$
\xi (a,\ova ,b,\ovb,\phi ,\bar{\phi} ;h,\ovh , j ,\ovj,\eta, \bar{\eta} )
$
with respect to the commuting ``sources"  $h^\ga$, $\ovh^{\dga}$
$j_\ga$, $\ovj_{\dga}$ and anticommuting ``sources"
$\bar{\eta}^i$, $\eta_i$. For the case of the flat space,
using (\ref{e0}), (\ref{gfc}) and the star product
(\ref{prod}) we obtain upon evaluation of
elementary Gaussian integrals
\bee
&{}&\gep_0 (a,\ova ,b,\ovb, \phi,\bar{\phi} ;h,\ovh ,
j ,\ovj,\eta, \bar{\eta}|x ) =\nn\\
&{}&\xi
\exp(h^\ga a_{\ga}+\ovh^\dga \ova_\dga +j_\gb b^{\gb}+\ovj_\dga \ovb^\dga
+\phi_i \bar{\eta}^i +\eta_i \bar{\phi}^i +j_\ga x^\ga{}_\dgb \ovb^\dgb
-a_\ga x^\ga{}_\dgb \bar{h}^\dgb
)\,.
\label{ez}
\eee
Substitution of $\gep_0$ into (\ref{gs}) gives
the global higher spin conformal symmetry transformations
induced by the parameter (\ref{xi})
\be
\delta
|\Phi(\ova , b,\bar{\phi} |x) \rangle =
\delta C(\ova , b,\bar{\phi}|x )* \vac \,,
\ee
where
\bee
\label{dec}
\delta C(\ova , b,\bar{\phi} |x ) &=& \xi
\exp(\ovh^\dga \ova_\dga +j_\gb b^{\gb}+
\eta_i \bar{\phi}^i -j_\ga x^\ga{}_\dgb \tilde{h}^\dgb
-\half \tilde{j}_\dga \tilde{h}^\dga +\half j_\ga h^\ga
- \half \eta_i \bar{\eta}^i )\nn\\
&{}&C(\ova_\dga - \tilde{j}_\dga -j_\gb x^\gb{}_\dga , b^\ga
+ h^\ga - x^\ga{}_\dgb \tilde{h}^\dgb ,
\bar{\phi}^i - \bar{\eta}^i |x)\,.
\eee
 Such a compact form of the higher spin
conformal transformations is a result of the
reformulation of the dynamical equations in the unfolded form of the
covariant constancy conditions, i.e.,
in terms of a flat section of the Fock fiber bundle.
Differentiating with respect to the sources
one derives explicit expressions for the particular
global higher spin conformal transformations.

For at most quadratic
conformal supergenerators acting on
$C(\ova , b,\bar{\phi}|x )$  one  obtains  with the help of
(\ref{deqr})
\be
P_\ga{}^\dgb = \f{\p}{\p x^\ga{}_\dgb}\,,\qquad
P_\un = \sigma^{\ga\dgb}_\un P_{\ga\dgb} = \f{\p}{\p x^\un }\,,
\ee
\be
D= 1 + x^\un \f{\p}{\p x^\un } +\half \left ( \ova_\dga \f{\p}{\p \ova_\dga}
+ b^\ga \f{\p}{\p b^\ga} \right )\,,
\ee
\be
K_\dga{}^\gb = \ova_\dga b^\gb - x^\gb{}_\dgd \ova_\dga \f{\p}{\p \ova_\dgd}
-x^\gga{}_\dga \f{\p}{\p b^\gga} b^\gb - x^\gga{}_\dga x^\gb{}_\dgd
\f{\p}{\p x^\gga{}_\dgd}\,,
\ee
\be
L_\ga{}^\gb = b^\gb \f{\p}{\p b^\ga } +x^\gb{}_\dga \f{\p}{\p x^\ga{}_\dga}
-\half \delta_\ga^\gb \left(
b^\gga \f{\p}{\p b^\gga } +x^\gga{}_\dga \f{\p}{\p x^\gga{}_\dga}\right )\,,
\ee
\be
\bar{L}_\dga{}^\dgb = -\ova_\dga \f{\p}{\p \ova_\dgb} - x^\gga{}_\dga
\f{\p}{\p x^\gga{}_\dgb} +\half\delta_\dga^\dgb\left (
\ova_\dgd \f{\p}{\p \ova_\dgd} + x^\gga{}_\dgd
\f{\p}{\p x^\gga_\dgd}\right )\,,
\ee
\be
T^j{}_i = \half \delta_i^j - \bar{\phi}^j \f{\p}{\p \bar{\phi}^i}\,,
\ee
\be
\label{Qa}
Q_\ga{}^i = \bar{\phi}^i \f{\p}{\p b^\ga}\,,
\ee
\be
\label{Qda}
Q_\dga{}^i = \bar{\phi}^i \left( \ova_\dga -x^\gb{}_\dga  \f{\p}{\p b^\gb }
\right ) \,,
\ee
\be
\label{bQa}
\bar{Q}_i{}^\ga = \f{\p}{\p \bar{\phi}^i} \left ( b^\ga -
x^\ga{}_\dgb \f{\p}{\p \ova_\dgb }              \right )\,,
\ee
\be
\label{bQda}
\bar{Q}_i{}^\dga = -\f{\p}{\p \bar{\phi}^i}\f{\p}{\p \ova_\dga }\,.
\ee
Here the $x$-independent supercharges (\ref{Qa}) and (\ref{bQda})
correspond to the $Q$-super\-symmetry while the $x$-dependent
supercharges (\ref{Qda}) and (\ref{bQa}) correspond to the
$S$-super\-symmetry.

$F$ is a module over the algebra $osp(2\N,8)$ which, together with the
$u(1)$ algebra generated by the unit element of the star product
algebra, forms a maximal finite-dimensional subalgebra of the higher spin
algebra \hsa. (\ref{ffe}) contains the equations for all
supermultiplets. The $osp(2\N, 8)$ invariance
links together all free $4d$ conformal supermultiplets. The
explicit transformation laws  derived from
(\ref{dec}) are
\bee
\label{U}
U_{\ga\gb} &=&\f{\p^2}{\p b^\ga \p b^{\gb}}\,,\\
U_{\ga\dgb} &=&\Big (\ova_{\dgb} - x^\gga{}_\dgb \f{\p}{\p b^\gga } \Big )
\f{\p}{\p b^\ga }\,,\\
U_{\dga\dgb} &=&
\Big (\ova_{\dga} - x^\gga{}_\dga \f{\p}{\p b^\gga } \Big )
\Big (\ova_{\dgb} - x^\ga{}_\dgb \f{\p}{\p b^\ga } \Big )\,,
\eee
\bee
\label{V}
V^{\dga\dgb} &=&\f{\p^2}{\p \ova_\dga \p \ova_{\dgb}}\,,\\
V^{\ga\dga} &=&- \Big (b^{\ga} - x^\ga{}_\dgb \f{\p}{\p \ova_\dgb } \Big )
\f{\p}{\p \ova_\dga }\,,\\
V^{\ga\gb} & = &
\Big (b^{\ga} - x^\ga{}_\dga \f{\p}{\p \ova_\dga } \Big )
\Big (b^{\gb} - x^\gb{}_\dgb \f{\p}{\p \ova_\dgb } \Big )\,,
\eee
\bee
U_{ij} = \f{\p^2}{\p \bar{\phi}^i \p \bar{\phi}^j}\,,\qquad
U^{ij} = \bar{\phi}^i \bar{\phi}^j\,,
\eee
\bee
\label{R}
R_{\ga i}  &=&\f{\p^2}{\p b^\ga \p \bar{\phi}^{i}}\,,\\
R_{\dga i}  &=& \Big (\ova_{\dga} - x^\gga{}_\dga \f{\p}{\p b^\gga } \Big )
\f{\p}{ \p \bar{\phi}^{i}}\,,\\
R^{\ga i}&=& \Big (b^{\ga} - x^\ga{}_\dgb \f{\p}{\p \ova_\dgb } \Big )
\bar{\phi}^i\,,\\
R^{\dga i}&=& -\f{\p}{\p \ova_\dga}
\bar{\phi}^i\,.
\label{RP}
\eee

To obtain the variation $\delta C(\ova ,b,\bar{\phi}|x)$,
one has to apply these generators to $C(\ova ,b,\bar{\phi}|x)$.
Application of the formulas (\ref{cc}) and (\ref{ccc}) to
$\delta C(\ova ,b,\bar{\phi}|x)$ then gives the variation of the
particular dynamical higher spin fields.
The rule is that whenever the second
derivative $\f{\p^2}{\p b^\ga \p \ova_\dga } (C)$ appears, it has to be
replaced by the space-time derivative $\p_\un$ according to (\ref{deqr}).
As a result, a parameter of the higher spin conformal transformation
$\gep (a,b;\phi,\bar{\phi} |x )$ polynomial in $\ova$ and $b$  generates
a local transformation of a dynamical field
with a finite number of derivatives. In particular,
the usual $su(2,2;\N)$ conformal
transformations and their extension to the $osp(2\N, 8)$ transformations
contain at most first space-time derivatives of the dynamical fields.
Thus, $osp(2\N ,8)$ is shown to act by local transformations on the
massless fields of all spins in four dimensions.
That $osp(2\N ,8)$  must act on the $4d$ massless fields
was emphasized by Fronsdal \cite{Fp}. The reformulation of
the higher spin dynamics in terms of the flat sections of
the Fock fiber bundle allows us to derive simple and manifestly
local explicit formulas (\ref{U})-(\ref{RP}).

Analogously, one can derive from (\ref{dec})
the transformation laws for the
higher spin gauge symmetries associated with the whole
infinite-dimensional superalgebra \hsa$\!\!.$
Note that the specific form of the dependence on the
space-time coordinates $x^{\ga}{}_{\dgb}$ originates from the choice of
the gauge function (\ref{gfc}). The approach we use is applicable
to any other coordinate system
and conformally flat background (for example, $ AdS_4$).
Also, let us note that it is straightforward to realize $osp(L,8)$
supersymmetry with odd $L$ by starting with the Clifford algebra with
an odd number of generating elements. The
reason we mostly focused on the case $L=2\N$ was that we
started with $su(2,2;\N)$. For general $L$ the
maximal conformal embedding is $su(2,2;\half [L] ) \subset osp(L ,8)$.

\section{Unfolded Field Theory and Quantization}
\label{Field Theory - Doubleton Duality}

The formulation of the  higher spin dynamics
proposed in this paper operates
in terms of the Fock module $F$ over $su(2,2)$
induced from the vacuum (\ref{vac}).
This Fock module is analogous to the Fock
module $S$ over $su(2,2)$ that contains all
irreducible $4d$ massless unitary representations  of the
conformal algebra called doubletons in \cite{dou}.
In fact, $S$ is the so-called singleton module over $sp(8)$
that decomposes into irreducible doubleton modules over $su(2,2)$.
The difference is that the $sp(8)$ singleton module $S$ is unitary
while the Fock module $F$ is not. That there exists a mapping
between the doubleton and field-theoretical representations
of the conformal (or $AdS$) algebra was originally shown
in \cite{Gun}. The goal of this section is to demonstrate that,
analogously to the $3d$ case considered in \cite{SV},
in our approach the
duality between the two pictures has the simple interpretation
of a certain  Bogolyubov transform. Remarkably, this form of duality
is coordinate independent. The coordinate dependence
results from the gauge choice (\ref{wvg}) that fixes
a particular form of the background gravitational field.

That the module (\ref{PC}) is non-unitary is obvious from
 the fact that, as a result of the Lorentz invariance of
the vacuum $\vac$, the set of component fields (\ref{cdef})
decomposes into
the infinite sum of finite-dimensional representations of the
noncompact $4d$ Lorentz algebra $o(3,1)$.
(Recall that noncompact semisimple Lie algebras do not admit
finite-dimensional unitary representations.)
Also this  is in
agreement with the fact that the conjugated vacuum $\bvac$
(\ref{bvac}) is different from $\vac$.

The unitary Fock module $S$ over $sp(8)\supset su(2,2)$ is built
in terms of the oscillators
\be
\label{uosc}
[ e_{\nu A} \,, e_{\mu B} ]_* =0\,,\qquad
[ f^\nu_{A} \,, f^\mu_{B} ]_* =0\,,
\qquad  [ e_{\nu A} \,, f^\mu_B ]_* = \delta_{\nu}^\mu \kappa_{AB}\,,
\ee
where
$\mu,\nu=1,2$; $ A,B = 1,2,$ and
$\kappa_{11}=1$, $\kappa_{22}=-1$, $\kappa_{12}=\kappa_{21}=0$.
The oscillators obey the Hermiticity  conditions
\be
\label{bog1}
( e_{\nu A} )^\dagger = f^\nu_A \,.
\ee
The unitary Fock vacuum $\uvac$ is defined as
\be
\label{Uvac}
 e_{\nu1}* \uvac = 0\,, \qquad
 f^\mu_{2}* \uvac = 0\,,
\qquad \uvac * f^\nu_1 = 0\,,
\qquad \uvac * e_{\mu 2} = 0\,.
\ee
The compact subalgebra $u(2)\oplus u(2)$ of $u(2,2)$ is spanned by
\be
\tau_{A\nu}{}^\mu =e_{A\nu} f_A^\mu \qquad (A=1,2\quad
\mbox{no summation over}\quad A)\,.
\ee
Noncompact generators of $su(2,2)$ are
\be
t^-_\mu{}^\nu =
e_{1\mu} f_2^\nu\,, \qquad t^+_\mu{}^\nu =  e_{2\mu} f_1^\nu\,.
\ee
(Recall that we  use the Weyl star product notation, i.e. all bilinears
listed above are elements of the star product algebra.)
The superextension is trivially achieved by requiring
 \be
\phi_i * \uvac =0\,, \qquad  \uvac * \bar{\phi}^j =0\,.
\ee

The relationship between the two sets of oscillators is
\bee
\label{bog2}
e_{1,1} = \frac{1}{\sqrt{2}} (a_1 +i \ova_{\dot{2}})\,,\qquad
e_{2,1} &=& \frac{1}{\sqrt{2}} (\ova_{\dot{1}} +i a_{{2}})\,,\nn\\
e_{1,2} = \frac{1}{\sqrt{2}} (a_1 -i \ova_{\dot{2}})\,,\qquad
e_{2,2} &=& \frac{1}{\sqrt{2}} (\ova_{\dot{1}} - i a_{{2}})\,,
\eee
\bee
\label{bog3}
f^1{}_{1} = \frac{1}{\sqrt{2}} (b_2 +i \ovb_{\dot{1}})\,,\qquad
f^2{}_{1} &=& \frac{1}{\sqrt{2}} (\ovb_{\dot{2}} +i b_{{1}})\,,\nn\\
f^1{}_{2} = \frac{1}{\sqrt{2}} (- b_2 +i \ovb_{\dot{1}})\,,\qquad
f^2{}_{2} &=& \frac{1}{\sqrt{2}} (- \ovb_{\dot{2}} +i b_{{1}})\,.
\eee

The unitary Fock vacuum is realized  in terms of the
star product algebra (\ref{prod}) as
\be
\label{uvac}
\uvac =2^{4-\N} \exp{2\Big (
-e_{1\nu} f_1^\nu -  e_{2\nu} f_2^\nu
+\phi_i \bar{\phi}^i }\Big)\,.
\ee
The unitary left and right Fock modules $S$ and $\bar{S}$
built from the vacuum $\uvac$
identify with the direct sum of all superdoubleton
representations of $su(2,2)$ and their conjugates.
As in the non-unitary case, the irreducible components are singled out
by the condition (\ref{eig}).
In the unitary basis,  $N_0$ has the form
\be
N_0 = e_{\nu A} f^{\nu}{}_B \kappa^{AB} \,.
\ee
The Fock space $S$ forms a unitary module over $sp(8)$
called singleton. It contains two irreducible components spanned by
even and odd functions, respectively.

The dependence on the space-time coordinates of the elements of the
field $|\Phi (x) \rangle $ is determined completely by the equation
(\ref{ffe}) in terms of its value at any fixed point $x_0$. This means
that the module $|\Phi (x_0) \rangle $ contains the complete
information on the on-mass-shell dynamics of the 4d conformal fields.
Analogously, the doubleton module contains complete
information on the (on-mass-shell) quantum states of the
corresponding free field theory.
Let us note that the two types of modules have different
gradations associated with the respective definitions of the
creation and annihilation oscillators. In the unitary
case the gradation is induced by the $AdS$ energy operator which,
together with the maximal compact subalgebra, spans the grade zero
subalgebra. In the field-theoretical case the gradation is induced
by the $o(1,1)$ dilatation generator which together with the
Lorentz algebra spans the (non-compact) grade zero subalgebra.

 We conclude that
there is a natural duality between the field-theoretical module
$F$ used in the unfolded formulation of the  conformal dynamics
and the unitary module $S$. This duality has
the simple form of the Bogolyubov
transform (\ref{bog2}), (\ref{bog3}). As a result, although being
unitary inequivalent, the modules associated with the classical
and quantum pictures become equivalent upon complexification.
The important consequence of this fact is that the values of the
Casimir operators of the symmetry algebras in the two pictures coincide.
Indeed, the values of the Casimir operators in the corresponding
irreducible representations (e.g., of $sp(8)$ in $F$ or $S$) are determined
by the fact of the realization of the algebras in terms of oscillators
rather than the particular conditions (\ref{Uvac}) or (\ref{xx})
on the vacuum state.
The duality map between the field-theoretical picture and the
unitary picture is essentially the quantization procedure.
The two modules are unitary inequivalent because the
 respective classes of functions associated with solutions
of the field equations are different.
We believe that this phenomenon is quite general, i.e. the
unfolded reformulation of  dynamical systems in the form of some
flatness (i.e., covariant constancy and/or zero-curvature) conditions
will make the duality between the classical and quantum descriptions
of  dynamical systems manifest for the general case.
Hopefully, the  Bogolyubov transform
duality between the classical and quantum
field theory descriptions can eventually shed some more
light on the nature of quantization and the
origin of quantum mechanics.

The classical-quantum duality of
the unfolded formulation of field-theoretical equations
allows a simple criterion for the compatibility of a field-theoretical
system with consistent
quantization. Namely, if a non-unitary module that
appears in the unfolded description of some classical dynamics
admits a dual unitary module with the same number of states
(i.e., generated with the same number of oscillators) we interpret
this as an indication that the dynamical system under
consideration admits a consistent quantization. Since
every dynamical system admits some unfolded formulation,
this provides us with a rather general criterion. Moreover,
this technique can be used in the opposite direction to derive
field-theoretical differential equations compatible with
unitarity such as those associated with the cohomology group
$H^1 (\sigma_- )$ of the unfolded systems that admit consistent
quantization. We now apply this idea to the derivation
of the compatible with unitarity $sp(2M)$ invariant equations
in generalized space-times.

\section{Conformal Dynamics in $osp(L,2M)$ Superspace}
\label{CD}
The unfolded formulation of the field-theoretical
dynamical systems allows one to extend the equations to
superspace and spaces with additional coordinates in a rather
straightforward way.  In this section
we apply this formalism to the $4d$ $\N$-extended superspace
and to superspaces with ``central charge coordinates" in four
and higher dimensions.  As a result, we shall be able to
formulate appropriate equations of motion in  generalized
(super)spaces. The
 manifest Bogolyubov transform duality between the field-theoretical picture
and the singleton pictures will guarantee that the proposed equations in the
generalized space-times correspond to the unitary quantum picture.

The main idea is simple. In the section
\ref{$4d$ Conformal Field Equations} we have shown that
the dynamics of $4d$ free massless fields is described
in terms of the generating function
$|\Phi (\ova ,b,\bar{\phi} |x)\rangle$
satisfying (\ref{ffe}). The equation (\ref{ffe}) can be interpreted
in two ways. The $\sigma_- -$picture
used in the section \ref{Generalities} implies
that (\ref{ffe})  imposes the
equations (\ref{4deq}) associated with the first
cohomology group $H^1 (\sigma_- )$
on the dynamical fields
associated with the  cohomology group $H^0 (\sigma_- )$.
All other  (auxiliary) components in
$|\Phi (\ova ,b,\bar{\phi} |x)\rangle$ are expressed via space-time
derivatives of the dynamical fields by virtue of (\ref{xeq}).
The $d-$picture used in the section \ref{Generic Solution}
implies that the equation(\ref{ffe}) allows one to reconstruct
the $x-$dependence of
$|\Phi (\ova ,b,\bar{\phi} |x)\rangle$ in terms of the
``initial data"
$|\Phi (\ova ,b,\bar{\phi} |x_0 )\rangle$ taken at some
particular point of space-time $x_0$. The $d-$picture is local.

Suppose now that we have a manifold $M^{p,q}$ with
a larger set of $p$ even and $q$ odd coordinates
$X^\uA$ that contains the original $4d$
coordinates $x^\un$ as a subset, i.e.$X^A = (x^\un,  y^\unu )$,
where $y^\unu$ are additional coordinates. Let $\hat{d}$
be the de Rahm differential on $M^{p,q}$
\be
\hd = dX^\uA \f{\p}{\p X^\uA}=dx^\un \f{\p}{\p x^\un}
+dy^\unu \f{\p}{\p y^\unu} \q \hd^2 =0
\ee
and $\hat{\go}_0$ be a zero-curvature connection in the
(appropriate fiber bundle over) $M^{p,q}$
\be
\hat{\go}_0(a,b,\phi,\bar{\phi}|X) =dX^\uA
\hat{\go}_{0\,\uA} (a,b,\phi,\bar{\phi}|X) \q
d\hat{\go}_0=\hat{\go}_0*\wedge \hat{\go}_0
\ee
such that its pullback to the original $4d$ space-time $M^4$
equals to the $4d$ connection $\go_0$, i.e.
\be
\hat{\go}_{0\,\un} (a,b,\phi,\bar{\phi}|X) =
{\go}_{0\un} (a,b,\phi,\bar{\phi}|x) \,.
\ee
Replacing the $4d$ equation (\ref{ffe}) with
\be
\hd |\Phi \rangle -\hat{\go}_0  * |\Phi \rangle =0 \q
|\Phi \rangle =| \Phi  (\ova,b,\bar{\phi}|x,y) \rangle
\label{effe}
\ee
one observes that the extended system is formally consistent,
while its restriction  to $M^4$ coincides with the original
system (\ref{ffe}). As a result, it turns out that
the system (\ref{effe}) is equivalent to the original $4d$
system (\ref{ffe}) at least locally in the additional coordinates.
 Indeed, as is obvious in the $\hd-$picture,
the equations in (\ref{effe}) different from those in (\ref{ffe})
just reconstruct the dependence of
$|\Phi  (\ova,b,\bar{\phi}|x,y)\rangle$
on the additional coordinates  $y^\unu$
of the $4d$ field $|\Phi (\ova,b,\bar{\phi}|x,y_0)\rangle $
for some $y_0$ (e.g, $y_0=0$).
Let us note that to link the global symmetries
associated with the Lie superalgebra in which $\hat{\go}_0$ takes
its values to the symmetries of the extended space $M^{p,q}$
one has to find such an extension of the space-time
that a frame field in the
generalized space-time is invertible.
In the $\sigma_- -$picture this means that the cohomology group
$H^r (\sigma_- )$ is small enough. An important example of the
application of the proposed scheme is the usual superspace. An
additional simplification here is due to the fact that the
extension along supercoordinates is always global because
superfields are polynomial in the odd coordinates.

The extension of the unfolded dynamical
equations discussed in this section has some similarity
to the ``Group Manifold Approach" developed in the context of
 supersymmetry and supergravity (see \cite{GMA} and references
therein). As we shall see, the maximal natural extension of the
space-time corresponds to the situation when coordinates of the
extended space are associated with all generators of the gauge
Lie superalgebra that underlies the unfolded formulation.

\subsection{Superspace}
\label{Superspace}

As a useful illustration let us embed the
$4d$ dynamics of massless fields into superspace.
We introduce anticommuting coordinates $\theta^{\uga}_{\ui}$ and
$\bar{\theta}_\udgb^\uj$ associated with the $Q-$supersymmetry
supergenerators $Q^i_\ga$ and $\bar{Q}^\dgb_j$, so that
$X=(x,\theta,\bar{\theta})$ (to simplify formulas, in the rest of
this section we shall not distinguish between the underlined
and fiber indices).  The vacuum
connection 1-form satisfying the zero-curvature equation
(\ref{ezc}) can be chosen in the form
\be
\label{sgoo} \hat{\go}_0 =\left( dx^\ga{}_\dgb
+\half\Big((1 +\gga )d\theta^\ga_i \bar{\theta}_\dgb^i
+(1-\gga)d\bar{\theta}^i_\dgb \theta^\ga_i \Big )\right )
a_\ga\ovb{}^\dgb +d\bar{\theta}_\dgb^i
\ovb^\dgb \phi_i +d\theta^\ga_i a_\ga \bar{\phi}^i\,,
\ee
where $\gamma$ is an arbitrary parameter.
Spinor differentials $d\theta^\ga_i$ and
$d\bar{\theta}^i_\dgb $ are required to commute to each other but
anticommute to $dx^\ga{}_\dgb$, $\phi_i$, $\bar{\phi}^i$ and the
supercoordinates $\theta^{\ga}{}_{i}$, $\bar{\theta}_\dgb{}^j$.
$\hat{\go}_0$ admits the pure gauge representation,
$\hat{\go}_0 = - g^{-1} *dg$,
with the gauge function $g$ of the form
\be
\label{sg} g= \exp-\left(
\Big( x^\ga{}_\dgb +\half\gga \theta^\ga{}_i \bar{\theta}_\dgb^i
\Big ) a_\ga\ovb{}^\dgb +\bar{\theta}_\dgb^i
\ovb^\dgb \phi_i +\theta^\ga_i a_\ga \bar{\phi}^i \right )\,.
\ee
The dependence on the supercoordinates is reconstructed by the
formula (\ref{sol}) in terms of the initial data fixed at any point in
superspace.

The superfield equations of motion have the form (\ref{effe}).
The superspace formulation however  does not have the
decomposition (\ref{geneq}). Instead it has the
${\bf Z}\times {\bf Z}$ grading
\be
\label{sgm--}
(\hat{d} +\sigma_{--} +\sigma_{-0} +\sigma_{0-} ) | \Phi \rangle =0\,
\ee
associated separately with the elements $a_\ga$ and $\ovb^\dgb$.
This does not affect the interpretation of the dynamical
superfields as representatives of the zeroth cohomology group
$H^0(\sigma_{--},\sigma_{-0},\sigma_{0-} )$ with the
cohomologies of $\sigma_{-0}$ and $\sigma_{0-}$ computed
on the subspace of $\sigma_{--}-$ closed 0-forms on which
$\sigma_{-0}$ and $\sigma_{0-}$ anticommute to zero. As a result,
the dynamical superfields identify with $|\Phi (0,b,0|X )\rangle $ and
 with the field $|\Phi (\ova,0,\bar{\phi}|X )\rangle $
of maximal degree $\N$ in $\bar{\phi}$.
Thus, as expected, the free field dynamics is described by
general superfields carrying external dotted or undotted spinor indices
(contracted with $b^\ga$ or $\ova_\dgb$) that
characterize a spin of a supermultiplet.
Superfields of this type were used in \cite{GGRS} for the description
of on-mass-shell massless supermultiplets in terms of field
strengths. To extend our formalism to
the off-mass-shell description of massless supermultiplets \cite{KS,GKS}
one has to introduce the higher spin superconnections.

The cohomological
identification of the dynamical superspace equations is less
straightforward however in view of (\ref{sgm--}) although
the main idea is still the same: the superspace equations identify
with the null vectors of the operator
$\sigma_{--} +\sigma_{-0} +\sigma_{0-} $. One complication might be
that, as is typical for the superspace approach,
 it may not always be possible to distinguish between
dynamical equations and constraints in the absence of a
clear $\sigma_-$ cohomological interpretation of the dynamical
equations. We hope to come back to the analysis of this
interesting issue elsewhere.

\subsection{$sp(2M)$ Covariant Space-Time}
\label{$sp(2M)$ Covariant Equations}

As shown in  section
\ref{4d Conformal Higher Spin Symmetries}, the set of
$4d$ conformal equations for all spins is invariant under the
$sp(8)$ symmetry that extends the $4d$ conformal symmetry
$su(2,2)$. This raises the problem of
an appropriate extension of the space-time that would
allow $sp(8)$ symmetry in a natural way. The question of
possible extensions of the space-time beyond the
traditional Minkowski-Riemann extension to higher dimension
has been addressed by many authors (see e.g.,
\cite{Fp},\cite{Wein}-\cite{Dev}).
In particular, a very
interesting option comes from the Jordan algebras
\cite{MC,GJ}. However, to the best of our knowledge,
no dynamical analysis of possible equations was done so far.
One important and difficult issue to
be addressed in such an analysis is whether  the
proposed equations give rise to consistent quantum
mechanics, and, in particular, allow one to get rid of
negative norm states.

More specifically, the analysis of $sp(8)$ invariant extended space-time
was originally undertaken by Fronsdal in \cite{Fp} just in the
context of a unified description of $4d$ massless higher spins.
It was argued in \cite{Fp} that the simplest appropriate extension of
the usual space-time is a certain $sp(8)$ invariant ten-dimensional space.
As shown in this section, our approach  leads to the same
conclusion. The new result will consist of the formulation of
compatible with unitarity
local covariant field equations in this generalized space.

The unfolded formulation of the dynamical equations in the
form of covariant constancy conditions  is ideal for the
analysis of this kind of questions for several reasons:
\begin{itemize}
\item It allows                  to define
an appropriately extended space-time in a natural way
via the (locally equivalent) extension of the
known conformal $4d$ equations of motion.
\item It suggests that the resulting equations
are compatible with unitarity once there is  Bogolyubov transform
duality with some unitary module.
\item Starting from the infinite unfolded system of $sp(8)$ invariant
equations of motion (\ref{effe}) we identify the
finite system of $sp(8)$ invariant dynamical differential
equations as the $\sigma_-$
cohomology $H^1 (\sigma_- )$. Being equivalent to the original
$4d$ conformal unfolded system of equations, the resulting $sp(8)$
invariant differential equations inherit all its
properties such as symmetries and compatibility with unitarity.
\end{itemize}

The approach we use is applicable to any algebra $sp(2M)$.
We therefore consider the general case.
In this subsection we suppress the dependence on the Clifford
elements $\bar{\phi}^i$ and $\phi_j$ which are inert in our
consideration of the purely bosonic space. They will play a role
in the superspace consideration of the next subsection.

Let us introduce the oscillators
\be
\label{hosc}
[ \ga_\hga \,,\gb^\hgb ]_* =\delta_\hga^\hgb\q
[ \ga_\hga \,,\ga_\hgb ]_* =0\q
[ \gb^\hga \,,\gb^\hgb ]_* =0\,.
\ee
We still use the Weyl
star product (\ref{prod}) for the oscillators
$\ga_\hga$ and $\gb^\hgb$
instead of $a_\hga$ and $b^\hgb$ but now we allow
the indices $\hga$ and $\hgb$ to range from 1 to $M$ where $M$ is
an arbitrary positive integer. (The normalization factor in (\ref{prod})
has to be changed appropriately: $\pi^8 \to \pi^{2M}$ ).

The generators of $sp(2M)$ are spanned by various
bilinears built from the oscillators
$\ga_\hga$ and $\gb^\hgb$
\be
\label{gen}
T_\hga{}^\hgb = \ga_\hga \gb^\hgb\q
P_{\hga\hgb} = \ga_\hga \ga_\hgb\q
K^{\hga\hgb} = \gb^\hga \gb^\hgb\,.
\ee
We interpret the generators $P_{\hga\hgb}$ and $K^{\hga\hgb}$ as
$sp(2M)$ ``translations" and ``special conformal transformations",
respectively. The $gl(M)$ generators $T_\hga{}^\hgb$ decompose
into the $sl(M)$ ``Lorentz" and $o(1,1)$ ``dilatation" generators
\be
L_\hga{}^\hgb = \ga_\hga \gb^\hgb - \f{1}{M}\delta_\hga{}^\hgb
\ga_\hgga \gb^\hgga\,,
\ee
\be
D = \half \ga_\hga \gb^\hga\,.
\ee
Note that $D$ is the gradation operator
\be
[D, P_{\hga\hgb}]_* = -P_{\hga\hgb}\q
[D, K^{\hga\hgb}]_* = K^{\hga\hgb}\q
[D, L_{\hga}{}^{\hgb}]_* = 0\,.
\ee
$P_{\hga\hgb}$ and $K^{\hga\hgb}$ generate Abelian subalgebras
\be
[K^{\hga\hgb}\, ,K^{\hgga\hat{\delta}}]_* =0\q
[P_{\hga\hgb} \,, P_{\hgga\hat{\delta}}]_* =0\,.
\ee
Together with $sp(2M)$ ``Lorentz rotations", $sp(2M)$
``translations" span the $sp(2M)$ ``Poincare subalgebra"
\be
[ L_{\hga}{}^{\hgb}\,,P_{\hgga\hat{\delta}}]_* =
-\delta_\hgga^\hgb P_{\hga\hat{\delta}}
-\delta_{\hat{\delta}}^\hgb P_{\hga\hgga}
+\f{2}{M} \delta_{\hga}^{\hgb}P_{\hgga\hat{\delta}}\,.
\ee
Analogously,
\be
[ L_{\hga}{}^{\hgb}\,,K^{\hgga\hat{\delta}}]_* =
\delta_\hga^\hgga K^{\hgb\hat{\delta}}
+\delta_\hga^{\hat{\delta}} K^{\hgb\hgga}
-\f{2}{M} \delta_{\hga}^{\hgb}K^{\hgga\hat{\delta}}\,.
\ee

The superextension to $osp(1,2M)$ is achieved by adding the
supergenerators
\be
Q_\hga = \ga_\hga \q S^\hgb = \gb^\hgb\,.
\ee
According to (\ref{gen}), we have
\be
T_\hga{}^\hgb \equiv L_\hga{}^\hgb +\f{1}{M}\delta_\hga^\hgb D
= \half \{Q_\hga, S^\hgb\}_* \,,
\ee
\be
P_{\hga\hgb} = \half\{ Q_\hga ,Q_\hgb\}_* \q K^{\hga\hgb} =\half\{
S^\hga , S^\hgb\} \,.
\ee

To compare with the $4d$ case, let us note that the
operators  $\gb^\hga$ and  $\ga_\hga$
are to be identified with the pairs
$\ova_\dga$, $b^\gb$, and  $a_\ga$, $\ovb^\dgb$, respectively.
The $4d$ notation used so far was convenient in the $su(2,2)$
framework because of the simple form of the operator $N_0$
singling out $su(2,2)$ as its centralizer in $sp(8)$.
Since $N_0$ does not play
a role in the manifestly $sp(2M)$ invariant setting, it is now more
convenient to have a simple form of the gradation operator $D$.

The Hermiticity conditions are introduced via the involution
$\dagger$ as in section \ref{Reality Conditions and Reduction}
with
\be
\label{rcC}
\ga_\hga^\dagger = i C_\hga{}^\hgb \ga_\hgb\q
(\gb^\hga )^\dagger = i C_\hgb{}^\hga \gb^\hgb\,,
\ee
where $C_\hga{}^\hgb$ is some real involutive matrix
(i.e., $C^2 = Id$). In particular, one can fix
$C_\hga{}^\hgb=\delta_\hga{}^\hgb $ that makes all the $sp(2M)$ generators
manifestly real.
For even $M$ we shall sometimes use another form of
$C_\hga{}^\hgb$ analogous to the $4d$ decomposition
of a real four-component Majorana spinor into two pairs
of mutually conjugated complex two-component spinors.
Namely, we decompose $\ga_\hga$ and $\gb^\hgb$ into two pairs of mutually
conjugated oscillators $\ga_\ga$, ${\bar{\ga}}_\dga$ and
$\gb^\ga$, ${\bar{\gb}}^\dga$ with
$\ga,\dga = 1\div \f{M}{2}$.

By analogy with the usual Minkowski space-time we introduce
$\half M(M+1)$  coordinates $X^{\uhga\uhgb}=X^{\uhgb\uhga}$,
de Rahm differential
\be
\hat{d} = dX^{\uhga\uhgb} \f{\p}{\p  X^{\uhga\uhgb}}\q\hat{d}^2 =0
\ee
and flat frame
\be
\hat{\go}_0 =
dX^{\uhga\uhgb} h_{\uhga\uhgb}{}^{\hga\hgb} \ga_\hga \ga_\hgb\,,
\ee
where $h_{\uhga\uhgb}{}^{\hga\hgb}$ is some constant nondegenerate
matrix so that
\be
\hat{d} \hat{\go}_0 = 0\,.
\ee
For example, one can set
\be
\label{hde}
h_{\uhga\uhgb}{}^{\hga\hgb}=\half \Big (
\delta_{\uhga}{}^{\hga}\delta_{\uhgb}{}^{\hgb}
+
\delta_{\uhga}{}^{\hgb}\delta_{\uhgb}{}^{\hga} \Big )\,.
\ee
$\hat{\go}_0$ satisfies the zero curvature equation
\be
\hat{d}\hat{\go}_0 = \hat{\go}_0\wedge * \hat{\go}_0
\ee
because the $sp(2M)$ translations are commutative and, therefore,
$\hat{\go}_0 \wedge * \hat{\go}_0 =0$. The pure gauge representation
(\ref{wvg}) for $\hat{\go}_0$ is given by
\be
g=\exp -
X^{\uhga\uhgb} h_{\uhga\uhgb}{}^{\hga\hgb} \ga_\hga \ga_\hgb\,.
\ee
For (\ref{hde}) we get
\be
\label{ogs}
g=\exp -
X^{\hga\hgb} \ga_\hga \ga_\hgb\,.
\ee

In terms of dotted and undotted indices (for even $M$), there are
$\f{M^2}{4}$ Hermitian coordinates $X^{\ga\dgb}$  and
$\f{M(M+2)}{4}$ coordinates parametrized by the complex matrix
$X^{\ga\gb}$ and its complex conjugate $X^{\dga\dgb}$.
For $M=2$ our approach is equivalent to the
standard treatment of the $3d$ conformal theory with the conformal
symmetry $sp(4)\sim o(3,2)$. Here  $X^{\hga\hgb}$ parametrize
 the three real coordinates. Therefore the $3d$ approach of
\cite{SV} was equivalent to a particular $M=2$ case of the general
$sp(2M)$ invariant approach.
For the  case of $sp(8)$ (i.e., $M=4$), $X^{\ga\dgb}$ identify
with the usual space-time coordinates $x^{\ga\dgb}$ while
$X^{\ga\gb}$ and $X^{\dga\dgb}$ parametrize
six additional real coordinates $y^\unu$. Altogether we have
ten-dimensional extended space in accordance with
the proposal of Fronsdal \cite{Fp}.

Let us now introduce the left Fock module
\be
\label{OPC}
|\Phi(\gb  |X) \rangle =
C(\gb|X )* \vac \,
\ee
with the vacuum state
\be
\label{ovac}
\vac = \exp{-2\ga_\hga \gb^\hga}\,,
\ee
satisfying
\be
\ga_\hga *\vac =0\q \vac*\gb^\hga =0\q \hat{d} \left (\vac
\right ) =0\,.
\ee

The $sp(2M)$ unfolded equation is
\be
\label{offe}
(\hat{d} - \hat{\go}_0 ) * |\Phi (\gb|X)\rangle =0\,.
\ee
It is $sp(2M)$ (in fact, $osp(1,2M)$) invariant according to the general
analysis of the section \ref{4d Vacuum}. Moreover, this equation has the
infinite-dimensional higher spin symmetry $hu(1,1|2M)$.

The duality to the unitary singleton module over $sp(2M)$
in the basis with the real matrix
$C_\hga{}^\hgb=\delta_\hga^\hgb$ (\ref{rcC})
is achieved by the Bogolyubov transform
\be
\gga^-_\hga= \f{1}{\sqrt{2}}(\ga_\hga +i\gb^\hga )\q
\gga^{+\hga}= \f{i}{\sqrt{2}}(\ga_\hga -i\gb^\hga )\,,
\ee
\be
[\gga_\hga^- ,\gga^{+\hgb} ]_* = \delta_\hga^\hgb\q
(\gga^{+\hga} )^\dagger = \gga^-_{\hga}\,.
\ee
The unitary vacuum
\be
\uvac = \exp-2\gga^-_\hga \gga^{+ \hga}
\ee
satisfies
\be
\gga^-_\hga *\uvac =0\q \uvac*\gga^{+\hga} =0\,.
\ee
As a result of this duality, the equation (\ref{offe})
is expected to admit consistent quantization.

The equation (\ref{offe}) has the form
\be
\label{eqc}
\left (\f{\p }{\p X^{\hga\hgb}} - \f{\p^2}{\p \gb^\hga \p \gb^\hgb}
\right ) C(\gb |X) =0\,.
\ee
For the particular case of $sp(8)$, in the sector of ordinary
coordinates $X^{\ga\dgb}$ it reduces to the $4d$ conformal
higher spin equations (\ref{ffe}).
The equation (\ref{eqc}) has the  form (\ref{geneq}) with
\be
\label{sig-}
\sigma_- =- dX^{\hga\hgb}\f{\p^2}{\p \gb^\hga \p \gb^\hgb}\q \sigma_+ =0
\q \D=\hat{d}\,.
\ee
Its content can therefore be analyzed in terms of the $\sigma_-$
cohomology. The cohomology group $H^0 (\sigma_-)$ is parametrized by
the solutions of the equation $\sigma_- (C(\gb,X))=0$ which consists
 of a scalar function $c(X)$ and a linear
function $c_\hga (X) \gb^\hga$. These are the dynamical fields
of the $sp(2M)$ setup. We shall call $sp(2M)$ vectors $c^\hga (X)$
``svectors" to distinguish them from the vectors of the Minkowski
space-time. Svectors are fermions (i.e., anticommuting
fields being spinors with respect to the usual space-time symmetry
algebras). Scalar and svector
form an irreducible supermultiplet of $osp(1,2M)$
dual to its unitary supersingleton representation.

We see that the number of dynamical fields in the $sp(8)$ invariant
generalized space is much smaller than in the standard $4d$ approach.
Instead of the infinite set of $4d$ massless fields of all spins
we are left with only two $sp(8)$ fields, namely, scalar
$c(X)$ and svector $c_\hga (X)$ that form an irreducible
supermultiplet of $osp(1,8)$. {}From this perspective, the
situation in all generalized $sp(2M)$ invariant symplectic
spaces is analogous to that of the $3d$ model of \cite{SV}
containing the massless scalar and spinor being the only
$3d$ conformal fields. The $4d$ fields now
appear in the expansion of the scalar and svector in powers of
the extra six coordinates
\be
\label{cexp}
c(X) =\sum_{m,n}
c(x)_{{\ga_1\gb_1}\ldots {\ga_n\gb_n},{\dga_1\dgb_1}\ldots {\dga_m\dgb_m}}
X^{\ga_1\gb_1}\ldots X^{\ga_n\gb_n}
X^{\dga_1\dgb_1}\ldots X^{\dga_m\dgb_m}\,,
\ee
\be
\label{sexp}
c_{\hat{\gga}} (X) =\sum_{m,n}
c(x)_{\hat{\gga}\,
{\ga_1\gb_1}\ldots {\ga_n\gb_n},{\dga_1\dgb_1}\ldots {\dga_m\dgb_m}}
X^{\ga_1\gb_1}\ldots X^{\ga_n\gb_n}
X^{\dga_1\dgb_1}\ldots X^{\dga_m\dgb_m}\,,
\ee
where $x^{\ga\dgb} = X^{\ga\dgb}$ are the $4d$ coordinates.
It has been argued by Fronsdal \cite{Fp} that such an expansion is appropriate
for the description of the set of all $4d$ massless fields. Another
important point discussed in \cite{Fp} was that the
analytic expansions in the extra coordinates in
(\ref{cexp}) and (\ref{sexp}) are complete in the generalized
symplectic spaces. Once this is true,
the local equivalence of the equation (\ref{offe}) to the original
$4d$ system extends to the full (global) equivalence.

For $sp(2M)$ with $M>4$ the interpretation in terms
of the  Minkowski picture is
less straightforward because the set of hermitian coordinates
$X^{\ga\dgb}$ becomes larger than the usual set of
Minkowski coordinates.  To this end one has to identify
the usual coordinates with the appropriate projection
of $X^{\ga\dgb}$ with the gamma matrices $\Gamma^n_{\ga\dgb}$
that is possible for $M=2^p$. It is not clear however
how important it is at all to describe $sp(2M)$ invariant
phenomena in terms of Minkowski geometry beyond $d=4$.
{}From this perspective, it looks like
the usual Minkowskian supergravity and superstring
models in higher dimensions might be
some very specific reductions of the new class of models
in generalized $sp(2M)$ invariant space-times underlying the
(generalized beyond $d=4$) higher spin dynamics.

Note that, geometrically, the generalized space-time considered
in this section is the coset space $P_M /SL_M$,
where $P$ is the $Sp(2M)$ analogue of the Poincare
group with the generators $L_\hga{}^\hgb$ and $P_{\hga\hgb}$
while $SL_M$ is the $Sp(2M)$ analogue of the Lorentz algebra with the
generators $L_\hga{}^\hgb$ isomorphic to $sl_M ({\bf R})$.
The $sp(2M)$ conformal
transformations of the generalized symplectic space-time
are realized by the following vector fields
\be
\label{PS}
P_{\hga\hgb} = \f{\p}{\p X^{\hga\hgb}}\,,
\ee
\be
T_\hga{}^\hgb =
2 X^{\hgb\hgga} \f{\p}{\p X^{\hga \hgga}}\,,
\ee
\bee
\label{PK}
K^{\hga\hgb}= 4 X^{\hga\hgga}
X^{\hgb\hat{\eta}}\f{\p}{\p X^{\hgga\hat{\eta}}}\,.
\eee

To derive the independent equations on the dynamical
conformal fields $c(X)$ and $c_\hga (X) $
in the $sp(2M)$ invariant conformal space,
the cohomology group $H^1 (\sigma_- )$ has to be studied
for $\sigma_-$ of the form (\ref{sig-}).
An elementary exercise with Young diagrams
shows that $H^1 (\sigma_- )$ is parametrized by the 1-forms that are
either linear or bilinear in the oscillators,
\be
dX^{\uhga\uhgb} h_{\uhga\uhgb}{}^{\hga\hgb}\Big (
F_{\hga\hgb\,,\hat{\gamma}}\gb^{\hat{\gamma}} +
B_{\hga\hgb\,,\hat{\gamma}\hat{\delta}}\gb^{\hat{\gamma}}\gb^{\hat{\delta}}
\Big )\,,
\ee
where
$F_{\hga\hgb\,,\hat{\gamma}}$ has the symmetry properties of
the three-cell hook diagram, i.e.
\be
F_{\hga\hgb\,,\hat{\gamma}}+
F_{\hga\hat{\gamma}\,,\hgb}+
F_{\hgb\hat{\gamma}\,,\hga}=0\q
F_{\hga\hgb\,,\hat{\gamma}}=
F_{\hgb\hga\,,\hat{\gamma}}\,,
\ee
while
$B_{\hga\hgb\,,\hat{\gamma}\hat{\delta}}$
has the symmetry properties of the four-cell square diagram, i.e.,
it is symmetric within each pair of indices
$\hga,\hgb$ and $\hat{\gamma},\hat{\delta}$ and vanishes upon
symmetrization over any three indices,
\be
B_{\hga\hgb\,,\hat{\gamma}\hat{\delta}}+
B_{\hga\hat{\gamma}\,,\hgb\hat{\delta}}+
B_{\hgb\hat{\gamma}\,,\hga\hat{\delta}}=0\,.
\ee
Note that the trivial cohomology class of $H^1 (\sigma_- )$
is parametrized by the totally symmetric (i.e., one-row)
diagrams of an arbitrary length.

This structure of $H^1 (\sigma_- )$ implies that the only nontrivial
differential equations on the dynamical fields
$c(X)$ and $c_\hga (X)$ hidden in the infinite system
of equations (\ref{offe}) are
\be
\label{oscal} \Big (
\f{\p^2}{\p X^{\hga\hgb} \p X^{\hgga\hgd}} - \f{\p^2}{\p X^{\hga\hgga} \p
X^{\hgb\hgd}}\Big ) c(X) =0
\ee
for the $sp(2M)$ scalar and
\be
\label{ofer} \f{\p}{\p X^{\hga\hgb}} c_\hgga(X)       -
\f{\p}{\p X^{\hga\hgga}} c_\hgb(X)      =0
\ee
for the $sp(2M)$ svector.  The equations (\ref{oscal}) and (\ref{ofer}) are
dynamically equivalent to the system of equations (\ref{offe}) and therefore
inherit all symmetries of the latter. Note that in agreement with the analysis
of \cite{SV}, because antisymmetrization of any two-component indices
$\hga$ and $\hgb$ is equivalent to their contraction with $\gep^{\ga\gb}$, for
the case of $3d$ conformal dynamics, the equations (\ref{oscal})
and (\ref{ofer}) coincide with the $3d$ massless Klein-Gordon and Dirac
equations, respectively.  {}From the $4d$ perspective the meaning of the
equations (\ref{oscal}) and (\ref{ofer}) is twofold. They imply that
the expansions (\ref{cexp}) and (\ref{sexp}) contain only totally symmetric
multispinors and that the latter satisfy the $4d$ massless equations.

The infinitesimal global symmetry transformation that leaves the
 equations (\ref{oscal}) and (\ref{ofer}) invariant
is given by the formula (\ref{gs}) with the global symmetry parameter
$\gep_0$ (\ref{e0}). Let us choose the symmetry generating
parameter in (\ref{e0})  in the form
\be
\label{sxi}
\xi (\ga,\gb;h ,j  )
=\xi
\exp(h^\hga \ga_{\hga} +j_\hgb \gb^{\hgb} )\,,
\ee
where $\xi$ is an infinitesimal parameter.
The polynomial  symmetry
parameters can be obtained via differentiation of
$
\xi (\ga,\gb;h, j  )
$
with respect to the commuting ``sources"  $h^\hga$ and
$j_\hga$. Using (\ref{e0}), (\ref{gs}) and the star product
(\ref{prod}) we obtain upon evaluation of the
elementary Gaussian integrals
\be
\gep_0 (\ga,\gb;h,j|X  ) = \xi
\exp(h^\hga \ga_{\hga}+j_\hgb \gb^{\hgb} +2X^{\hga\hgb}j_\hga \ga_\hgb )\,.
\label{sez}
\ee
Substitution of $\gep_0$ into (\ref{gs}) gives
the global higher spin conformal symmetry transformations
induced by the parameter $\xi$ (\ref{xi})
\be
\delta
|\Phi( \gb,|X) \rangle =
\delta C( \gb|X )* \vac \,,
\ee
where
\be
\label{sdec}
\delta C(\gb|X ) = \xi
\exp(j_\hgb \gb^{\hgb}+ \half j_\hgb h^\hgb +X^{\hga\hgb}j_\hga j_\hgb )
C(\gb^\hgga +h^\hgga +2X^{\hgga\hat{\delta }}j_{\hat{\delta}}|X )\,.
\ee
Differentiating with respect to the sources
one derives explicit expressions for the particular
global higher spin conformal transformations.

The physical fields are
\be
c(X) = C(0|X)\q c_\hga (X) = \f{\p}{\p \gb^\hga} C(\gb|X)\Big|_{\gb=0}\,.
\ee
All higher derivatives with respect to $\gb^\hga$ are expressed
via the derivatives in $X^{\hga\hgb}$ by the equation (\ref{eqc}).
For example, for $c(X)$ we obtain
\be
\delta c(X ) = \xi
\exp( \half j_\hgb h^\hgb +X^{\hga\hgb}j_\hga j_\hgb )
C(h^\hgga +2X^{\hgga\hat{\delta }}j_{\hat{\delta}}|X )\,.
\ee

For at most quadratic supergenerators of $osp(1,2M)$ acting on
$C( \gb|X )$  one  finds
\be
\label{XXX}
P_{\hga\hgb} = \f{\p}{\p X^{\hga\hgb}}\,,
\ee
\be
\label{YYY}
T_\hga{}^\hgb = \half \delta_\hga^\hgb +\gb^\hgb \f{\p}{\p \gb^\hga}
+2 X^{\hgb\hgga} \f{\p}{\p X^{\hga \hgga}}\,,
\ee
\bee
\label{ZZZ}
K^{\hga\hgb}= \gb^\hga \gb^\hgb +2 X^{\hga\hgb} +4 X^{\hga\hgga}
X^{\hgb\hat{\eta}}\f{\p}{\p X^{\hgga\hat{\eta}}}
 +2 X^{\hga\hgga}\gb^\hgb \f{\p}{\p \gb^\hgga}
 +2 X^{\hgb\hgga}\gb^\hga \f{\p}{\p \gb^\hgga}\,,
\eee
\be
Q_\hga = \f{\p}{\p \gb^\hga}\,,
\ee
\be
S^\hga = \gb^\hga +2 X^{\hga \hgb} \f{\p}{\p \gb^\hgb}\,.
\ee
{}From here one derives in particular that the fields $c(X)$
and $c_\hga (X)$ form a supermultiplet with respect to the
$Q-$supersymmetry transformation
\be
\label{Qc}
\delta c(x) = \gvep^\hga c_\hga (x)\q
\delta c_\hga (x) = \gvep^\hgb \f{\p}{\p X^{\hga \hgb}} c (x)\,,
\ee
where $\gvep^\hga$ is a $X-$independent global supersymmetry parameter.
The $S-$supersymmetry with a constant superparameter $\gvep_\hga$ has
the form
\be
\delta c(x)= 2\gvep_\hga X^{\hga \hgb} c_\hgb (X) \q
\delta c_\hga (X) = 2
\gvep_\hgga  X^{\hgga \hgb} \f{\p}{\p X^{\hgb\hat{\hga}}} c(X) \,.
\ee
Note that the (symplectic) conformal transformations of the scalar field
are described by the transformations (\ref{XXX})-(\ref{ZZZ}) at
$\beta^\hgb =0$.
The $T$ and $K$ transformation law of the svector $c_\hga$ gets
additional ``spin" terms from the $\gb-$dependent part of the generators.

The $sl_M$ generalized Lorentz transformations with the traceless
infinitesimal parameter $\gvep_\hgb{}^{\hga}$, $\gvep_\hga{}^{\hga} =0$
act as follows
\be
\label{lorsc}
\delta^{lor} c(X) = 2\gvep_\hgb{}^{\hga}
X^{\hgb\hgga} \f{\p}{\p X^{\hga \hgga}}c(X)\,,
\ee
\be
\label{lorfe}
\delta^{lor} c_\hga(X) = 2\gvep_\hgb{}^{\hgd}
X^{\hgb\hgga} \f{\p}{\p X^{\hgd \hgga}}c_\hga (X)
+  \gvep_\hga{}^{\hgb} c_\hgb (X)\,.
\ee
The dilatation transformations associated with the trace
part $D=\half T_\hga{}^\hga$ are
\be
\label{dilsc}
\delta^{dil} c(X) = \gvep
X^{\hga\hgga} \f{\p}{\p X^{\hga \hgga}}c(X) +\f{M}{4} c(X)\,,
\ee
\be
\label{dilfer}
\delta^{dil} c_\hga (X) = \gvep
X^{\hgb\hgga} \f{\p}{\p X^{\hgb \hgga}}c_\hga(X) +
\Big (\f{M}{4}+\half \Big ) c_\hga(X)\,.
\ee

Since the equations (\ref{oscal}) and (\ref{ofer}) are derived from
the unfolded system that admits a dual unitary formulation, they are
expected to admit consistent quantization. In a separate publication
\cite{qeq}, where the equations in the
generalized space-times are studied within the traditional field
theoretical approach, we show that they indeed admit a consistent
quantization.
A nontrivial question for the future is what is a lagrangian
formulation that might lead to the equations (\ref{oscal})
and (\ref{ofer}). It is clear that in order to solve this problem
some auxiliary fields have to be introduced in analogy with the
Pauli-Fierz program \cite{PFi} for the usual higher spin
fields.

\subsection{$osp(L,2M)$ Superspace}
\label{???}

To describe $osp(2\N,2M)$ we re-introduce the Clifford elements
$\phi_i$ and $\bar{\phi}^j$ and add the bosonic generators
(\ref{T}) and (\ref{UVij}) along with the supergenerators
\be
\label{QG}
Q_\hga^i = \ga_\hga \bar{\phi}^i\q
Q_{\hga i} = \ga_\hga {\phi}_i\,.
\ee
\be
\label{SG}
S^\hga{}_i = \gb^\hga \bar{\phi}^i\q
S^{\hga i} = \gb^\hga {\phi}_i\,.
\ee
In particular, the following anticommutation relations are true
\be
\label{QQ}
\{Q_\hga^i\,,Q_\hgb{}_j\} = \delta_j^i P_{\hga\hgb}\q
\{Q_{\hga i}\,,Q_{\hgb j}\} = 0\q
\{Q_\hga^i\,,  Q_\hgb^j\} =0\,,
\ee
\be
\{S^\hga{}^i\,,S^\hgb_j\} = \delta_j^i K^{\hga\hgb}\q
\{S^\hga{}^i\,,S^\hgb{}^j\} = 0\q
\{S^\hga_i\,,  S^\hgb_j\} =0\,.
\ee
We introduce the Grassmann odd coordinates
$\theta^\hga_i$ and $\theta^{\hga i}$
and differentials
$d\theta^\hga_i$ and $d\theta^{\hga i}$
associated with the $Q-$supergenerators. It is convenient to
define the differentials $d\theta^\hga_i$ and $d\theta^{\hga i}$
to commute to each other but anticommute to $dX^{\hga\hgb}$ and
the Grassmann coordinates
$\theta^\hga_i$ and $\theta^{\hga i}$.

The vacuum 0-form is defined as
\be
\hat{\go}_0 = \left ( dX^{\hga\hgb}
+\half \Big (
(1+\gga) d\theta^{\hga}{}^i\theta^\hgb_i
+(1-\gga) d\theta^\hga_i\theta^\hgb{}^i\Big )\right ) P_{\hga\hgb}
+d\theta^{\hga}{}^i Q_\hga{}_i+d\theta^{\hga}_i Q_\hga^i\,.
\ee
The gauge function analogous to (\ref{SG}) is
\be
\label{osg}
g= \exp-\left(
\Big( X^\hga{}^\hgb
+\half\gga \theta^\hga{}_i {\theta}^\hgb{}^i
\Big )
\ga_\hga \ga_\hgb +{\theta}^\hgb{}^i
\ga_\hgb \phi_i +\theta^\hga{}_i \ga_\hga \bar{\phi}^i
\right )\,.
\ee
The left Fock module $|\Phi (\gb,\bar{\phi}|X, \theta)\rangle$
satisfies the $osp(2\N,2M)$ supersymmetric equations
\be
\label{soffe}
(\hat{d} - \hat{\go}_0 )|\Phi (\gb,\bar{\phi}|X,\theta )\rangle=0.
\ee

Let us note that these formulas are trivially generalized to
the case of $osp(L,2M)$ with odd $L$ by writing
\be
Q_\hga^i = \ga_\hga \psi^i\,,\q S^{j\hgb}=b^\hga \psi^j
\ee
with
\be
\{ \psi^i \,,\psi^j \}_* = \delta^{ij}
\ee
so that
\be
\{Q_\hga^i \,,Q_\hgb^j\} = \delta^{ij}P_{\hga\hgb}\q
\{S^{\hga i} \,,S^{\hgb j}\} = \delta^{ij}K^{\hga\hgb}\,
\ee
and
\be
\hat{\go}_0 = \left ( dX^{\hga\hgb}
+\half
 d\theta^{\hga}_i \theta^{\hgb i} \right ) P_{\hga\hgb}
+d\theta^{\hga}_i Q_\hga^i\,,
\ee
\be
g= \exp-\Big ( X^\hga{}^\hgb
\ga_\hga \ga_\hgb +{\theta}^\hgb_i \ga_\hgb \psi^i
\Big )\,.
\ee

The equation (\ref{soffe}) still makes sense with the only comment
that the Fock vacuum has to be defined in such a way that it is
annihilated by the $\half (L-1)$ annihilation Clifford elements
and is an eigenvector of the central element $\psi_1 \ldots
\psi_L$.

\subsection{Higher Spin (Super)Space}
\label{Higher Spin (Super)Space}

One can further extend the base manifold description of the
$osp(L,2M)$ conformal dynamics by introducing the
higher spin coordinates $X^{\hga_1\ldots\hga_{2n}}$
and Grassmann odd supercoordinates
$\theta_i^{\hga_1\ldots\hga_{2n+1}}$
associated with the mutually commuting higher spin
generators
\be
P_{\hga_1\ldots\hga_{2n}} = \ga_{\hga_1}\ldots\ga_{\hga_{2n}}
\ee
and supercharges
\be
Q^i{}_{\,\hga_1\ldots\hga_{2n+1}} ={\psi}^i
\ga_{\hga_1}\ldots\ga_{\hga_{2n+1}}\q
\{ \psi^i \,,\psi^j \}_* = \delta^{ij}\,,
\ee
which satisfy the higher spin superPoincare algebra with the
nonzero relationships
\be
\{Q^i{}_{\,\hga_1\ldots\hga_{2n+1}} \,,
Q^j{}_{\,\hgb_1\ldots\hgb_{2m+1}}\} =
\delta^{ij} P_{\hga_1\ldots\hga_{2n+1} \hgb_1\ldots\hgb_{2m+1}}\,.
\ee
The zero-curvature vacuum 1-form is
\bee
\hat{\go}_0 &=& \sum_n\Big ( \f{1}{(2n)!}       dX^{\hga_1\ldots\hga_{2n}}
P_{\hga_1\ldots\hga_{2n}} +\f{1}{(2n+1)!}
d\theta_i^{\,\hga_1 \ldots \hga_{2n+1}}Q^i_{\,\hga_1\ldots
\hga_{2n+1}}\Big )\nn\\
\ls &+& \half\sum_{q,p}\f{1}{(2p+1)!(2q+1)!}
P_{\hga_1\ldots\hga_{2p+1}\,\hgb_1\ldots\hgb_{2q+1}}
d\theta_i{}^{\,\hga_1 \ldots \hga_{2p+1}}
\theta^{i\,\hgb_1\ldots\hgb_{2q+1}} \,.
\eee

Let us note that the higher spin (super)coordinates
introduced here are to some extent reminiscent of the
$4d$ higher spin coordinates discussed in
\cite{Dev}, although the particular realization is different.  The unfolded
equations of the form (\ref{soffe}) reconstruct the dependence on the
higher spin coordinates in terms of (usual) space-time derivatives of the
massless higher spin fields. In principle, one can extend the formalism
to the maximal case in which every element of the infinite-dimensional
higher spin algebra (say, $hu(m,n|2M)$) has a coordinate counterpart.
This is analogous to the description on the group
manifold. Let us note that any further extension would imply
a degenerate frame field and, therefore does not lead to interesting
equations. The equations with a fewer coordinates corresponding to
reductions to some coset spaces are possible, however. Let us note that
the unfolded formulation in these smaller spaces is
reminiscent of the group manifold approach  \cite{GMA}.

\section{World Line Particle Interpretation}
\label{World Line Particle Interpretation}

Free field equations of motion in the unfolded form
admit a natural interpretation in terms of a world line
particle dynamics. The free field equation (\ref{ffe})
is interpreted as an invariance condition
\be
\label{QPhi}
Q_0|\Phi\rangle =0
\ee
with a BRST operator built from some first-class constraints.
The zero-curvature condition (\ref{ezc}) takes the form
\be
\label{QQ2}
Q_0^2 =0\,.
\ee
To make contact with some world-line particle dynamics one has
to find a world-line model that gives rise to an operator $Q_0$
associated with the unfolded  equations under consideration.
Usually it is a simple exercise.

The literature on  world line (super)particle dynamics appeared
after the classical works \cite{BVH,Cas,BM,BS} is enormous.
The twistor reformulation was
initiated in \cite{Ferber,Shir} and further developed in
\cite{BBCL,BC,STV,ES,Eis,Tow1,FZ}. The idea that additional
(often  called central charge) coordinates have to be introduced
to extend the twistor approach beyond four dimensions was
exploited in \cite{ES,SR,BL,BLS,BDM,bars}.

The $sp(2M)$  invariant equation (\ref{eqc}) can be obtained
as a result of quantization of the following Lagrangian
\be
L = \dot{X}^{\hga\hgb} \ga_\hga \ga_\hgb  +
\ga_\hga  \dot{\gb}^\hga\,,
\ee
where the overdot
denotes the derivative with respect to the world line
parameter. Indeed, the primary constraints are
\be
\label{con}
0=\chi_{\hga\hgb} = \pi_{\hga\hgb} - \ga_\hga \ga_\hgb \,,
\ee
and
\be
\label{con2}
0= \chi_\hga = \pi_\hga - \ga_\hga \q
0=\chi^\hga = \pi^\hga\,,
\ee
where $\pi_{\hga\hgb},$ $\pi_{\hga}$ and $\pi^{\hga}$ are momenta conjugated
to $X^{\hga\hgb},$ $\gb^{\hga}$ and $\ga_{\hga}$, respectively.
The constraints (\ref{con2}) are second-class.
It is elementary to compute the corresponding Dirac brackets.
The only important fact, however, is that within the set  of
variables
$X^{\hga\hgb}$, $\pi_{\hga\hgb}$,
$ \gb^\hga$ and $ \pi_\hga$
the Dirac brackets coincide
with the Poisson ones,
\be
\label{brak}
\{ X^{\hga\hgb}\,,\pi_{\hgga\hat{\delta}}\} =\half \left (
\delta^{\hga}_{\hgga} \delta^{\hgb}_{\hat{\delta}}+
\delta^{\hgb}_{\hgga}\delta^{\hga}_{\hat{\delta}}\right )
\q \{ \gb^\hga\,, \pi_\hgb \} =
\delta^\hga_\hgb\,.
\ee
This allows one to get rid of the variables
$\ga_\hga$ and $\pi^\hga$
expressing them in terms of
$X^{\hga\hgb}$, $\pi_{\hga\hgb}$, $ \gb^\hga$ and $ \pi_\hga$
with the help of the second-class constraints (\ref{con2}).
The leftover constraints (\ref{con}) acquire the form
\be
\label{conm}
0_0 =\chi_{\hga\hgb} = \pi_{\hga\hgb} - \pi_\hga \pi_\hgb \,,
\ee
and are obviously  first-class.
Interpreting the space-time differentials as ghost fields
$c^{\hga\hgb}$ one arrives at the BRST operator
\be
Q=c^{\hga\hgb} \left ( \pi_{\hga\hgb} - \pi_\hga \pi_\hgb \right )
\ee
which, upon quantization,
reproduces the equations (\ref{eqc}) in the form
(\ref{QPhi}).

The superextension is straightforward:
\bee
L &=& \dot{X}^{\hga\hgb} \ga_\hga \ga_\hgb  +
\ga_\hga  \dot{\gb}^\hga -{\bar{\phi}}^i \dot{\phi}_i \nn\\
&+&\dot{\theta}^{\hga i}\left( \ga_\hga \phi_i +
 \half (1+\gga) \theta_i^\hgb \ga_\hga\ga_\hgb\right )
+\dot{\bar{\theta}}{}^\hga_i \left (\ga_\hga\bar{\phi}^i +\half (1-\gga)
\theta^{\hgb i}\ga_\hga \ga_\hgb \right )\,.
\eee
(The variables $\theta^{\hgb i}$, $\phi_i$ and $\bar{\phi}^i$ are
anticommuting and are assumed to have symmetric Poisson brackets $\{,\}$
with their momenta.)
Excluding by virtue of the second class constraints
the variables $\ga_\hga$,
their conjugated momenta $\pi^\hga$ and
the fermionic variables $\phi^i$ with their conjugated momenta,
one is left with the
conjugated pairs of variables ($X^{\hga\hgb}$, $\pi_{\hga\hgb}$),
($\gb^\hga$, $\pi_\hga$), ($\theta^{\hga i}$, $\pi_{\hga i}$)
($\theta^\hga_i$, $\pi_\hga^i$) and ($\bar{\phi}^i$ , $\pi_i$)
and the first-class constraints (\ref{conm}) along with
\bee
\label{oconf}
\chi_{\hga i} = \pi_{\hga i}-
\left( \ga_\hga \pi_i
 +\half (1+\gga) \theta_i^\hgb \pi_\hga\pi_\hgb\right )\,\nn\\
\chi_\hga^i = \pi_\hga^i -
 \left (\ga_\hga \bar{\phi}^i +\half (1-\gga)
\theta^{\hgb i}\pi_\hga \pi_\hgb \right )\,.
\eee
Altogether, these first-class constraints form the  supersymmetry
algebra with the only nonzero relation
\be
\{\chi_{\hga i}\,,  \chi_\hgb^j \} = \delta^j_i \chi_{\hga\hgb}\,.
\ee

Quantum-mechanical models containing
 ``central charge" coordinates associated with
symplectic algebras,  analogous to the
coordinates $X^{\hga\hgb}$, were considered in
 \cite{BL,BLS,BLPS}. However, to the best of our knowledge,
the particular Lagrangians were different from those proposed
above.

Analogously one can consider the model with the
Lagrangian
\bee
\label{4dl}
L &=& \dot{X}^{\ga\dgb} \ga_\ga \ga_\dgb  +
\ga_\ga  \dot{\gb}^\ga +
\bar{\ga}{}_\dga  \dot{\bar{\gb}}{}^\dga
-{\bar{\phi}}^i \dot{\phi}_i \nn\\
&+&\dot{\theta}^{\ga i}\left( \ga_\ga \phi_i +
 \half (1+\gga) \theta_i^\dgb \ga_\ga\bar{\ga}_\dgb\right )
+\dot{\bar{\theta}}{}^\dga_i \left (\bar{\phi}^i\bar{\ga}_\dga
+\half (1-\gga)\theta^{\gb i}\bar{\ga}_\dga \ga_\gb \right )\,
\eee
(hopefully, the overdotted indices cause no confusion with
the world-line parameter derivative).
In the $4d$ case this model gives rise to the
$4d$ conformal equations of motion of the section
\ref{Fock Space Realization}. The $4d$ Lagrangian
(\ref{4dl}) with $\gga =0$ was introduced in \cite{ES} and
was then shown to give rise to the massless equations in \cite{Eis}
(more precisely, the Lagrangians of \cite{ES,Eis} contained
additional constraints  giving rise to the irreducibility
condition (\ref{eig})). The important difference from many other
world-line twistor Lagrangians is that no twistor
relationship between the space-time coordinates and spinor
variables is imposed, which instead are regarded
as independent dynamical variables.

The generalization to the higher spin coordinates is
described by the Lagrangian
\bee
L &=& \sum_n \f{1}{(2n)!} \dot{X}^{\hga_1\ldots\hga_{2n}}
\ga_{\hga_1}\ldots\ga_{\hga_{2n}}+
\ga_\hga  \dot{\gb}^\hga -{\bar{\phi}}^i \dot{\phi}_i\nn\\
&+&\sum_n \f{1}{(2n+1)!}\Big (
\dot{\theta}^{i\,\hga_1 \ldots \hga_{2n+1}}\phi_i \ga_{\hga_1}\ldots
\ga_{\hga_{2n+1}}+
\dot{\theta}_i^{\,\hga_1 \ldots \ga_{2n+1}}\bar{\phi}^i
\ga_{\hga_1}\ldots\ga_{\hga_{2n+1}}\Big )\nn\\
\ls &+& \half\sum_{q,p}\f{1}{(2p+1)!(2q+1)!}
\ga_{\hga_1}\ldots\ga_{\hga_{2p+1}}\,\ga_{\hgb_1}\ldots
\ga_{\hgb_{2q+1}}\nn\\
&{}&\times \Big (
(1+\gga)
\dot{\theta}^{i\,\hga_1 \ldots {\hga_{2p+1}}}
\theta^{\hgb_1\ldots\hgb_{2q+1}}_i
+(1-\gga)
\dot{\theta}_i{}^{\,\hga_1 \ldots {\hga_{2p+1}}}\theta^{i\,\hgb_1\ldots
\hgb_{2q+1}}\Big )\,.
\eee

All world-line  particle Lagrangians discussed in this
section have the general form
\be
\label{genpl}
L = \dot{X}^\uA \hat{\go}_{0\uA} (\ga,\gb,\phi, \bar{\phi}|X )
+\ga_\hga  \dot{\gb}^\hga -{\bar{\phi}}^i \dot{\phi}_i \,,
\ee
where $X^{\uA}$ denotes the whole set of the supercoordinates
while
$d{X}^\uA\hat{\go}_{0\uA} (\ga,\gb,\phi, \bar{\phi}|X ) =
\hat{\go}_0 (\ga,\gb,\phi, \bar{\phi}|X )$ is some vacuum 1-form
satisfying the zero curvature equation (\ref{ezc}). Let us stress
that (\ref{ezc}) is supposed to be
true in the quantum regime, i.e. with respect to the star product.
In the classical approximation, the star product
has to be replaced by the Poisson (in fact, Dirac) brackets, which
usually makes sense for the BRST interpretation (\ref{QQ2})
of the vacuum condition (\ref{ezc}) but not necessarily for the
dynamical field equations in the essentially ``quantum" form
(\ref{QPhi}).

The constraints have the form
\be
\chi_\uA = \f{\p}{\p X^{\uA}} - \go_{0\,\uA} (X)\,.
\ee
They are first class as a consequence of the
flatness condition (\ref{ezc}). We see that this construction  indeed
leads to the BRST realization of the linearized unfolded dynamics
in the form (\ref{QPhi}), (\ref{QQ2}).

The Lagrangian (\ref{genpl}) is universal in the sense that it
gives rise to the unfolded equations of the conformal
higher spin fields interpreted as the first-class constraints
independently of a particular form of the vacuum 1-form
$\hgo_0$ once it satisfies the zero-curvature equation
(\ref{ezc}). The ambiguity in
$\hgo_0$ parametrizes the ambiguity in the choice of particular
geometry and/or
coordinate system.  For the particular case of the conformal algebra, any
conformally flat geometry is available. For example, $AdS_4$ geometry is
described by the vacuum 1-form (\ref{goads}). Note that it is well-known
that the zero curvature (=left invariant Cartan) forms play the key role in the
formulation of the (super)particle and brane dynamics because they possess
necessary global symmetries (namely, the symmetries (\ref{ipgs})).
The fact that  $\hgo_0$ satisfies the zero-curvature condition
guarantees that the Lagrangian (\ref{genpl}) has necessary local symmetries
(i.e., first class constraints).
Note that some examples of the zero-curvature 1-forms of
$osp(1,2n)$ are given in \cite{BLPS}.

Applying the Stokes theorem and using the
zero-curvature condition for $\hat{\go}_0$,
the particle action (\ref{genpl}) can be rewritten in the
topological string form as an integral over a two-dimensional
surface bounded by a particle trajectory and parametrized by $\sigma^l$
\bee
\label{sact}
S&=&\int_{\Sigma^2}\Big (
\hat{\go}_0 (\ga,\gb,\phi, \bar{\phi}|X )* \wedge
\hat{\go}_0 (\ga,\gb,\phi, \bar{\phi}|X )
+d\ga_\hga \wedge  d{\gb}^\hga -d{\bar{\phi}}^i\wedge d{\phi}_i
\nn\\
&+&
\Big ( d\ga_\hga \f{\p}{\p \ga_\hga }+
d\gb^\hga \f{\p}{\p \gb^\hga }       +
d\phi_i \f{\p}{\p \phi_i }            +
d\bar{\phi}^i \f{\p}{\p \bar{\phi}^i } \Big ) \wedge
\hat{\go}_0 (\ga,\gb,\phi, \bar{\phi}|X ) \Big ) \,,
\eee
where
\be
\hat{\go}_0 (\ga,\gb,\phi, \bar{\phi}|X )
=   d\sigma^l \f{\p X^\uA}{\p \sigma^l} \hat{\go}_{0\,\uA}
(\ga,\gb,\phi, \bar{\phi}|X ) \,,
\ee
\be
d\ga_\ga = d\sigma^l \f{\p \ga_\ga}{\p \sigma^l}\q
d\gb^\ga = d\sigma^l \f{\p \gb^\ga}{\p \sigma^l}\q
d\phi_i = d\sigma^l \f{\p \phi_i}{\p \sigma^l}\q
d\bar{\phi}^i = d\sigma^l \f{\p \bar{\phi}^i} {\p \sigma^l}\,.
\ee

Keeping in mind that the theory of higher spin gauge fields is expected
to be related to a symmetric phase of the superstring theory,
let us speculate that this topological action can be related
to the superstring actions in the framework of some perturbative
expansion relevant to the usual string picture which, however, breaks
down the manifestly topological form of the whole action defined in the
generalized target superspace.

Note that the  action (\ref{sact}) can be rewritten as
\be
\label{strac}
S= \int_{\Sigma^2}\Big (
w_0 (\ga,\gb,\phi, \bar{\phi}|X ) * \wedge \,
w_0 (\ga,\gb,\phi, \bar{\phi}|X ) \Big ) \,,
\ee
where
\be
w_0 = \go_0 +d\gb^\hga \ga_\hga - d\ga_\hga \gb^\hga
+d\phi_i \bar{\phi}^i  -{\phi}_id\bar{\phi}^i          \,
\ee
with the convention that the star product
in (\ref{strac}) acts on the components of the differential form
$w_0$
but not on the differentials $d\ga_\hga$, $d\gb^\hga$, $d\phi_i $
and $d\bar{\phi}^i$.

A few comments are now in order.

It is important that the ``quantization" is performed in such a way
that the equations like (\ref{eqc}) contain
 differential rather than  multiplication  operators. This allows
to express all higher order polynomials in the twistor variables
via higher space-time derivatives of the physical fields. Note
that the ``coordinate" and ``momentum" representations are not equivalent
in the framework of  nonunitary modules underlying (classical)
field theory dynamics. One way to see this is to observe that the
dualization (Fourier transform) that interchanges twistors with their
conjugate momenta, interchanges the translations
$P_{\hga\hgb}$  and the special conformal transformations
$K^{\hga\hgb}$.

The conversion procedure
applied in the paper \cite{BLS} to get rid of the complicated
second-class constraints in a particle-type
twistor model  based on the $osp(2,8)$ superalgebra
led to the first-class constraints analogous to
(\ref{conm}) and (\ref{oconf}) modulo
exchange of the twistor variables with their momenta.
It was concluded in \cite{BLS}
that the space of quantum states of this model consists of
the massless fields of all spins (every spin appears in
two copies), i.e. it is identical  to the spectrum of massless
higher spin fields associated with the simplest $\N =2$
supersymmetric conformal higher spin algebra $hu(1,1;8)$.
Since the approach of \cite{BLS} was insensitive to the
difference between the twistor variables and their momenta,
the one-to-one
correspondence between the spectrum of $4d$ massless
higher spin excitations found in
\cite{BLS} and in this paper is not accidental.

Beyond the linearized approximation the
world-line quantum mechanical interpretation of the unfolded dynamics
becomes less straightforward.
Indeed, the interaction problem consists of searching for a
consistent deformation of the equations (\ref{QPhi}) and (\ref{QQ2})
with nonlinear contributions
to the equations (\ref{QPhi}) and (\ref{QQ2}) both from
the dynamical gauge fields $\go = \go_0 +\ldots$
and from the ``matter sector" $|\Phi\rangle $.
The modification due to the gauge fields admits interpretation
in terms of connection in the linear fiber bundle  with the
module $F$ of quantum states $|\Phi\rangle$ as a fiber.
The terms nonlinear in $|\Phi\rangle $ can, however, hardly be
interpreted in the usual quantum mechanical framework that
respects the superposition principle. Relaxing the superposition
principle one  arrives at the standard setting of the free
differential algebras (\ref{dFw}),
suggested originally in \cite{Ann} for the
analysis of the higher spin problem.
The world-line particle models
can be useful for the second quantized description
of the nonlinear higher spin dynamics in a form
analogous to the open string field theory functional of Witten
\cite{OST}
\be
S=\langle \bar{\Phi}| Q A | \Phi\rangle  + S^3\,,
\ee
where $A$ is some insertion needed to make the quadratic part
well-defined  and $S^3$ is the interaction part to be determined.

\section{$AdS / CFT$ Correspondence}
\label{$AdS / CFT$ Correspondence}

The classical result of Flato and Fronsdal
\cite{FF} states that the tensor product of two
singleton representations of $sp(4)$ amounts to the direct sum of all
unitary representations of $sp(4)$ associated with the massless
fields of all spins in $AdS_4$. Once the unfolded formulation of
massless dynamics exhibits Bogolyubov duality with the unitary
representations, there must be some field-theoretical dual version
of the Flato-Fronsdal theorem. This was confirmed by the analysis
of the boundary current and bulk gauge field representations
in \cite{FerFr}. It was also observed in
 \cite{SV} that for the $3d$ conformal theory
there is  one-to-one correspondence between
the tensor product of $3d$ boundary fields and the set of the
$AdS_4$ bulk higher spin gauge fields (and, therefore, conserved
higher spin currents of \cite{KVZ}). This statement is supposed to
underly the $AdS_4/CFT_3$ duality in the framework of the higher spin
theories.

A $AdS_5$ analog of the Flato-Fronsdal
theorem suggests \cite{Gun0,GM} that
the double tensor products
of the doubleton representations contain all massless unitary
representations of the $AdS_5$ algebra $o(4,2)\sim su(2,2)$.
It is interesting to see what is a field-theoretical
counterpart of this statement.

Consider first the self-conjugated
massless supermultiplets with  $\ga =0$. The corresponding
conformal higher spin gauge symmetry algebra
\hsao was argued
in the section
\ref{Self-conjugated  Supermultiplet Reduction}
to be spanned by the
elements of the star product algebra (i.e., polynomials of
oscillators) that commute to $N_\N$, are identified modulo
$N_\N$ and satisfy the reality  condition
(\ref{reco}). On the other hand, the elements
of the tensor product of the space of
states satisfying (\ref{eig}) with its conjugate
\be
\label{E}
E_{12}= |\Phi_1 \rangle \otimes \langle \bar{\Phi}_2 |\,
\ee
automatically satisfy these condition as a consequence of
(\ref{eig})
\be
[N_\N ,E_{12} ]_* = 0 \,,\qquad N_\N  *E_{12} = 0\,.
\ee
Also, it is consistent with the conditions
(\ref{reco}), (\ref{arec}) after appropriate specification
of the action of the involution and antiautomorphism on the
tensor product symbol to compensate the insertions of the
products of elements $\phi_i$ or $\bar{\phi}^j$ in (\ref{fac1}),
 (\ref{fac2}).

The $4d$ conformal higher spin algebras
$hu_0 (2^{\N -1}, 2^{\N -1}|8)$
(being isomorphic to $AdS_5$ higher spin algebras) and their
further orthogonal or symplectic subalgebras
identify with the (sub)algebras of endomorphisms of the module $F_0$
spanned by the states satisfying (\ref{eig}) at $\ga =0$. Discarding the
(sometimes important) normalizability issues, it is a matter of basis
choice to realize this algebra in terms of either elements (\ref{E})
or polynomials of the
star product algebra\footnote{Note that the action of the operator
  (\ref{E}) in
$F_0$ is described by an infinite matrix having at most a  finite number of
non-zero elements, while the polynomial elements of the star product algebra
have the Jacobi form with an infinite number of non-zero elements but at most a
finite number of non-zero diagonals. This means that a polynomial in the star
product algebra is described by an infinite sum in the basis (\ref{E}).}.
Therefore, the tensor product of the $4d$ matter multiplets has the same
structure as the $AdS_5$ higher spin algebra $hu_0 (2^{\N -1}, 2^{\N -1}|8)$
in which $5d$ higher spin gauge fields
(equivalently, conserved currents \cite{KVZ}) take their values.
This fact provides the field-theoretical counterpart of the
statement on the structure of the tensor products of the unitary
doubleton representations of \cite{Gun0,GM}.
The non self-conjugated case is analogous except
that the reduction condition (\ref{arec}) is
inconsistent with the eigenvalues $\ga \neq 0$ and, therefore,
the subalgebras of symplectic and orthogonal types allowed for the
self-conjugated case are not allowed for  $\ga \neq 0$.  Note,
that it is also possible to relax the condition (\ref{arec}) in the
self-conjugated case which effectively
leads to the doubling of the self-conjugated multiplets.

Thus, the higher spin AdS/CFT correspondence suggests
that the $AdS_5$ higher spin algebra associated with
the boundary self-conjugated matter supermultiplets is one of
the subalgebras (\ref{sub1}) or (\ref{sub2}).
{}From the $AdS_5$ bulk perspective only
the purely bosonic case $\N =0$ was analyzed so far
at the level of cubic Lagrangian interactions \cite{5d}.
This analysis matches
the  consideration of the present paper since it has been
shown in \cite{5d} that the $AdS_5$ higher spin gauge fields
associated with the algebras $hu_0 (1,0|8)$ and $ho_0 (1,0|8)$
allow consistent cubic interactions.
In the forthcoming paper \cite{AV} we shall show that the same is
true for the $\N =1$ supersymmetric case. In the both of these cases
the situation is relatively simple because the corresponding
$AdS_5$ higher spin gauge fields correspond to the
totally symmetric (spinor-)tensors representations of the $AdS_5$ algebra.
The gauge field formalism for the description of these fields suitable for the
higher spin gauge problem in any dimension was elaborated in \cite{LV,VF}.
As shown in the recent publication
\cite{SSd} (see also \cite{5d}) for the bosonic case and in \cite{A} for the
fermionic case, the sets of gauge fields associated with the
$\N = 0 $ and $\N = 1$ $AdS_5$ higher spin algebras are just
what is expected from the perspective of
the approach of \cite{LV,VF}. Namely, the infinite-dimensional higher spin
algebras decompose under the adjoint action of its $AdS_5$ subalgebra
$o(4,2)\sim su(2,2)$ into an infinite sum of finite-dimensional
representations associated with various two-row tensors or
spinor-tensors of $o(4,2)$ \cite{5d,A}.

Starting from $\N = 2$ representations of $o(4,2)$
with three rows appear, however. The simplest way to see
this is to observe that, for increasing $\N$, the restriction
$[N_\N , f]_* =0$ on the types of
representations of $su(2,2)$ contained in the star product
element $f(a,b;\phi,\bar{\phi})$ becomes less and less
restrictive, rather imposing some
relationships between the types of $su(2,2)$ tensors and $u(\N )$
tensors in the supermultiplet. One can see that the three-row diagrams
of $so(4,2)$ appear whenever a number of oscillators $a$ and
$b$ in $f$ can differ by two that is possible starting from
$\N\geq 2$. As a result, the $\N \geq 2$
$AdS_5$ higher spin gauge theories based on the algebras
$hu_0 (2^{\N-1},2^{\N-1}|8)$ and their further reductions
will contain some mixed symmetry gauge fields. Because
the $5d$ massless little Wigner algebra is $o(3)$,
in the $5d$ flat space such fields
are equivalent to the usual totally symmetric higher spin fields.
This is not true however in the $AdS_5$ space where the
systematics of the massless fields is different from the flat one
\cite{BMV}. In particular, to every two-row  Young diagram
of the maximal compact algebra $so(4)\sim su(2) \oplus su(2)$
corresponds a particular  $AdS_5$ massless field. In the flat
limit such fields decompose into a
number of the flat space massless fields, each equivalent
(dual) to some totally symmetric field in the flat space.
So far, no systematic approach to the mixed symmetry
higher spin fields in the $AdS$ space has been elaborated
in the covariant approach underlying the unfolded dynamics,
although a considerable progress in the flat space was
achieved in \cite{labas,BPT}. To extend the results of \cite{5d,AV} to
$\N\geq 2$ it  is  first of all necessary to develop a
gauge formulation of the higher spin fields
carrying mixed symmetry representations of the
$AdS$ algebras $o(d-1,2)$. This problem is now under investigation.
\footnote{Note that after the original version of this paper has
been sent to {\tt hep-th} some progress in this direction was
achieved in \cite{SS2,M,BSi}.}

It is tempting to speculate that once the two-row mixed symmetry higher
spin $AdS_5$ fields are included, the condition that the elements of
the higher spin algebra have to commute to $N_\N$ can be relaxed and
the (symplectic)
$AdS_5$ dual versions of the $osp(2\N,8)$ conformal boundary models
might be constructed. These models are
expected to contain all types of gauge (massless) fields
in $AdS_5$ having one of the algebras $hu(n,m|8)$, $ho(n,m|8)$
or $husp (n,m|8)$ as the  gauge algebra.
In that case we arrive at the remarkable possibility that
the generalized $sp(8)$ $AdS_5/CFT_4$
correspondence will relate the bulk model that describes
$AdS_5$ massless fields of  all spins (types)
to the boundary conformal model describing $4d$ conformal massless
fields of all spins.  This is the $AdS_5/CFT_4$ analogue of the
Flato-Fronsdal theorem relating the $AdS_5$ massless fields with
the tensor product of the $sp(8)$ (super)singletons.
Once such a generalization is really possible, this
would lead to the surprising
conclusions on the higher spin AdS/CFT correspondence which, in fact,
would imply the space-time dimension democracy.

Indeed, the following  extension of the
Flato-Fronsdal theorem is likely to take place
\be
\label{S0S}
S_{osp(L,2M)}\otimes S_{osp(L,2M)}
=\sum_s m^{0\,s}_{osp(L,2M)} = S_{osp(2L,4M)}\,,
\ee
where $S_{osp(L,2M)}$ denotes the (super)singleton representation of
${osp(L,2M)}$ while $m^{0\,s}_{ osp(L,2M)} $ denotes all massless unitary
representations of $ osp(L,2M) $ characterized by the spin parameters
$s$.  The chain of identities can be continued
to the left provided that $L$ and $M$ are even. For $L=2^q$ and
$M=2^p$ the chain continues down to the case of $sp(2)$ or $sp(4)$
with the appropriate truncations in the Clifford sector associated
with $L$ if necessary (say, by singling out the bosonic or fermionic
constituents of some of the supersingletons). Since the tensor product of the
representations is associated with the bilinear currents built from the
boundary fields, the conclusion is that the generalized (symplectic)
higher dimensional models are expected to be dual to the nonlinear
effective theories built from the lowest dimensional (higher spin)
models.

The equality $S_{osp(L,2M)}\otimes S_{osp(L,2M)}= S_{osp(2L,4M)}$
is obvious because the supersingleton $S$ of $S_{osp(L,2M)}$
is the Fock module generated by $L$ fermionic and
$2M$ bosonic oscillators. By definition, its tensor square is the
Fock module generated by two sorts of the same oscillator which is
equivalent  to the supersingleton module of $S_{osp(2L,4M)}$. The fact that
$S_{osp(2L,4M)}$ is equivalent to the sum of all massless representations
of $ osp(L,2M)$ is less trivial. It is in agreement with the definition
of masslessness given by G\"unaydin in \cite{Gun0,GM}. However, to make
this definition consistent with the property that massless fields
(except for scalar and spinor) are gauge fields, it is necessary
\cite{metr,BMV,FeF} to prove that the unitary representations corresponding
to the gauge massless higher spin fields are at the boundary of the unitarity
region of the modules of $ osp(L,2M)$, thus being associated with
certain singular vectors, decoupling of which manifests the gauge
symmetry\footnote{Let us note that beyond the  $AdS_3$ and $AdS_4$ cases
in which the symplectic and orthogonal tracks are
equivalent, the concept of masslessness may be different for,
say, symplectic $AdS^M$ (i.e., symplectic bulk)
and orthogonal $AdS_d$ (i.e., usual bulk) theories. For the
symplectic algebras $osp(L,2^p)$, which contain the (maximally
embedded) $AdS$ subalgebras $o(2p,2)$ or $o(2p +1,2)$, the values of the
lowest energies  compatible with  unitarity are expected
to be higher than the lowest energies of
the lowest weight unitary  representations of their $AdS_d$ subalgebras.
(I am grateful to R.Metsaev for a useful discussion of this point.)
In fact, there is nothing special in this phenomenon, which
would just signal that the extra symplectic dimensions play a real
role. Very much the same story happens for the usual $AdS_d$
algebras $o(d-1,2)$: the lowest energies of $o(d-1,2)$
are higher than those of its
lower-dimensional subalgebra $o(d-2,2)$ \cite{metr,FeF}.
Let us note that from this perspective,
G\"unaydin's identification \cite{Gun0}
of the massless representations
of $AdS$ algebras with those that belong to the tensor product
of the singleton and doubleton representations is likely to be true
for the symplectic track rather than for the usual $AdS_d$ one.}.

As conjectured in \cite{W,Sun}, the higher spin  AdS/CFT
correspondence is expected to correspond to the limit $g^2 n \to 0$,
where $n$ is the number of the boundary conformal supermultiplets
and $g$ is the boundary coupling constant.
An interesting related question is whether the free $4d$ boundary
theories
discussed in this paper admit nonlinear deformations preserving the
infinite-dimensional higher spin symmetries
$hu (2^{\N -1},2^{\N -1} |8)$ (or some their deformations).
Let us argue that, most probably,
these symmetries are broken by interactions to lower
symmetries\footnote{I am grateful to E.Witten for a stimulating
discussion of this issue.}. One argument is based on the knowledge
\cite{more,Gol} of the full nonlinear higher spin dynamics in $d=4$.

The $4d$ conformal system analyzed in the section
\ref{$4d$ Conformal Field Equations} describes a set of $4d$
massless fields of all spins which decomposes into
irreducible representations of $sp(8)$. {}From \cite{FV1,more} it is known
that such sets of massless fields admit consistent interactions in
$AdS_4$ but not in the flat space. The interactions are introduced
in terms of higher spin potentials rather than in terms of the
(higher spin) Weyl tensors discussed in this paper. This breaks
down the usual $4d$ conformal symmetry. The breaking of the
conformal symmetry is expected to be
of spontaneous type via vacuum
expectation values of certain auxiliary fields needed to provide
consistent higher spin dynamics. This results in the
$CFT_d \rightarrow  AdS_d$  deformation with respect to the $d$-dimensional
coupling constant $g^2 \sim \Lambda \kappa^{d-2}$, where $\Lambda$ and
$\kappa$ are the
cosmological constant and the gravitational constant, respectively.
Let us note
that by $AdS_d$ we assume the universal covering of the anti-de Sitter
space-time
(or an appropriate its symplectic generalization discussed below), which
although being curved, is topologically $R^d$. Note that since the $AdS_d$
geometry is conformally flat it should be possible to have the $AdS/CFT$
correspondence with the boundary $CFT$ theory formulated in the $AdS$
space-time
rather than in the Minkowski one.  (To the best of our knowledge this
technically more involved possibility has so far not been investigated.) As a
result, in the framework of the higher spin gauge theories the $
AdS^{2M}/{AdS^M} $ correspondence is likely to replace
the usual $AdS/CFT$ correspondence.  (Abusing notation, we use the notation
$AdS^M$ for the generalized space-time identified below with
$Sp(M)$).  Perhaps the breakdown of the
conformal higher spin symmetries down to the $AdS$ higher spin
symmetries can be
understood as a result of the conformal anomaly arising in the process of
approaching the conformal infinity \cite{MS}.  Also, let us note that since the
AdS/CFT correspondence refers to the conformal boundary of the bulk space a
possible argument against the infinite chain of AdS/CFT dualities
(\ref{chain})
based on the fact that the boundary of a boundary is zero is
avoided just because the full conformal symmetry is expected to be broken.

The formulation of the full nonlinear $4d$  higher spin dynamics of
\cite{more} provides us with some hints on the character of
the breaking of the ``conformal" $sp(2M)$ by interactions.
The full nonlinear formulation of the $4d$ higher spin dynamics
was given in terms of the star product algebra with eight spinor
generating elements. In other words, the construction of \cite{more}
has explicit local $hu(1,1|8)$ symmetry (extension to $hu(n,m|8)$ is
trivial by considering matrix versions of the model along the lines
of \cite{KV1}) and, in particular, $sp(8)$
as its finite-dimensional subalgebra. These local symmetries are
broken by the vacuum expectation values of the auxiliary fields
called $S$ to $hu(1,1|4) \oplus hu(1,1|4) $ containing
$sp(4)\oplus sp(4)$. (The doubling is due to the Klein operators.)
The lesson is that the higher spin interactions
 break the conformal $hu(n,m|2M)$ symmetry to
$hu(n^\prime,m^\prime|M )$ (for $M$ even).

This conclusion fits the analysis of the embedding
of the generalized $AdS$ algebra into the conformal algebra $sp(2M)$.
Indeed,
to embed the usual $AdS_d$ algebra $o(d-1,2)$ into the $d-$dimensional
conformal
algebra $o(d,2)$ one identifies the $AdS_d$ translations with a mixture of the
translations and special conformal transformations in the conformal algebra
$P^a_{AdS_d} =P^a_{d-conf} +\lambda^2 K^a_{d-conf}$.  Commutators of such
defined $AdS_d$ translations close to $d-$dimensional Lorentz transformations
$L^{ab}$. $P^a_{AdS}$ and $L_{ab}$ form the $AdS_d$ algebra
$o(d-1,2)\subset o(d,2)$ (cf. eq. (\ref{goads}) for the particular case of
$AdS_4$). This embedding breaks down the explicit $o(1,1)$
dilatational covariance because it mixes
the operators $P^a$ and $K^a$, which have different scaling dimensions.

Let us now analyze the analogous embedding of a generalized $AdS$ subalgebra
into the             conformal algebra $sp(2M)$ in the
$\half M(M+1)$ -dimensional generalized  space-time.
 Since we want to keep the dimension of the generalized
space-time intact, the generators of $AdS$ translations have to be
of the form $P^{AdS}_{\hga\hgb} =P_{\hga\hgb} +\lambda^2
\eta_{\hga\hgb\,\hgga\hgd}K^{\hgga\hgd}$ with some bilinear
form  $\eta_{\hga\hgb\,\hgga\hgd}$. To allow embedding of the
generalized $AdS$ superalgebra into the conformal superalgebra with the
$AdS$ supercharges being a mixture of the $Q$ and $S$ supercharges of
the conformal algebra i.e.,
$Q_\hga^{AdS}=Q_\hga +\lambda  V_{\hgb\hga}S^\hgb$,
the form  $\eta_{\hga\hgb\hgga\hgd}$ has to
have  a factorized form, i.e.,
\be
\label{PPK}
P^{AdS}_{\hga\hgb} =P_{\hga\hgb} +\lambda^2
V_{\hga\hgga}V_{\hgb\hgd}
K^{\hgga\hgd}\,
\ee
with some antisymmetric bilinear form $V_{\hga\hgb}$. We require
$V_{\hga\hgb}$ to be non-degenerate, which assumes that $M$ is even
(for the case of odd $M$ the resulting generalized $AdS$ algebra
is not semisimple). The commutator of such defined generalized
$AdS$ translations closes to the subalgebra $sp(M)$ of
$sl_M\subset sp(2M)$, which leaves invariant
the antisymmetric bilinear form $V_{\hga\hgb}$.
The full generalized $AdS$ subalgebra is
\be
sp(M)\oplus sp(M) \subset sp(2M)\,.
\ee
Its Lorentz subalgebra  $sp^{l}(M)$ identifies with the diagonal $sp(M)$
while $AdS$ translations belong to the coset space
$sp(M)\oplus sp(M) /sp^{l}(M)$. For $M=2$ one recovers the usual
$3d$ embedding $o(2,2)\sim sp(2)\oplus sp(2) \subset sp(4)\sim o(3,2)$.
Analogously to the $3d$ case, the $\half M(M+1)$-dimensional
space-time where the generalized $AdS$ algebra
$sp(M)\oplus sp(M) $ acts is  the group manifold $Sp(M)$,
while the two $sp(M)$ symmetry algebras are induced by its
left and right actions on itself. In particular, the ten-dimensional
generalized space-time associated with the $AdS$ phase of $4d$ massless
fields of all spins is $Sp(4)$.

Thus, for even $M$ we obtain that the $AdS$ subalgebra of the conformal
algebra acting in the $\half M(M+1)$ dimensional space-time is
isomorphic to the
direct sum of the two conformal algebras of the generalized
$\f{M(M+2)}{8}-$dimensional space-time. The process can be continued
to the lower dimensions provided that $M=2^q$. Let us note that
the fact that the $AdS$ algebra is semisimple may indicate that the
corresponding reduced higher spin algebra acquires more supersymmetry.
A particularly nice scenario would be that the $AdS$ reduction
of the $\N$ extended
conformal higher spin algebra $hu(2^{\N-1},2^{\N-1}|2 M )$
in the generalized space-time $Sp(M)$ is
$hu(2^{\N},2^{\N}|M )$. In that case, the extension
$\N-1 \to \N$ would imply the doubling of the even sector because
of the new unimodular bosonic element
$\phi_{\N+1} \bar{\phi}^{\N+1}$ built from the additional Clifford
elements\footnote{
Let us note that this scenario does not sound too unrealistic taking
into account that
the reduction of the star product sector algebra allows for introducing
unimodular Klein-type operators built from the bosonic oscillators.}.
Then, the breaking of the free field conformal
symmetry $hu(2^{\N-1},2^{\N-1}|2 M )$ to the $AdS^M$ one by interactions
would imply
\be
\label{chhu}
hu(2^{\N-1},2^{\N-1}|2M )\to hu(2^{\N},2^{\N}| M ) \,,
\ee
which would lead along with (\ref{S0S})
to the chain of correspondences
\be
\label{chc}
\ldots AdS^{2M,\N }/ AdS^{M,\N+1} \to AdS^{M,\N +1}/AdS^{\half M,\N+2 }\to
AdS^{\half M,\N+2 }/AdS^{\f{ M}{4},\N+3 }\to \ldots\,
\ee
with
$hu(2^{\N-1},2^{\N-1}|M )$ realized either
as $AdS^M$ higher spin  algebra in the
 generalized space-time $Sp(M )$ or as the conformal higher spin algebra
in the generalized space-time $Sp(\half M )$.
(We assume that the proposed scenario is going to
work when all relevant algebras
$sp(m)$ have even $m$. The chain of correspondences
continues down to the lowest dimensions for $M=2^q$.)

Let us stress  that this  scenario is mainly
justified by the observation
that the full $4d$ $sp(8)$ conformal massless higher spin
multiplets expected to provide a boundary theory for the $AdS_5$
bulk higher spin theory have the spectra
identical to those of $AdS_4$ higher spin theories thus requiring
the deformation of the flat boundary geometry to the anti-de Sitter
one in the phase
with higher spin interactions respecting higher spin gauge symmetries.
(Note that an analogous
 observation was made in the paper \cite{SV}, where it was found
that the $3d$ free conformal higher spin theories describe the
same sets of massless fields (scalar and spinor)
as the nonlinear
$AdS_3$ higher spin theories constructed in \cite{AdS3}.)
Since the standard $AdS / CFT$ duality
is a nonlinear mapping of the
bulk fields to the boundary currents bilinear in the elementary
boundary fields \cite{FFA,AdS/CFTW},
the resulting generalized space-time dimension democracy
suggests the chain of nonlinear mappings with the
higher dimensional models equivalent to the theories of
composite fields of the lower dimensional ones.

The suggested chain of $AdS/CFT$ correspondences can
be true for the full higher spin theories based on the algebras \hsa
(say, as conjectured in (\ref{chc})) but makes no sense
for the reduced theories based on the algebras \hsan
and their further reductions. Once  a theory is
truncated to the subsector singled out by the condition (\ref{eig}),
say, to the $\N=4$ $SYM$ theory, no full $CFT_d \rightarrow
AdS_d$ deformation correspondence can be expected.
In other words, a reduction to the usual space-times and symmetries
is expected to break the correspondence chain (\ref{chain})
at some point. Note that such a reduction is likely to result
from some sort of spontaneous breaking mechanism with a Higgs
type field $\varphi$ acquiring a vacuum expectation value
proportional to $N_\N$, thus reducing the full
higher spin algebra \hsa to its subalgebra being the centralizer
of $N_\N$.

The argument against a nontrivial deformation of the full higher spin
conformal symmetries to a nonlinear theory, based on the peculiarities
of the higher spin dynamics requiring the $AdS$ geometry,
fails to be directly applicable to the models based on
the algebras $hu_0 (2^{\N -1},2^{\N -1} |8)$ with $\N \leq 4$ because the
corresponding supermultiplets do not contain higher spins.
Although the problem is formulated
in flat space-time, this possibility is not strictly speaking
ruled out by
the Coleman-Mandula type theorems because conformal theories do not admit
a well-defined S-matrix. Indeed, some of the models of interest were
argued to admit a conformal quantum phase compatible with the
higher spin symmetries \cite{Ans}.
In the framework of the classical field theory,
 the problem is to find a nonlinear deformation of
the equations (\ref{ezc}), (\ref{ffe}) with the matter field
$|\Phi\rangle$ contributing to the right-hand-side of the equation
(\ref{ezc}). Provided that the deformed equations are formally consistent,
the appropriately deformed conformal higher spin symmetries will
also be guaranteed. It is {\it a priori} not excluded
that a nonlinear deformation of the
free field dynamics compatible with the conformal higher spin
symmetries, e.g. in the $\N=4$ SYM theory, may exist.
On the other hand, a potential difficulty is due to a possible
anomaly resulting from the divergency
of the star product of the Fock vacua (\ref{vac}) and (\ref{bvac})
in the $|\Phi \rangle * \langle {\Psi} |$ - like bilinear terms.

\section{Conclusions and Outlook}
\label{Conclusions and Perspectives}

In this paper, the infinite-dimensional $4d$ conformal higher spin
symmetries have been realized on the free massless
supermultiplets. The explicit form of the higher spin
transformations is given by virtue of the unfolded formulation
of the equations of motion for massless fields in the form of
the covariant constancy condition for the appropriate Fock fiber bundle.
Such conformal field theories were
conjectured to be boundary
dual to the nonlinear higher spin theories
in the bulk $AdS$ space \cite{KVZ}. In \cite{W,Sun} it was conjectured
that the $AdS/CFT$ duality for higher spin theories
should correspond to the weak coupling
regime $g^2 n\to 0$ in the superstring picture. To
verify these conjectures it is now necessary to build the $AdS_5$ higher spin
theory. A progress in this direction for the simplest case of $\N =0$
higher spin theory is achieved in \cite {5d} where some cubic
higher spin interactions were found. To extend these results to
$\N \neq 0$ and, in particular, to $\N=4$ it is necessary to
extend the results of \cite{5d} to higher spin gauge fields carrying
mixed symmetry massless representations of the $AdS_5$ algebra
associated with the two-row Young diagrams.

As a by-product of our formulation it is shown how the $osp(L, 8)$
symmetry is realized on the infinite set of free
boundary conformal fields of all spins.
This result is interesting from various points of view. First of all,
it was argued by many authors \cite{Tow,Ba,GM,G,FP}
that the algebras $osp(m , 2^n )$ and, in particular, $osp(1,32)$ and
$osp(1,64)$ play a fundamental role for  the $M$ theory interpretation
of the superstring theory. It is usually believed that the
related symmetries are broken by the brane charges.
{}From the results of this paper it follows that the algebras of this type
can be unbroken if an infinite number of massless fields of all spins
is allowed. A natural mechanism of spontaneous breaking
of the symplectic symmetries to the usual ($AdS$ or conformal)
symmetry algebras might result from a scalar
field $\varphi$ in the (bulk or boundary) theory, which acquires a non-zero
vacuum expectation value
$\varphi = N_\N + \ldots$, where $N_\N$ is the operator (\ref{Nb}),
that breaks $osp(\N, 8)$ to $su(2,2| 2\N )$ and the
higher spin algebra \hsa to \hsao.
In that case the breaking of the symmetries associated with the
so called central charge coordinates results from a condensate of
the higher spin fields.

The new equations (\ref{oscal}) and
(\ref{ofer}) for the scalar and
svector (symplectic vector) fields in the
manifestly $sp(2M)$ conformally invariant
$\half M (M+1)-$dimensional extended space-time are formulated.
These equations encode in a concise form the dynamical equations
for all types of massless fields
in the $3d$ and $4d$ cases for $M=2$ and $M=4$, respectively.
Remarkably, the proposed $sp(2M)$ invariant equations are
compatible with unitarity as it follows from the Bogolyubov
transform duality of their unfolded formulation to the unitary singleton
representation of $sp(2M)$. The superextension of these equations is
also given in the form of an infinite chain  of the equations in the
extended superspace associated with $osp(L,2M)$.

This result can affect dramatically
our understanding of the nature of extra dimensions. In fact,
we argue that, from the perspective of the higher spin gauge theory,
 the proposed  symplectic higher
dimensional space-times
have a better chance of describing appropriately
higher-dimensional extensions of the
space-time geometry than the traditional Minkowski extension.
Among other things, this improves the situation with
supersymmetry. Indeed, the main reason why supersymmetry singles
out some particular dimensions in the Minkowski track is that the
dimension of the spinor representations of the Lorentz algebra
increases exponentially with the space-time dimension
(as $2^{[\f{d}{2}]}$) while dimensions of its tensor representations
increase polynomially. This implies mismatch between
the numbers of bosonic and fermionic coordinates, thus singling out
some particular dimensions $d\leq11$
where the number of spinor coordinates is not too high due to imposing
appropriate Majorana and/or Weyl conditions. If our conjecture is true,
the higher-dimensional models considered so far would correspond to some
specific truncations of the hypothetical symplectic theories.
The crucial ingredient underlying the ``symplectic track"
conjecture is  that the generalized symplectic conformal
equations (\ref{oscal}) and (\ref{ofer}) admit consistent quantization.

We argued that the generalized symplectic space-time is
the group manifold $Sp(M)$ that has the conformal
(boundary) symmetry $Sp(2M)$ and $AdS$ (bulk) symmetry
$Sp(M)\times Sp(M)$ ($M$ is even). The generalized superspace
is $OSp(L,M)$. The usual $3d$ case corresponds to the case of
$M=2$, while the usual $4d$ geometry is
embedded into the ten-dimensional generalized space-time $Sp(4)$.
The fact that the generalized space-time is the group manifold is
interesting from various points of view and, in particular, because
the generalized superstring theories may admit a natural formulation
in terms of the appropriate $WZWN$  models.

The algebras $sp (2^p )$ and the related generalized
space-times play a distinguished role in many respects.
The odd elements of $osp(L, 2^p )$ can be interpreted as forming
the spinor representations of the usual Lorentz algebras in
$d=2p$ or $d=2p+1$ dimensional space-times,
so that the theories of this class admit an interpretation in
terms of the usual Minkowski track space-time symmetries and
 supersymmetries. In particular, the generalized space-time
coordinates $X^{\hga\hgb}$ are equivalent to a set
of antisymmetric tensor coordinates $x^{a_1\ldots a_n}$
\be
X^{\hga\hgb} =\sum_{n=0}^{d}
(\Gamma^{\hga\hgb}_{a_1\ldots a_n}+
\Gamma^{\hgb\hga}_{a_1\ldots a_n}) x^{a_1\ldots a_n}
\ee
associated with all those antisymmetrized combinations of the
$\Gamma-$matrices $\Gamma^{\hga\hgb}_{a_1\ldots a_n}$ which are
symmetric in the indices $\hga$ and $\hgb$. The
dynamical equations (\ref{oscal}) and (\ref{ofer}) amount to some sets
of differential equations with respect to the generalized coordinates
$x^{a_1\ldots a_n}$. An interesting possibility consists of the
interpretation of the dynamics of branes in the Minkowski track picture
as point particles in the generalized spaces of the symplectic track.

Another exciting possibility is that in the framework of the full
(i.e., symplectic) higher spin theories the chain of $AdS/CFT$
correspondences can be continued (\ref{chain}) to link together
higher spin theories in symplectic space-times of various dimensions
$\half M(M +1)$ via a nonlinear  field-current correspondence
\cite{FFA,AdS/CFTW}.
The dramatic effect of this would be ``space-time dimension democracy"
establishing duality between higher spin gauge theories in different
dimensions.   Since higher spin gauge theory is expected to
describe a symmetric phase of the theory of fundamental interactions,
like superstring theory and $M$-theory, this would imply
that the analogous dualities are to be expected in the superstring
theory, although in a hidden form as a result of spontaneous breakdown
of the higher spin symmetries and, in particular, the $osp(L ,2M)$
supersymmetry. {}From this perspective the dimensions $M=2^p$ again
play a distinguished role because the analogue of the
Flato-Fronsdal theorem (\ref{S0S}) is expected to be
true for the generalized
space-times $Sp(2^p )$ with all $p$. In other words the
 conjectured chain of dualities links all theories
that admit an interpretation in terms of usual space-time spinors and
tensors to
each other via a nonlinear generalized $AdS/CFT$ correspondence
(\ref{chain}).

\section*{Acknowledgments}
The author is grateful to I.Bandos, I.Bars, M.G\"unaydin, R.Metsaev,
V.Rubakov and E.Witten for useful discussions. I would like to
thank O.Shaynkman for careful reading of the manuscript and helpful
comments. This research was supported in part by INTAS, Grant No.99-1-590,
the RFBR Grant No.99-02-16207 and the RFBR Grant No.01-02-30024.

\end{document}